\documentclass[11pt]{memoir}

\usepackage[utf8]{inputenc}
\usepackage{amssymb,amsmath}
\usepackage{bm}
\usepackage{mathrsfs}
\usepackage[normalem]{ulem}
\usepackage{tikz} 
\usetikzlibrary{calc}
\usetikzlibrary{arrows}

\usetikzlibrary{external}
\tikzset{external/system call={latex \tikzexternalcheckshellescape
    -halt-on-error -interaction=batchmode -jobname "\image" "\texsource";
    dvips -o "\image".ps "\image".dvi; ps2eps "\image.ps" } }

\usepackage{multibib}
\usepackage[pdfauthor={Peder K. Sorensen},
            pdftitle={Three-body dynamics with zero-range potentials},
            pdfsubject={Few-body Physics},
            pdfkeywords={PhD Thesis},
            pdfproducer={Latex with hyperref and bibtex},
            pdfcreator={Latex\, bibtex\, dvips and ps2pdf},
            citecolor=brown,
            ]{hyperref}

\usepackage{memhfixc}

\setsecnumdepth{subsection}
\maxtocdepth{subsection}

\newcommand{\eqr}[1]{eq.~\eqref{#1}}
\newcommand{\eqrs}[1]{eqs.~\eqref{#1}}
\newcommand{\fig}[1]{Figure~\ref{#1}}
\newcommand{\tab}[1]{Table~\ref{#1}}

\captionnamefont{\small}
\captiontitlefont{\small}

\newcommand{\Cs}{${}^{133}$Cs }
\newcommand{\K}{${}^{39}$K }
\newcommand{\Rb}{${}^{85}$Rb }
\newcommand{\Li}[1]{${}^{#1}$Li}

\newcommand{\Vim}{V_\textnormal{imag}}
\newcommand{\rim}{r_\textnormal{imag}}

\newcommand{\op}{\textnormal{open}}
\newcommand{\cl}{\textnormal{closed}}
\newcommand{\p}{\partial}
\newcommand{\rc}{\rho_\textnormal{cut}}
\newcommand{\rec}{\alpha_\textnormal{rec}}
\newcommand{\vdW}{r_\textnormal{vdW}}
\newcommand{\aminus}{a^{(-)}}
\newcommand{\R}{\Upsilon}

\newcites{publications}{List of publications}

\begin{document}

\frontmatter


\begin{titlingpage}
  \centering
  \hrule height 1pt\par
  \vspace{1.0em}\par
  \textbf{\HUGE Three-Body Recombination}\par
  \vspace{0.5em}\par
  \textbf{\HUGE in Cold Atomic Gases}\par
  \vspace{1.0em}\par
  \vspace{1.0em}\par
  \hrule height 1pt\par
  \vspace{4.0em}
  \textbf{\Large Peder Klokmose S{\o}rensen}\par
  \vspace{1.0em}\par
  {\Large Department of Physics and Astronomy\\
  Aarhus University, Denmark}\par
  \strut\vfill\par
  \includegraphics[width=8cm]{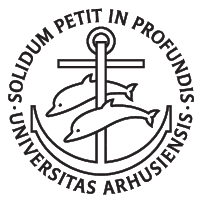}
  \strut\vfill\par
  \textbf{\Large Dissertation for the degree of\\
    Doctor of Philosophy}\par
  \vspace{1.0em}\par
  \Large July 2013

  \newpage
  \phantom{42}
  \vfill
  \begin{flushleft}
    \noindent Copyleft \reflectbox\textcopyright{} 2013
      Peder Klokmose S{\o}rensen\\
    Department of Physics and Astronomy\\
    Aarhus University\\
    Ny Munkegade, building 1520\\
    DK-8000 Aarhus C\\
    Denmark\\

    1st edition, July 2013.


    \vspace{2em}
    Typesetting done using \LaTeX{}, BibTex{} and the memoir class.\\
    Figures made with gnuplot 4.6 and TikZ/PGF.\\
    Printed by SUN-Tryk, Aarhus University.
  \end{flushleft}

  \vspace{2em}
  \begin{center}
    \includegraphics{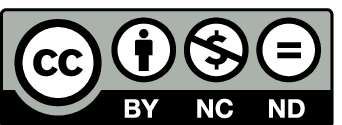}
  \end{center}
  \label{fig:cc}
  \small{This thesis is licensed under a
  \href{http://creativecommons.org/licenses/by-nc-nd/3.0/}{Creative
  Commons Attribution-Non Commercial-No Derivs 3.0 Unported License.}}

\end{titlingpage}

\newpage

\begin{center}
  \framebox{
    \parbox{0.95\textwidth}{
    \noindent This dissertation has been submitted to the Faculty of
    Science and Technology at Aarhus University, Denmark, in fulfillment
    of the requirements for the PhD degree in physics. The work presented
    has been performed under the supervision of Dmitri V. Fedorov and
    Aksel S. Jensen in the period of August 2009 to July 2013.
  }
}
\end{center}
\vfill
\verb'E13:'
\begin{flushright}
  \verb'Fbzr snapl naq vagryyvtrag pvgngvba ol n snzbhf crefba.'\\
  \verb'gur_snzbhf_crefba (n_lrne)'
\end{flushright}

\cleardoublepage

\tableofcontents*




\chapter{Resume / Resumé}
\section{English}
Systems of three particles show a surprising feature in their bound state
spectrum: a series of geometrically scaled states, known as Efimov states.
These states have not yet been observed directly, but many recent
experiments show indirect evidence of their existence via the so-called
recombination process. The theories that predict the Efimov states also
predicts either resonant enhancement of the recombination process or
suppression by destructive interference, depending on the sign of the
interaction between the particles. The theories predict universal features
for the Efimov states, for instance that the geometric scaling factor is
  22.7, meaning that one state is 22.7 times larger than its lower lying
  neighbour state. This thesis seeks to investigate non-universal effects
  by incorporating additional information about the physical interactions
  into the universal theories.

\section{Dansk}
Systemer af tre partikler viser en overraskende effekt i spektret for
bundne tilstande: en række geometrisk skalerede tilstande, kaldet Efimov
tilstande. Disse tilstande er endnu ikke blevet observeret direkte, men
mange nyere eksperimenter viser indirekte evidens for deres eksistens
gennem den såkaldte rekombinationsprocess. Teorierne der forudsiger Efimov
tilstandene, forudsiger også enten resonant forstærkning af
rekombinationsprocessen eller undertryk\-kelse pga. destruktiv
interferens, afhængig af fortegnet på vekselvirkningen mellem partiklerne.
Teorierne forudsiger en række universelle kendetegn, for eksempel at den
geometriske skalafaktor har værdien 22.7, hvilket betyder at en given
tilstand er 22.7 gange større end dens lavest liggende nabotilstand. Denne
afhandling undersøger ikke-universelle effekter ved at inkorporere
yderligere information om de fysiske vekselvirkninger.

\chapter{Acknowledgements}

\begin{flushright}
\emph{"I would like to thank the Academy\ldots. Wait\ldots. What\ldots?\\\
Oh\ldots wrong speech. Let me start over."}
\end{flushright}

\noindent\rule{\textwidth}{0.4pt}

I would first and foremost like to thank my supervisors Dmitri Fedorov and
Aksel Jensen for all the help, support, guidance and advice they have
provided me with during the past four years. Their insight in everything
from the physical understanding of interesting problems to practical
computer implementations over written and verbal presentations to
publishing in the academic community and a multitude of order major and
minor things, has proven invaluable to my studies as a PhD student at
Aarhus University.

I also very much enjoyed working with Nikolaj Zinner with whom I had the
privilege of sharing an office for some time. This has resulted in some
great papers that would probably not have come about without his
contributions.

The rest of the sub-atomics group: Karsten Riisager, Hans Fynbo, Kasper
Lind, Gunvor Koldste, Oleksandr Marchukov, Artem Volosniev and Jakob
Pedersen also deserve honourable mention. Many a lunch break has been
filled out by interesting discussion topics covering most aspects of life
both inside and outside the yellow walls.

A special thank you goes to Artem Volosniev for reading through the
manuscript of this thesis and pointing out all the weird things, minor
mistakes and genuine errors all over the place.

Finally a thank you to my girlfriend Frederikke who has also read through
the manuscript but more importantly helped me keep my spirits high when it
all seemed pointless.

\nocitepublications{*}
\renewcommand\bibname{List of publications}
\bibliographystylepublications{thesisstyle}
\bibliographypublications{data/publications.bib}

\mainmatter

\chapter{Introduction}
\label{introduction1}
\section{Few-body physics and the Efimov effect}
The quantum mechanical three-body problem has been investigated heavily
since the birth of quantum mechanics. As with the classical counterpart
there are no general analytical solutions to the problem. Many features of
the system of three interacting bodies are known nonetheless. For
instance, given a system of three identical interacting particles where
any subsystem of two particles supports a bound state with infinite
scattering length, the total system has a spectrum of infinitely many
bound states with a characteristic scaling relation between successive
states. The size of a state is a factor of $22.7$ times larger than the
previous state and the energy is a factor $22.7^2=515$ times smaller.

This is known as the Efimov effect which was predicted in
1970~\cite{efi70} but remained unobserved for many years. The first
attempts to discover the effect was in nuclear physics, however, without
success~\cite{jen04}. It was in the realm of atomic physics that the first
observation was made in 2006 using a gas of cold Cs
atoms~\cite{kraemer2006,ferlaino2010}.

\section{Cold atoms and recombination}
Cold atoms as a research area has exploded in the past two decades. The
experimental realization of Bose Einstein Condensates (BEC's), a
macroscopic collective of thousands, up to millions \cite{AIP.78.013102}
of atoms at $\mu K$ to nK temperatures, has sparked a revolution in cold
gas physics. An essential experimental tool for this progress is that
of Feshbach resonances \cite{chin2010}, without which cold atomic gas
experiments would probably be quite different today.

\begin{figure}
  \centering
  \tikzsetnextfilename{fig1_1}
  \begin{tikzpicture}[scale=1,inner sep=3mm]
\tikzstyle{particle} = [circle,fill,shading=ball,ball color=blue!50!red]
\tikzstyle{tarrows}  = [->,>=triangle 45,very thick]

\draw[particle] 
                 (0,0)   node[particle] (d1) {}
                 (0,2)   node[particle] (d2) {}
                 (3,1)   node[particle] (d3) {}
                 (7.8,1) node[particle] (c1) {}
                 (8,0.5) node[particle] (c2) {}
                 (10,1)  node[particle] (c3) {};

\draw[thick,->] (d1) -- +(45:1.2cm);
\draw[thick,->] (d2) -- +(-30:1.0cm);
\draw[thick,->] (d3) -- +(175:1.01cm);

\draw[thick,->] ($0.5*(c1)+0.5*(c2)$) -- +(165:1.75cm);
\draw[thick,->] (c3) -- +(20:1.5cm);

\node at (1.5,3) {$A+A+A$};
\node at (9,3) {$A_2+A$};
\draw[very thick] (4,1) -- (5.3,1);
\draw[very thick] (4,0.9) -- (5.3,0.9);
\draw[very thick] (5,0.8) -- (5.5,0.95) -- (5,1.1);
\end{tikzpicture}
  \caption[]{It's all about recombination.}
  \label{figure:1.1}
\end{figure}
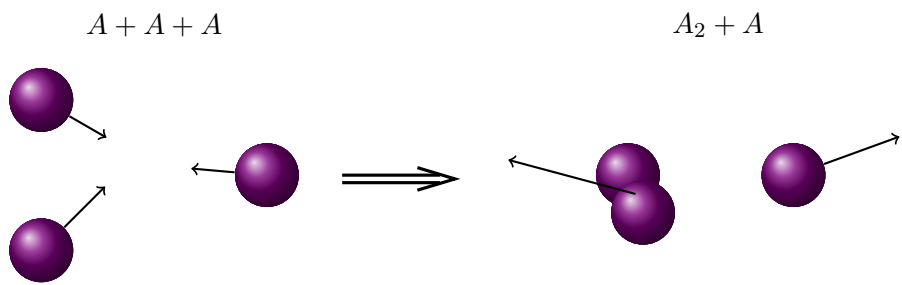
Three-body states are an important part of cold gas atomic physics because
the gases consist of interacting particles. Collisions can lead to loss of
particles via the process known as recombination, illustrated in
\fig{figure:1.1}. Three particles, $A+A+A$, may interact in such a way
that two of them form a bound state, known as a dimer, $A_2$, while the
third atom, $A$, carries away excess energy and momentum. The products
will have an increased kinetic energy due to the increased binding energy
in the dimer. The result is that all three atoms are lost from the trap
since it is typically to shallow to withhold the energetic products. Thus
particles are lost from the trap leading to a finite lifetime of such
experimental set-ups \cite{esry1999, Macek1999, braaten2001}.

\section{The Efimov effect as indirect observation}
Recombination leads to a loss rate of the form $\dot n = -\rec n^3$ where
$n$ is the number density of the particles, the dot denotes the temporal
derivative and $\rec$ is known as the recombination coefficient. The power
3 is due to the number of triplets in a gas of $N$ atoms scaling as $N^3$
for large $N$. The Efimov effect is observed as a characteristic series of
peaks and troughs in the recombination coefficient as a function of the
scattering length, see for instance Figures
\ref{figure:3.8} to \ref{figure:3.11} for positive scattering length and
\fig{figure:5.3} for negative scattering length. This is caused by the
existence of the before mentioned Efimov states.

Similar processes can occur for four, five, six, \ldots particles.
However, such recombination events are progressively less likely to occur
since more particles are required to be in the same small volume of space.
Correspondingly, a much higher density of particles is expected to be
required in order to see these higher order effects, which is not the case
for current experiments.

\section{Feshbach resonances}
The interaction strength between atoms in cold gases is usually described
by the scattering length $a$, which is the lowest order measure of the
strength of the interaction potential. Using only the scattering length is
mostly sufficient when dealing with low-energy $s$-wave scattering. For a
given atomic system the scattering length is fixed by nature. However, an
experimental tool exists that allows experimentalists to tune the
scattering length to practically any desired value. This tool is known as
a Feshbach resonance. It utilises that the difference in magnetic momenta
in two different reaction channels makes it possible to tune the energy
difference between these channels by applying an external magnetic field
to the system. If the energy of the interacting particles is close to the
energy of a bound state in the upper of the interaction channels a
resonant behaviour is seen where the scattering length diverges. In
principle any desired value of the scattering length can be obtained, both
positive and negative, by tuning the magnetic field,

From dimensional analysis one can easily show that the recombination
coefficient, $\rec$, goes roughly as $a^4$, however, with some
modifications depending on the sign of $a$. Hence the form $\rec=C(a)a^4$
where $C(a)$ is a log-periodic function of $a$. For three identical
particles the function $C(a)$ obeys $C(22.7a)=C(a)$. A fundamental
conclusion of this thesis is that the factor $22.7$ changes when the
effective range is included.

The sign of the scattering length $a$ indicates whether a given two-body
system is governed by attractive or repulsive interactions and hence
whether the system supports a so-called shallow dimer. A shallow dimer is
characterized by the binding energy of the order $a^{-2}$. This is the
case when $a$ is positive. As mentioned in the first paragraph, when a
two-body subsystem of three particles supports a dimer at zero energy,
i.e. $a=\infty$, the Efimov effect occurs. This is seen in the
recombination coefficient as a characteristic series of troughs in the
spectrum at certain values of the scattering length, $a$. The ratio of the
$a$-values of these troughs is precisely the Efimov scaling factor $22.7$.
The existence of troughs in the spectrum, i.e. a lowering of the
recombination rate as a function of the scattering length, is due to the
existence of the trimer system and hence a reduced probability to
recombine into the dimer plus free particle state.

For negative $a$, a similar tendency is observed, however, with peaks
instead of troughs in the spectrum. This suggests that the mechanics for
recombination is quite different for the case of positive scattering
length. Indeed there are no shallow dimers for negative $a$ so
recombination goes into a deeply bound dimer and a free particle. The
origin of the peaks is also quite different. In the interaction potential
for three particles with negative scattering length there is a centrifugal
barrier that the three-body wave function must tunnel through (there is no
such barrier for positive scattering lengths). The barrier height and
location depend on the scattering length. When the scattering length is
tuned such that the energy of the incoming wave function matches with the
energy of a resonance behind the barrier, recombination is strongly
increased.

\section{Methods of describing three-body physics}
Real inter-atomic potentials are quite complicated if all details are
included. Therefore it is desirable to use simplified models that
nevertheless carry some of the same properties as the full potential. In
low-energy physics the scattering length mentioned above is exactly one
such property. Two potentials with the same scattering length would yield
the same results, to leading order, for scattering observables, provided
that the scattering energy is small (on some appropriate scale). Therefore
choosing the simpler potential is beneficial.

The simplest potential possible is, in some sense, the $\delta$-function
potential, also known as the zero-range or contact potential. The
zero-range function potential has the scattering length as the single
parameter. It has been proven to be a quite accurate tool for describing
low-energy three-body physics. Furthermore, models based on zero-range
potentials provide reasonably easy and straightforward numerical
calculations without too many complications and with fast computational
run-times. They also allow for clear and intuitive understanding of the
physical processes.

However, it is desirable to get a feeling for higher order effects, that
is, take into account not only the scattering length but higher order
parameters as well. The next order succeeding the scattering length is
known as the effective range. A physical potential in cold atomic gasses
has some length scale outside of which the potential is practically $0$,
this can be quantified as the range of the potential. While the scattering
length generally is not related to the physical range of the potential,
but can span many orders of magnitude in both positive and negative
directions, the effective range is a much better measure for the physical
range of the potential.

Inclusion of the effective range can be done by choosing a finite range
potential instead of the above mentioned zero-range potential. However,
this adds quite a bit of complexity to the calculations. Retaining the
computational simplicity of the zero-range potentials is thus be
desirable. This can in fact be done by using the so-called coupled
channels approach. By describing the interaction, not with a single
zero-range potential, but as a system of two coupled components with two
zero-range potentials, an effective range can be extracted. This still
rather simple set-up allows range effects to be investigated while
retaining computational simplicity and interpretative elegance.

\section{Three-body parameter}
A problem with pure zero-range potentials is that the bound state spectrum
does not have a lower bound. The Efimov effect equally well allows
downscaling with $22.7$, making the bound system smaller while increasing
the energy. This can be done indefinitely, yielding tighter and tighter
bound systems. This is called the Thomas effect \cite{thomas} and is
caused by a breakdown in the assumption of using contact potentials for
the interactions.

A way of avoiding the Thomas effect is to artificially introduce a short
range length scale that acts as a regularization and sets a lower bound
for the bound state energies \cite{FedorovJensen2001}. This parameter is
known as the three-body parameter and was long thought to be related to
the short-range details of the atomic potentials and as such expected to
differ greatly from one system to another. In recent years it has come to
attention that the three-body parameter, when divided by the van der Waals
length, a length scale related to the long-range behaviour of neutral atom
potentials, has a seemingly universal value of $\sim9.8$ for several
different atomic species. This curiosity has led to a lot of theoretical
activity trying to explain this phenomena \cite{chin2011, naidon2012,
wang2012, schmidt2012}.

\section{Thesis outline}

\hspace{6mm}\emph{Chapter 1}

\noindent This current introductory chapter that you are now reading and
have completed by about $79.9\%$.

\emph{Chapter 2}

\noindent
Here most of the theoretical and numerical groundwork is laid out. Low
energy scattering theory is shortly revised and the most fundamental
quantities for this thesis, the scattering length, $a$, and the effective
range, $R$, are defined. Then the hyper-spherical coordinates are
introduced as a scheme for practically and succinctly describing systems
of three particles. The zero-range models are then introduced and woven
into the hyper-spherical formalism. Methods for calculating recombination
rates and bound state energies for three-body systems are then discussed.
This involves going into the complex plane and returning on a different
level than what you started out on, having in the meantime passed around a
hidden crossing. This does not work for negative scattering lengths, so
instead the differential equation is solved directly, however, again with
the aid of complex \sout{wonders} numbers.

\emph{Chapter 3}

\noindent
Here we present some effects of the effective range. In the trimer bound
state spectrum it is seen how the three-body parameter depends on the
effective range. This holds also for the bound state energies on
resonance. Recombination for positive scattering length also shows clear
dependency of the effective range. Calculations are compared to the
experimental evidence, unfortunately the presently available experimental
data exists only for systems of broad resonances where the effective range
is small. The range effect are thus not easily observed in these systems.
It is shown that the effects of the effective range show up in a
non-trivial manner in the recombination coefficient.

\emph{Chapter 4}

\noindent
The apparent universality of the three-body parameter, when given in units
of the van der Waals length, is given yet another possible explanation.
This is done by the relating three-body quantities, i.e. the three-body
parameter as a function of the regularization cut-off, to appropriate
two-body equivalents. We then look at how these parameters relate to
typical two-body potentials and the number of bound states in these. We
find reasonably good agreement between the model and available
experimental data.

\emph{Chapter 5}

\noindent
In this chapter we revisit the three-body recombination rate but for
negative scattering lengths. The method outlined in chapter 2 is not
applicable in this case and instead we turn to the radial differential
equation directly. By superposing the potential with an optical potential,
i.e. a potential that has a complex value, we obtain a recombination
coefficient that models experimental data quite well, given its
simplicity. We furthermore include the effects of finite temperature in
the experimental system.

\emph{Chapter 6}

\noindent
Here some, as of this writing still unpublished, results for
mass-imbalanced systems are presented. Systems of mixed species of atoms
may provide a key insight into Efimov physics since the main feature,
namely the geometric scaling of states, persists in these systems. We show
that the frequency of resonance peaks or troughs is increased. This could
allow for better and more accurate determination of the scaling factor and
therefore allow for better testing of predictions involving the effective
range.

\emph{Chapter 7}

\noindent
Finally we provide a summary and look ahead to possible future projects.

\chapter{Theory and Methods}
\label{theory}
\emph{The theoretical groundwork is laid out and the different models and
methods that will be used in this thesis are described.}

\noindent\rule{\textwidth}{0.4pt}

\noindent In this chapter the theoretical models and methods that are used
in this thesis will be introduced. Initially the basic two-body scattering
concepts are described and terminology is defined. Still in the two-body
regime the ubiquitous zero-range model is presented along with some
extension models known as the two-channel model and the effective range
expansion model. The experimentally important tool of Feshbach resonances
is introduced and related to these models. Then we go into the three-body
sector where the formalism just described for the two-body physics is
stated in the three-body formalism. Some important three-body results,
like the Efimov effect and three-body recombination rates, are discussed
and related to the experimental observables. Finally the numerical
methods, implemented on top of these models, are presented.

\section{Two-body physics}
\label{two_body_physics}
\subsection{Two-body scattering}
\label{basic_scattering}
The problem is to predict the outcome of colliding two particles. Given an
initial state of particles, their energy and momentum, the aim is to
calculate the energy- and momentum distribution of the products. This is a
well-established discipline in both classical and quantum mechanics.

The system of two particles with coordinates $\bm r_1$ and $\bm r_2$ and
masses $m_1$ and $m_2$ is described by the Schrödinger equation
\begin{equation}
  \left[\frac{-\hbar^2}{2m_1} \nabla_1^2+\frac{-\hbar^2}{2m_2}
  \nabla_2^2 +V(\bm r_1,\bm r_2)\right]\Psi(\bm
r_1,\bm r_2) = E\Psi(\bm r_1,\bm r_2)\;,
  \label{eq:2.1}
\end{equation}
where $V(r_1,r_2)$ is the interparticle interaction. We will assume that
no external forces act on the system.

If a potential depends on the distance between particles, $r=|\bm r|$,
only, where $\bm r=\bm r_1-\bm r_2$, the center of mass motion is easily
separated out by introducing the center-of-mass coordinate $\bm R =
\dfrac{m_1\bm r_1+m_2\bm r_2}{m_1+m_2}$. For spherical potentials the
angular part of the wave function is given by the spherical harmonics
$Y_l^m(\theta,\phi)$ \cite{griffiths} (the superscript $m$ is the magnetic
quantum number, not to be confused with any mass). This leaves us with the
relative radial wave function $\psi(r)$ described by
\begin{equation}
  \left[ \frac{-\hbar^2}{2m} \frac{d^2}{dr^2}
    +\left(V(r)+\frac{\hbar^2l(l+1)}{2mr^2}\right) \right]\psi(r) =
  E\psi(r)\;,
  \label{eq:2.2}
\end{equation}
where $l$ is the angular momentum eigenvalue from the angular equation
that leads to the spherical harmonics and $m=\dfrac{m_1m_2}{m_1+m_2}$ is
the reduced mass. The full radial wave function is $\psi(r)/r$. $E$ is now
the energy of the relative motion since the center-of-mass motion has been
separated out.

Before we go on to the scattering, some simplifying assumptions are in
order. The cold atomic gases, that constitute the physical system under
observation, are at very low temperature, and hence the particles have low
kinetic energy. This leads to two important simplifications. First, since
the centrifugal barrier will suppress contributions from anything higher
than $s$-waves, only $s$-waves need to be considered and we can put $l=0$,
simplifying \eqr{eq:2.2} quite a bit. Second, and perhaps most important,
the use of $s$-waves only, allows us to choose the potential $V(r)$ almost
as we wish, provided some simple quantities are retained. Physical
inter-particle potentials are quite complicated if all details are
included. If we could somehow choose simpler potentials with some of the
same merits the analytical and numerical results might be easier to
obtain.

A very practical result from elementary scattering theory is the first
Born-approximation expression for the scattering amplitude
$f^{(1)}(\theta)$ at low incoming energy \cite{sakurai}
\begin{equation}
  f^{(1)}(\theta)=-\frac{1}{4\pi}\frac{2m}{\hbar^2}\int V(r)\;
  \textnormal d^3\bm x\;.
  \label{eq:2.3}
\end{equation}
This is essentially the 3D Fourier transform of the potential in the limit
of the energy going to zero, the superscript 1 indicates that this is a
first order approximation. The scattering amplitude $f^{(1)}(\theta)$ is a
measure of how much of the incoming wave gets deflected an angle $\theta$
from the original trajectory through the differential cross section
$d\sigma/d\Omega=|f^{(1)}(\theta)|^2$. In the present limit of small
energies the integral is independent of $\theta$ and the incoming wave
gets scattered equally in all directions. The loss of particles from the
initial beam is neatly summarized in the total cross section $\sigma$ as
the integral over the unit sphere of the differential cross section. It is
often written as

\begin{equation}
  \sigma = \int \frac{d\sigma}{d\Omega}\,\textnormal d\Omega= 4\pi a^2\;,
  \label{eq:2.4}
\end{equation}
where $a$ is known as the scattering length\footnote{In the case of
distinguishable particles, for identical bosons an additional factor 2 is
needed while for identical fermions the total cross section vanished in
this limit.}, in this case we have simply $a=|f^{(1)}(\theta)|$. The
scattering length is a lowest order measure of the strength of the
potential. Two potentials with the same scattering length will scatter an
incoming wave equally in the low-energy limit. Thus given a complicated
potential (either with a cumbersome analytical expression or perhaps only
known phenomenologically) it can be replaced by a simpler potential with
the same scattering length if only low energy is considered.

The physical reasoning behind these results is that at low energy the wave
length of the wave function is very large, larger even than the spatial
extent of the potential. Small details of the potential thus cannot be
probed by the wave function and only the overall cumulative effects are
measured.

If we can measure the scattering length of a physical potential it will
suffice to use a simpler potential with the same scattering length in
calculations. Any result must agree, to leading order, with similar but
much more complicated calculations using the full potential.

A short geometrical interpretation is in order which will also reveal an
important feature of the scattering length. Assume that the potential is
zero outside some finite range $r_0$ and that the energy, $E$, is positive
such that the wave function for $r>r_0$ takes the free form
\begin{equation}
  \psi(r) = C\sin(kr+\delta(k))\;,\quad\textnormal{for}\quad r>r_0\;,
  \label{eq:2.5}
\end{equation}
where $\delta(k)$ is the energy dependent phase shift and the wave number
$k$ is defined by $k^2=2mE/\hbar^2$. The phase shift depends on the
short-range details of the potential.

In the limit of very low energy, essentially $E = 0$, the solution to
\eqr{eq:2.2} is a simple linear function which can be obtained from
\eqr{eq:2.5} by
\begin{equation}
  \psi(r)\overset{k\rightarrow0}{\approx}
  C(\sin\delta+kr\cos\delta)=K\left(1-\frac{r}{a}\right)\;.
  \label{eq:2.6}
\end{equation}
Here the scattering length enters in the form
\begin{equation}
  \lim_{k\rightarrow0}k\cot\delta(k)=-\frac{1}{a}\;.
  \label{eq:2.7}
\end{equation}

\begin{figure}[ht]
  \centering
  \tikzsetnextfilename{fig2_1}
  \begin{tikzpicture}[scale=1.5]
	\pgfmathsetmacro{\r}{3}
	\pgfmathsetmacro{\V}{.21}
	\pgfmathsetmacro{\s}{sqrt(2*\V)*\r}
	\pgfmathsetmacro{\a}{1-sin(\s r)/cos(\s r)/\s}
	\pgfmathsetmacro{\A}{-1/(\s*cos(\s r)-sin(\s r))}
	\definecolor{orange}{HTML}{FF7F00}

	\node at (1,1) {$a>0$};

\begin{scope}[xshift=1cm]
	\draw[blue,very thick] (1.5,-0.5) -- (2.0,-0.5) node[anchor=west,black] {$\psi(r)$};
	\draw[blue!70,thick] (1.5,-0.5) -- (2.0,-0.5);

	\draw[orange,very thick] (1.5,-0.8) -- (2.0,-0.8) node[anchor=west,black] {$V(r)$};
	\draw[orange!50,thick] (1.5,-0.8) -- (2.0,-0.8);
\end{scope}

	\draw[->] (-.1,0) -- (1.2*\a,0) node[anchor=south] {$r$};
	\draw[->] (0,-1) -- (0,1);
	\draw (\a,-.1) -- (\a,.1) node[anchor=south] {$a$};

	\draw[orange,very thick] (0,-\s*\s/4) node[anchor = east,black] {$-V_0$} -- (1,-\s*\s/4) --
		(1,0) node[anchor = south,black] {$r_0$} -- (1.15*\a,0);
	\draw[orange!50,thick] (0,-\s*\s/4) -- (1,-\s*\s/4) -- (1,0) -- (1.15*\a,0);

	\draw[blue,very thick] plot[domain=0:1] (\x,{\A*sin(\x*\s r)}) --
		plot[domain=1:1.2*\a]  (\x,{1-\x/\a});
	\draw[blue!70,thick] plot[domain=0:1] (\x,{\A*sin(\x*\s r)}) --
		plot[domain=1:1.2*\a]  (\x,{1-\x/\a});

	\pgfmathsetmacro{\Vb}{.09}
	\pgfmathsetmacro{\sb}{sqrt(2*\Vb)*\r}
	\pgfmathsetmacro{\ab}{1-sin(\sb r)/cos(\sb r)/\sb}
	\pgfmathsetmacro{\Ab}{-1/(\sb*cos(\sb r)-sin(\sb r))/3}
	\pgfmathsetmacro{\dx}{\ab+6.5}

	\begin{scope}[xshift=\dx cm]
		\node at (-1.25,1) {$a<0$};
		\draw[->] (-.1,0) -- (1.5,0) node[anchor=south] {$r$};
		\draw[->] (0,-1) -- (0,1);
		\draw (\ab,-.1) -- (\ab,.1) node[anchor=south] {$-|a|$};
		\draw[orange,very thick] (\ab*1.2,0) -- (0,0) -- (0,-\sb*\sb/4) node[anchor = east,black]
			{$-V_0$}	 -- (1,-\sb*\sb/4) -- (1,0) node[anchor = south,black] {$r_0$} -- (1.4,0);
		\draw[orange!50,thick] (\ab*1.2,0) -- (0,0) -- (0,-\sb*\sb/4) -- (1,-\sb*\sb/4) -- (1,0) --
			(1.4,0);
		\draw[blue!50,thick,dashed] plot[domain=1:1.2*\ab]  (\x,{(1-\x/\ab)/3});
		\draw[gray!50,dashed] plot[domain=1:1.2*\ab]  (\x,{(1-\x/\ab)/3});
		\draw[blue,thick] plot[domain=0:1] (\x,{\Ab*sin(\x*\sb r)}) --
			plot[domain=1:1.5]  (\x,{(1-\x/\ab)/3});
		\draw[blue!70,thick] plot[domain=0:1] (\x,{\Ab*sin(\x*\sb r)}) --
			plot[domain=1:1.5]  (\x,{(1-\x/\ab)/3});
	\end{scope}
\end{tikzpicture}
  \caption[Square well potential]{The zero energy solution to the
    Schrödinger equation for the finite square well. The left well is deep
    enough to support a bound state and has a positive scattering length
    whereas the right potential is too shallow and has a negative
    scattering length.}
  \label{figure:2.1}
\end{figure}
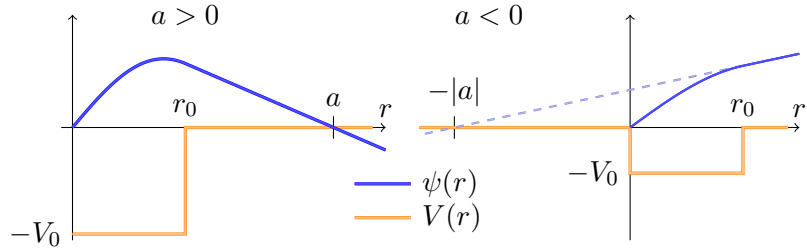

The low-energy limit of \eqr{eq:2.5} is shown in \fig{figure:2.1} for the
finite square well, for which an analytical solution is readily available
\cite{Martin2009}. The left potential is deep enough that the zero-energy
scattering wave function intersects the positive $x$-axis. This is
equivalent to the potential being able to support a bound state
\cite{sakurai}. The right potential is too shallow to support a bound
state. Correspondingly the zero energy wave function does not intersect
the $x$-axis at a positive value. The linear extrapolation of the
asymptotic wave function, however, intersects the negative $x$-axis at
the, now negative, scattering length.

Increasing the potential depth $V_0$ from the right scenario to the left
must yield a critical value for which the scattering length will diverge
to $\pm\infty$. At precisely this value the potential is deep enough to
support yet another bound state with energy zero.

\subsection{Effective range expansion}
\label{effective_range_expansion}
Equation \eqref{eq:2.7} looks quite like a Taylor expansion of
$k\cot\delta(k)$ around $k=0$. Thus it seems natural to include another
term in the expansion
\begin{equation}
  k\cot\delta(k)=-\frac{1}{a}+\frac{1}{2}Rk^2\;.
  \label{eq:2.8}
\end{equation}
Here $R$ is known as the effective range. The value of the effective range
is typically of the order of the actual physical range of the potential
($r_0$ in \fig{figure:2.1}). Whereas the scattering length, $a$, can vary
between $\pm\infty$, the effective range, $R$, will vary much less.

\subsection{The zero-range model}
\label{zero_range_model}

The derivation of \eqr{eq:2.7} from \eqr{eq:2.5} can for the zero-range
interaction (where essentially $r_0=0$ and \eqr{eq:2.5} is valid
everywhere) be formulated as \cite{FedorovJensen2001}
\begin{equation}
  \frac{1}{\psi}\frac{d\psi}{dr}\bigg|_{r=0}=
    k\cot\delta(k)\overset{k\rightarrow0}=-\frac{1}{a}\;.
  \label{eq:2.9}
\end{equation}
Assume that $V(r)=0$ everywhere except the origin. Then the wave function
is $\psi=\sin(kr+\delta(k))$. Imposing the boundary condition
\eqr{eq:2.9} yields
\begin{equation}
  \frac{1}{\sin(kr+\delta(k))}k\cos(kr+\delta(k))\bigg|_{r=0}=
    k\cot\delta(k)\;.
  \label{eq:2.10}
\end{equation}
Correspondingly, a bound state is given by $\psi(r)=\exp(-\kappa r)$
(disregarding normalization) yielding
\begin{equation}
  \frac{1}{\exp(-\kappa r)}(-\kappa)\exp(-\kappa r)=
    -\kappa=-\frac{1}{a}\;,
  \label{eq:2.11}
\end{equation}
where $\kappa^2=-2mE/\hbar^2$. Only for positive scattering lengths does a
bound state exists, since $\kappa$ is defined to be positive and the wave
function has to be normalizable.

\subsection{Feshbach resonances}
The important experimental tool of Feshbach resonances lies at the heart
of a lot of the cold atomic physics experiments since its discovery and
realization in 1998 \cite{stenger2} in a gas of Sodium atoms. Having since
been realized in almost at alkali atoms (Na \cite{stenger}, Li
\cite{pollack2009}, K \cite{Zaccanti}, Cs \cite{kraemer2006}) its
usefulness lies in the ability to tune the interaction strength simply by
applying an external magnetic field.

\begin{figure}[ht]
  \centering
  \input{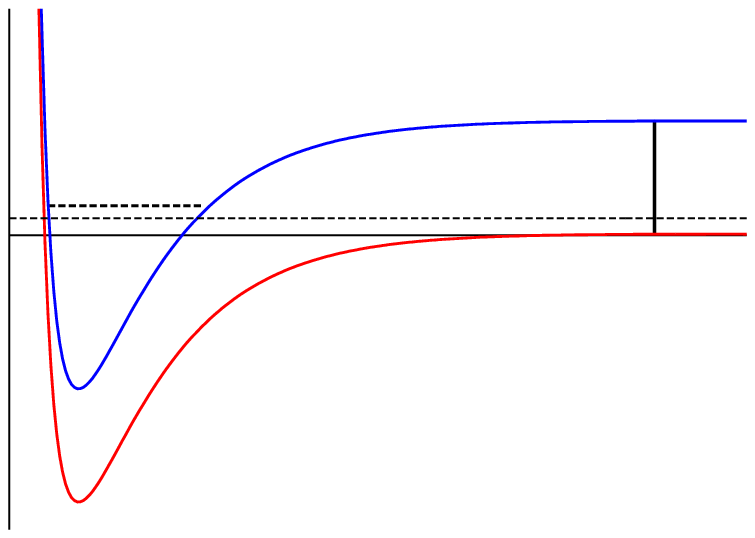}
  \caption[Schematic model of Feshbach resonance system]{A schematic
  drawing of a system with a system with a Feshbach resonance. The open
  and closed channels denote two different interaction channels for the
  system. When the relative kinetic energy, $E$, corresponds to the energy
  of a bound state in the closed channel, $E^*$, the scattering length is
  resonantly enhanced due to the degeneracy of the states. The Feshbach
  tuning is enabled using the Zeeman effect by changing the value of
  $\epsilon$ and thus the value of $E^*$.}
  \label{figure:2.2}
\end{figure}

\fig{figure:2.2} shows a schematic set-up of a system with a Feshbach
resonance. The interaction is described using two channels, the open
channel where the potential value at large distances is smaller than the
relative kinetic energy and the closed channel where the kinetic energy is
lower than the asymptotic value of the potential. Both the open and closed
channels may support a number of bound states. If the closed channel
happens to have a bound state at $E^*$ that corresponds to the value of
relative incoming kinetic energy, $E$, there is resonant coupling between
the channels and the scattering cross section is greatly enhanced.

The bound state at $E^*$ may not be near $E$ due to the temperature at
which the experiment is carried out. Changing the incoming energy $E$ by
changing the temperature may not be practical/possible. However, changing
the energy levels $E^*$ can be done simply by applying an external
magnetic field. Due to the Zeeman effect an external magnetic field will
change the energy levels of both the open and the closed channels. If the
magnetic momenta in the open and closed channels differ, then the energy
levels in the two channels will change relative to each other when the
magnetic field strength is changed, thus changing the value of $\epsilon$
and therefore also of $E^*$ and finally $a$. A phenomenological
expression for the scattering length as a function of an externally
applied magnetic field of strength $B$ is \cite{chin2010}
\begin{equation}
  a(B)=a_\textnormal{bg}\left(1-\frac{\Delta B}{B-B_0}\right)\;.
  \label{eq:2.12}
\end{equation}
Here $B_0$ is the field strength at which the scattering length diverges,
$\Delta B$ is the width of the resonance and $a_\textnormal{bg}$ is the
scattering length far from resonance. The effective range near a Feshbach
resonance is given by \cite{Martin2009}
\begin{equation}
  R(a) = R_0\left(1-\frac{a_\textnormal{bg}}{a}\right)^2\;,
  \label{eq:2.13}
\end{equation}
where $R_0$, the effective range on resonance, is given by the width of
the resonance \cite{bruun2005}
\begin{equation}
  R_0 =-\frac{2}{m\delta\mu a_\textnormal{bg}\Delta B}\;,
  \label{eq:2.14}
\end{equation}
where $\delta\mu$ is the difference in magnetic momenta between the two
channels.

\subsection{The two-channel model}
\label{two_channel_model}
The scattering length is the sole parameter of the zero-range model. This
section adds to this model and incorporates the effective range using
another simple model. We build the model upon the two-channel set-up of
Feshbach resonances. The two-component wave function is


\begin{equation}
  \psi(r)=\begin{bmatrix}u_\cl(r)\\ u_{\op\;\;}(r)\end{bmatrix}\;,
  \label{eq:2.15}
\end{equation}
where $u_\op$ describes the open channel, where both atoms are in their
ground state, while the closed channel, $u_\cl$, has one of the atoms in
an excited state.

The Schrödinger equation for this system is
\begin{subequations}\label{eq:2.16}
\begin{align}
  -\frac{\hbar^2}{2m}u_\cl'' &= (E-\epsilon)u_\cl \;,
  \label{eq:2.16a}\\
  -\frac{\hbar^2}{2m}u_{\op\;\;}'' &= Eu_\op\;,
  \label{eq:2.16b}
\end{align}
\end{subequations}
where primes denote differentiation with respect to $r$, $m$ is the
reduced mass of the two atoms, $E$ is the relative energy and $\epsilon$
is the excitation energy of the closed channel with respect to the open
channel. The coupling between channels is obtained through the boundary
condition \eqr{eq:2.9}, that we generalize to a two-level system in the
following way
\begin{equation}
  \begin{bmatrix}u_\cl'\\u_{\op\;\;}'\end{bmatrix}_{r=0} =
  \begin{bmatrix}-a_\cl^{-1}&\beta\\\beta&-a_\op^{-1}\end{bmatrix}
  \begin{bmatrix}u_\cl\\u_{\op\;\;}\end{bmatrix}_{r=0} \;,
  \label{eq:2.17}
\end{equation}
where the constant $\beta$ parametrizes the coupling between the channels,
and $a_\op$ and $a_\cl$ are the respective scattering lengths in the two
channels.

Assume that the energy is below the threshold for excitation,
$0<E<\epsilon$, then the solution to \eqrs{eq:2.16} is
\begin{subequations}\label{eq:2.18}
\begin{align}
  u_\cl &= A_\cl\exp(-\kappa_\cl r)\;,
  \label{eq:2.18a}\\
  u_{\op\;\;} &= A_{\op\;\;}\sin(k_\op r+\delta)\;,
  \label{eq:2.18b}
\end{align}
\end{subequations}
where $A_\cl$ and $A_\op$ are constants, $k_\op=\sqrt{2mE/\hbar^2}$ and
$\kappa_\cl=\sqrt{2m(\epsilon-E)/\hbar^2}$.

Inserting this solution into the boundary condition \eqr{eq:2.17} yields a
system of linear equations for the constants $A_\cl$ and $A_\op$
\begin{equation}
  \begin{bmatrix}-\beta\sin\delta & a_\cl^{-1}-\kappa_\cl\\
    k_\op\cos\delta+a_\op^{-1}\sin\delta & -\beta\end{bmatrix}
  \begin{bmatrix}A_\cl\\A_{\op\;\;}\end{bmatrix}=0\;.
  \label{eq:2.19}
\end{equation}
This system of equations has a non-trivial solution only if the
determinant of the $2\times2$ matrix vanishes, yielding
\begin{equation}
  \beta^2\sin\delta-\left(k_\op\cos\delta+\frac{\sin\delta}{a_\op}\right)
  \left(\frac1{a_\cl}-\kappa_\cl\right)=0 \;.
  \label{eq:2.20}
\end{equation}
Isolating $k_\op\cot\delta$, as in \eqr{eq:2.7} and \eqr{eq:2.8}, gives
\begin{equation}
  k_\op\cot\delta=-\frac{1}{a_\op}+
    \frac{\beta^2}{a_\cl^{-1}-\kappa_\cl}\;.
  \label{eq:2.21}
\end{equation}
Taylor expansion of the right hand side around $k_\op=0$ yields the
effective scattering length, $a$, and effective range, $R$, of this
two-level system
\begin{align}
  \frac1a&=\frac{1}{a_\op}+\frac{\beta^2}{\kappa-a_\cl^{-1}}\;,
  \label{eq:2.22} \\
  R&=-\frac{\beta^2}
  {\kappa\left(\kappa-a_\cl^{-1}\right)^2}\;,
  \label{eq:2.23}
\end{align}
where $\kappa^2=2m\epsilon/\hbar^2$. The two-channel model can thus
emulate a system with finite effective range yet it consists only of
contact interactions. Note that the effective range in this model is
always negative.

The effective range \eqr{eq:2.23} can be written in terms of the
scattering length in the clearer way
\begin{equation}
  R(a) = R_0 \left(1-\frac{a_\op}{a}\right)^2\;,
  \label{eq:2.24}
\end{equation}
where
\begin{equation}
  R_0=\frac{-1}{\kappa_0a_\op^2\beta^2}\;,
  \label{eq:2.25}
\end{equation}
is the effective range on resonance, $a=\infty$, and where $\kappa_0$ is
the value of $\kappa$ in \eqr{eq:2.22} that yields $a=\infty$. This agrees
with \eqr{eq:2.13}.

This simple two-level system will be used to investigate finite range
effects without using finite range potentials, which are quite a bit more
cumbersome to work with. The downside to this approach is that only
negative effective ranges can be modelled. This, however, is not a problem
as we will be looking at Feshbach systems at or near resonance where the
effective range is always negative \cite{chin2010}.

\subsubsection{The two-channel model relation to Feshbach resonances}
\label{feshbach_resonances}
The effective scattering length of the two-channel model is fixed by the
parameters of the two-level system, i.e. the scattering lengths in each
sub-system, $a_\op$ and $a_\cl$, and the coupling between them, $\beta$.
To be able to tune the scattering length as desired the Zeeman effect is
now added. An external magnetic field of strength $B$ changes the energy
splitting $\epsilon$ of the two-level system by
\begin{equation}
  \epsilon\rightarrow \epsilon-~\delta\mu B\;,
  \label{eq:2.26}
\end{equation}
where $\delta\mu$ is the difference in magnetic moments of the atom in the
ground and the excited state. The scattering length from \eqr{eq:2.22} is
then a function of the magnetic field,
\begin{equation}
  a(B)=a_\op\frac{\kappa(B)-a_\cl^{-1}}
    {\kappa(B)-a_\cl^{-1}+\beta^2a_\op}\;,
  \label{eq:2.27}
\end{equation}
where $\kappa(B)=\sqrt{2m(\epsilon-\delta\mu B)/\hbar^2}$. The
scattering length diverges at the critical value of the magnetic field,
$B_0$, given by
\begin{equation}
  \kappa_0\equiv\kappa(B_0)=\frac{1}{a_\cl}-\beta^2a_\op\;,
  \label{eq:2.28}
\end{equation}
which gives
\begin{equation}
  B_0=\frac1{\delta\mu}\left(\epsilon-\frac{\hbar^2\kappa_0^2}
    {2m}\right)\;.
  \label{eq:2.29}
\end{equation}
Expanding $a(B)$ in the vicinity of $B_0$ gives precisely \eqr{eq:2.12}
with $a_\textnormal{bg}=a_\op$ and
\begin{equation}
  \Delta B = \frac1{\delta\mu} \frac{\hbar^2\kappa_0\beta^2a_\op}{m}\;.
  \label{eq:2.30}
\end{equation}
On resonance the effective range, $R$, is inversely proportional to the
width of the resonance, $\Delta B$,
\begin{equation}
  R(B_0)=-\frac1{\kappa_0\beta^2a_\op^2}=
    -\frac1{a_\op}\frac{\hbar^2}{m\delta\mu\Delta B}\;,
  \label{eq:2.31}
\end{equation}
as in \eqr{eq:2.14}.

\begin{figure}[ht]
  \centering
  \input{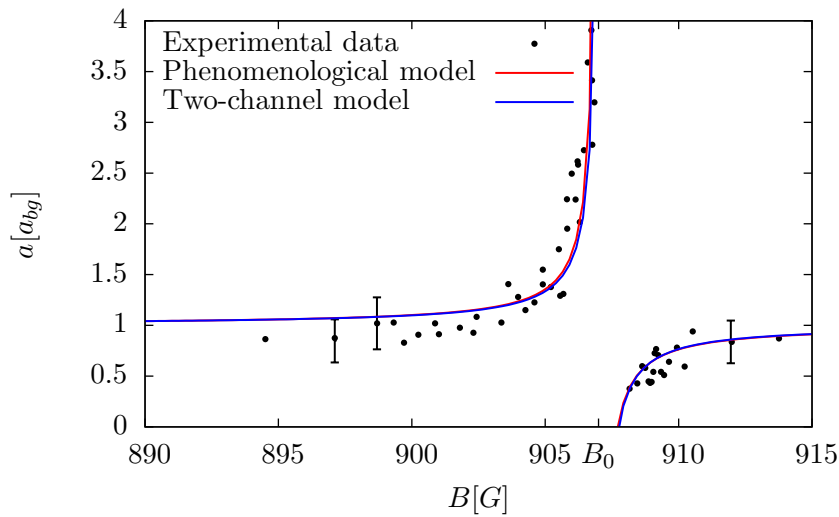}
  \caption[Feshbach resonance in ${}^{23}$Na]{The scattering length, $a$,
  between two ${}^{23}$Na atoms as a function of the external magnetic
  field, $B$, with two-channel \eqr{eq:2.22} and empirical \eqr{eq:2.12}
  fits to the experimental data from \cite{stenger}. The Feshbach
  resonance location is at $B_0=907$ G with width $\Delta B=0.7$ G.}
  \label{figure:2.3}
\end{figure}

Given $\Delta B,\;B_0$ and the background scattering length
$a_\textnormal{bg}$ from experiment, \eqr{eq:2.22}, \eqr{eq:2.29} and
\eqr{eq:2.30} can be solved for the model parameters $a_\op,\;a_\cl$ and
$\beta$
\begin{subequations}\label{eq:2.32}
  \begin{align}
    a_\op&=a_\textnormal{bg}\;,
      \label{eq:2.32a}\\
    a_\cl&=\frac{2\cdot\textnormal{sign}(\Delta B)
      \sqrt{\tilde\varepsilon}}
    {\frac{\delta\mu\Delta B}{E_0}+2\tilde\varepsilon}a_\op\;,
      \label{eq:2.32b}\\
    \beta^2&=\frac{1}{2a_\op^2}\frac{1}{\sqrt {\tilde\varepsilon}}
      \frac{\delta\mu|\Delta B|}{E_0}\;,
    \label{eq:2.32c}
  \end{align}
\end{subequations}
where
\begin{equation}
  \tilde\varepsilon=\frac{\epsilon-\delta\mu B_0}{E_0},\quad
  E_0=\frac{\hbar^2}{2ma_\textnormal{bg}^2}\;.
  \label{eq:2.33}
\end{equation}
The value of $\epsilon$ cannot be determined uniquely from these
equations. It can, however, be found by fitting \eqr{eq:2.27} to
experimental data, $a(B)$, as shown in \fig{figure:2.3}. However, the
value of $\epsilon$ does not affect the final observables significantly,
provided it is greater than $\delta\mu B_0$ and is of the order the
hyperfine splitting.

\fig{figure:2.3} shows experimental data for a Feshbach resonance in
${}^{23}$Na along with fits from \eqr{eq:2.12} and \eqr{eq:2.22}. The
phenomenological expression \eqr{eq:2.12} gives $B_0=907.0$ G, $\Delta
B=0.71$ G, while the two-channel expression \eqr{eq:2.22} gives
$B_0=907.1$ G, $\Delta B=0.69$ G when using \eqr{eq:2.29} and
\eqr{eq:2.30}. Choosing $\epsilon=23\;\mu$eV provides a good fit. Varying
$\epsilon$ has relatively little influence on the values of $B_0$ and
$\Delta B$ provided it is within this order of magnitude.

The phenomenological and the two-channel curves are virtually identical,
so from now on the experimental parameters $a_\textnormal{bg},\; \Delta B,
\;B_0$ and $\delta\mu$ are used to determine the parameters of the
two-channel model through \eqrs{eq:2.32}.

\section{Three-body physics}
\label{three_body_physics}
The previously described models rely on two-body physics alone. Applying
them to three-body systems requires a framework that efficiently takes
into account the extra degrees of freedom without adding too much
complexity such that the simple methods using contact interactions can
still be applied.

\subsection{Hyper-spherical coordinates}
\label{hyperspherical_coordinates}
A set of three particles labelled by indices $\{i,j,k\}$ can be described
either by their absolute Cartesian coordinates $\{\bm r_i,\bm r_j,\bm
r_k\}$ or by a linear combination of these. One such combination is the
Jacobi coordinates \cite{Nielsen2001}
\begin{align}
  \bm x_i&=\sqrt{\mu_i}(\bm r_j-\bm r_k)\;,\qquad
  \bm y_i=\sqrt{\mu_{jk}}\left(\bm r_i-
    \frac{m_j\bm r_j+m_k\bm r_k}{m_j+m_j}\right)\;,
  \label{eq:2.34}\\
  \mu_i&=\frac1m\frac{m_jm_k}{m_j+m_k}\;,\qquad
  \mu_{jk}=\frac1m\frac{m_i(m_j+m_k)}{m_i+m_j+m_k}\;,
  \label{eq:2.35}
\end{align}
where $m_{\{i,j,k\}}$ are the masses of particle $\{i,j,k\}$. Here
$\{i,j,k\}$ is a cyclic permutations of $\{1,2,3\}$. The mass scaling
parameter $m$ has no inherent meaning and can be chosen arbitrarily, since
it only serves as a scaling of the coordinates, it is not to be confused
with the two-body reduced mass of previous sections. For equal mass
particles the choice $m=m_i=m_j=m_k$ is obvious, such that
$\mu_i=\frac{1}{2}$ and $\mu_{jk}=\frac{2}{3}$. For chapter
\ref{mass_imbalanced_systems}, however, the masses will differ when a
mixed system will be studied.

\begin{figure}[ht]
  \centering
  \tikzsetnextfilename{fig2_4}
  \begin{tikzpicture}[inner sep=1mm,scale=1.0,inner sep=.25cm]
	\tikzstyle{particle} = [circle,fill,shading=call,ball color=blue!50!red]
	\tikzstyle{tarrows} = [->,>=triangle 45,thick]

  \node (p1) at (-1,2) [particle,label=1]       {};
  \node (p2) at (2,3)  [particle,label=2]       {};
  \node (p3) at (1,1)  [particle,label=right:3] {};

  \draw[tarrows] (p2) -- node[anchor = south east] {$r_{12}$}(p1);
  \draw[tarrows] (p3) -- node[anchor = north east] {$r_{13}$} (p1);
  \draw[tarrows] (p3) -- node[anchor = west] {$r_{23}$} (p2);

  \node (p1) at (5,2) [particle,label=1]       {};
  \node (p2) at (8,3) [particle,label=2]       {};
  \node (p3) at (7,1) [particle,label=right:3] {};
  \draw[tarrows] ($0.5*(p2)+0.5*(p3)$) -- node  [above] {$\bm y_1$}  (p1);
  \draw[tarrows] (p3) -- node  [right] {$\bm x_1$} (p2);
\end{tikzpicture}
  \caption[Jacobi coordinates]{To the left, three particles and their
    relative coordinates. To the right, the same three particles and the
    Jacobi coordinate set with index $i=1$.}
  \label{figure:2.4}
\end{figure}
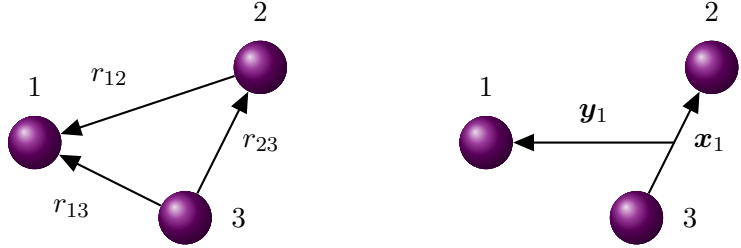

The choice of index $i$ yields a specific coordinate set, as illustrated
in \fig{figure:2.4}. The vector $\bm x_i$ is the relative coordinate
between particles $j$ and $k$ while the vector $\bm y_i$ is the relative
coordinate between the center-of-mass of particles $j$ and $k$, and
particle $i$.

The hyper-radial coordinates $\rho$ and $\alpha_i$ are defined from the
Jacobi coordinates as
\begin{equation}
  \rho^2=x_i^2+y_i^2\;,\qquad\rho\sin\alpha_i=x_i\;,
    \qquad\rho\cos\alpha_i=y_i\;,
  \label{eq:2.36}
\end{equation}
where $x_i=|\bm x_i|$ and likewise for $y_i$. $\rho$ is known as the
hyper-radius and $\alpha_i$ is one of five hyper-angles, the remaining
four being comprised of the directions of the vectors $\bm x_i$ and $\bm
y_i$. All five hyper-angles are collectively denoted as $\Omega_i$.

The hyper-radius is a measure of the overall size of the system, it is
independent of the choice of Jacobi index. For large $\rho$ either all
three particles are far one another or one particle is far from the other
two (i.e. when these form a bound dimer). Conversely, the hyper-radius is
small only if all three particles are close to one another.

The hyper-angle, $\alpha_i$, is small when particles $j$ and $k$ are close
together. When the particles are co-linear $\alpha_i$ attains its maximal
value of $\pi/2$.

The kinetic energy operator in the hyper-spherical coordinates is given by
\cite{FedorovJensen2001}
\begin{subequations}
  \label{eq:2.37}
  \begin{align}
    T=T_\rho&+\frac{\hbar^2}{2m\rho^2}\Lambda^2\;,\qquad
      T_\rho=-\frac{\hbar^2}{2m}\left(\rho^{-5/2}\frac{\p^2}{\p\rho^2}
      \rho^{5/2}-\frac1{\rho^2}\frac{15}{4}\right)\;,
    \label{eq:2.37a}\\
    \Lambda^2&=-\frac1{\sin(2\alpha_i)}\frac{\p^2}{\p\alpha_i^2}
      \sin(2\alpha_i)-4+\frac{l_{x_i}^2}{\sin^2\alpha_i}+
      \frac{l_{y_i}^2}{\cos^2\alpha_i}\;,
      \label{eq:2.37b}
  \end{align}
\end{subequations}
where $\Lambda^2$ is the grand angular momentum operator and $l_{x_i}$
and $l_{y_i}$ are the conjugate angular momenta to the Jacobi coordinates
$\bm x_i$ and $\bm y_i$.

For the wave function itself the hyper-radial adiabatic expansion is
applied
\begin{equation}
  \Psi_i(\rho,\Omega_i)=\rho^{-5/2}\sum_nf_n(\rho)\Phi_n(\rho,\Omega_i)\;,
  \label{eq:2.38}
\end{equation}
where $f_n(\rho)$ is a radial wave function and $\Phi_n(\rho,\Omega)$ is
an angular wave function. The radial dependence of $\Phi_n$ is slow when
compared to $f_n$.

Using the hyper-radial adiabatic expansion, \eqr{eq:2.38}, along with
\eqrs{eq:2.37} the Schrödinger equation $(T+V)\Psi=E\Psi$ yields a
hyper-angular equation
\begin{equation}
  \left(\Lambda^2+\frac{2m\rho^2}{\hbar^2}V\right)\Phi_n(\rho,\Omega_i)=
    \lambda_n(\rho)\Phi_n(\rho,\Omega_i)\;,
  \label{eq:2.39}
\end{equation}
and a set of coupled radial equations
\begin{multline}
  \left(-\frac{d^2}{d\rho^2}+\frac{\lambda_n+15/4}{\rho^2}-Q_{nn}(\rho)-
    \frac{2mE}{\hbar^2}\right)f_n(\rho)=\\
  \sum_{m\neq n}\left(2P_{nm}(\rho)\frac{d}{d\rho}+
    Q_{nm}(\rho)\right)f_{m}(\rho)\;,
  \label{eq:2.40}
\end{multline}
where the terms $P_{nm}$ and $Q_{nm}$, known as adiabatic potentials, are
given by
\begin{equation}
  P_{nm}(\rho)=\left\langle\Phi_n\left|\frac{\p}{\p\rho}
    \right|\Phi_{m}\right\rangle_{\Omega_i}\;,\qquad
  Q_{nm}(\rho)=\left\langle\Phi_n\left|\frac{\p^2}{\p\rho^2}
    \right|\Phi_{m}\right\rangle_{\Omega_i}\;,
  \label{eq:2.41}
\end{equation}
where angle brackets indicate integration over hyper-angles $\Omega_i$ and
$m$ is the mass scaling parameter from \eqr{eq:2.35}.

The approach to solving this system of equations involves first solving
the angular equation \eqref{eq:2.39} for fixed $\rho$ to obtain
$\lambda_n(\rho)$. Then the radial equations can be solved using
$\lambda_n(\rho)$ as part of the radial potentials. This is similar to the
Born-Oppenheimer approximation where the fast-moving dynamics is
integrated out yielding an effective potential for the slow-moving part.
The radial potential with $\dfrac{\lambda_n+15/4}{\rho^2}$ is in essence a
hyper-angular centrifugal barrier similar to the usual centrifugal barrier
of \eqr{eq:2.2}.

In this thesis we never solve the full system \eqref{eq:2.40}. First, we
find numerically that the parameters $P$ and $Q$ are very small compared
to the effective $\lambda$-potential when using the zero-range two-body
potentials. This is a great simplification as they are not easily
calculated. Second, the expansion in \eqr{eq:2.38} is terminated after the
first term, i.e. only $n=0$ is included in calculations involving the
radial equation directly. When dealing with Efimov physics this has been
proven to be quite a good approximation \cite{Nielsen2001} and leaves for
a much easier calculation of the results as well as an easier
interpretation of the model. Therefore, the actual equation that will be
used is
\begin{equation}
  \left(-\frac{d^2}{d\rho^2}+\frac{\lambda_0+15/4}{\rho^2}-
    \frac{2mE}{\hbar^2}\right)f_0(\rho)=0\;.
  \label{eq:2.42}
\end{equation}

\subsection{The zero-range model, hyper-spherical edition}
\label{zero_range_model_hyperspherical_edition}
In this section the hyper-angular equation \eqref{eq:2.39} is solved to
obtain the eigenvalue $\lambda_n$ as function of hyper-radius, $\rho$.
This is done by applying the zero-range model from section
\ref{basic_scattering} to three-body systems.

Two-body interactions in a three-body system is most easily dealt with
using Faddeev decomposition where the wave function is split into three
components, one for each two-body subsystem
\begin{equation}
  \Phi(\rho,\Omega)=
    \sum_{i=1}^3\frac{\varphi_i(\rho,\alpha_i)}{\sin(2\alpha_i)}\;,
  \label{eq:2.43}
\end{equation}
where $\varphi_i(\rho,\alpha_i)$ is the angular wave function for the
particle pair $\{j,k\}$. The subscript $n$ from \eqr{eq:2.39} has been
suppressed. The factor $\sin(2\alpha_i)$ is for convenience
in the final equations. The potential is equally described as a sum of
two-body potentials $V = \sum_i V_i$ where again $V_i$ is the interaction
between particles $j$ and $k$. With these decompositions \eqr{eq:2.39}
becomes a sum of three identical terms, one for each particle
\begin{equation}
  \left(\Lambda-\lambda(\rho)\right)\frac{\varphi_i(\rho,\alpha_i)}
    {\sin(2\alpha_i)}+\frac{2m\rho^2}{\hbar^2}V_i
    \frac{\varphi_i(\rho,\alpha_i)}{\sin(2\alpha_i)}=0,\quad i=1,2,3\;.
  \label{eq:2.44}
\end{equation}

As noted in section \ref{basic_scattering} it is not necessary to include
all angular momentum states since the energy is very low in our system. In
\eqr{eq:2.44} we will thus set $l_{x_i}=l_{y_i}=0$ and use only $s$-wave
states, greatly simplifying the grand angular momentum operator
$\Lambda^2$. For non-zero separation, i.e. $\alpha_i\neq0$, and since
zero-range potentials are used, \eqr{eq:2.44} simplifies further into
\begin{equation}
  \frac{\p^2}{\p\alpha_i^2}\varphi_i(\rho,\alpha_i)=
    -[4+\lambda(\rho)]\varphi_i(\rho,\alpha_i)=
    -\nu^2\varphi_i(\rho,\alpha_i)\;,
  \label{eq:2.45}
\end{equation}
where $\nu^2=4+\lambda$. Both $\nu$ and $\lambda$ will be referred to as
angular eigenvalues and will be used interchangeably depending on context.
The solution to \eqr{eq:2.45} is
\begin{equation}
  \varphi_i=\varphi(\rho,\alpha_i)=
    N_i(\rho)\sin\left(\nu(\rho)\left[\alpha_i-\frac\pi2\right]\right)\;,
  \label{eq:2.46}
\end{equation}
where $N_i(\rho)$ is a normalization constant. The choice of phase ensures
that the total wave function remains finite at $\alpha_i=\frac{\pi}{2}$.

The zero-range boundary condition from \eqr{eq:2.9} takes the following
form \cite{FedorovJensen2001}
\begin{equation}
  \frac{\p\ln(r_{jk}\Phi)}{\p r_{jk}}\bigg|_{r_{jk}=0}=-\frac{1}{a_i}\;,
  \label{eq:2.47}
\end{equation}
where $r_{jk}$ is the distance between particles $j$ and $k$ and $a_i$ is
the corresponding scattering length between them. In the limit
$r_{jk}\rightarrow0$, \eqr{eq:2.36} gives
\begin{equation}
  r_{jk}=\frac{x_i}{\sqrt{\mu_i}}\approx
    \frac{\rho\alpha_i}{\sqrt{\mu_i}}\;,
  \label{eq:2.48}
\end{equation}
for small $\alpha_i$ and fixed $\rho$. The boundary condition becomes
\begin{equation}
  \frac{\p(\alpha_i\Phi)}{\p\alpha_i}\bigg|_{\alpha_i=0}=-\frac{\rho}
    {\sqrt{\mu_i}}\frac1{a_i}\alpha_i\Phi\bigg|_{\alpha_i=0}\;.
  \label{eq:2.49}
\end{equation}

To apply this boundary condition to all three two-particle wave functions
they must all be expressed in the same Jacobi coordinate system, for
instance the one in \fig{figure:2.4}. This is done using the kinematic
rotation operator $\mathcal R$, that rotates the wave function from system
$k$ to system $j$ and is defined by \cite{FedorovJensen2001}
\begin{equation}
  \mathcal R[\varphi_k](\alpha_j)=
    \frac1{\sin(2\phi_{jk})}\int_{|\phi_{jk}-\alpha_j|}
    ^{\frac\pi2-|\frac\pi2-\phi_{jk}-\alpha_j|}
    \varphi_{k}(\alpha_k)d\alpha_k\;,
  \label{eq:2.50}
\end{equation}
where
\begin{equation}
 \phi_{jk}=\arctan\left(\sqrt{\frac{m_i(m_1+m_2+m_3)}{m_jm_k}}\right)\;.
  \label{eq:2.51}
\end{equation}
For three identical particles $\phi_{jk}=\frac{\pi}{3}$ and the following
results hold for equal mass particles only. In the boundary condition
\eqref{eq:2.49} the Faddeev components are replaced by
\begin{equation}
  \varphi_1(\alpha_1)+\varphi_2(\alpha_2)+\varphi_3(\alpha_3)\rightarrow
    \varphi(\alpha_i)+2\mathcal R[\varphi_k](\alpha_i)\;,
  \label{eq:2.52}
\end{equation}
and finally
\begin{equation}
  \frac{\p(\varphi_i+2\mathcal R[\varphi])}{\p\alpha_i}\bigg|_{\alpha_i=0}
    =-\frac{\rho}{\sqrt{\mu_i}}\frac1{a_i}(\varphi_i+2\mathcal R[\varphi])
    \bigg|_{\alpha_i=0}\;.
  \label{eq:2.53}
\end{equation}
With \eqr{eq:2.46} the rotated wave functions become
\begin{equation}
  \mathcal R[\varphi_k](\alpha_j)=\frac{4N}{\nu\sqrt3}
    \begin{cases}
      -\sin\left(\dfrac{\nu\pi}6\right)\sin(\nu\alpha_j)&,\;
      0\leq\alpha_j\leq\dfrac\pi3\\[.7em]
      \phantom{-}\sin\left(\dfrac{\nu\pi}3\right)
        \sin\left(\nu\left[\alpha_j-\dfrac\pi2\right]\right)&,\;
        \dfrac\pi3\leq\alpha_j\leq\dfrac\pi2
    \end{cases}
  \label{eq:2.54}
\end{equation}
and finally the boundary condition becomes
\begin{equation}
  \frac{\nu\cos\left(\dfrac{\nu\pi}{2}\right)-
    \dfrac{8}{\sqrt3}\sin\left(\dfrac{\nu\pi}{6}\right)}
    {\sin\left(\dfrac{\nu\pi}{2}\right)}=\frac{\rho}{\sqrt\mu a}\;,
  \label{eq:2.55}
\end{equation}
where the index on $a$ and $\mu$ has been removed since all particles are
equal and their interactions also.

This is the fundamental equation of this thesis, it is the foundation of
the two more elaborate models that include the effective range. Any
solution $\nu(\rho)$ to this equation will depend only on the ratio
$\dfrac{\rho}{\sqrt{\mu}a}$ so it is in principle necessary to solve
\eqr{eq:2.55} only once. In general it can be solved only numerically, but
in order to do this efficiently some analytical properties are important
to know.

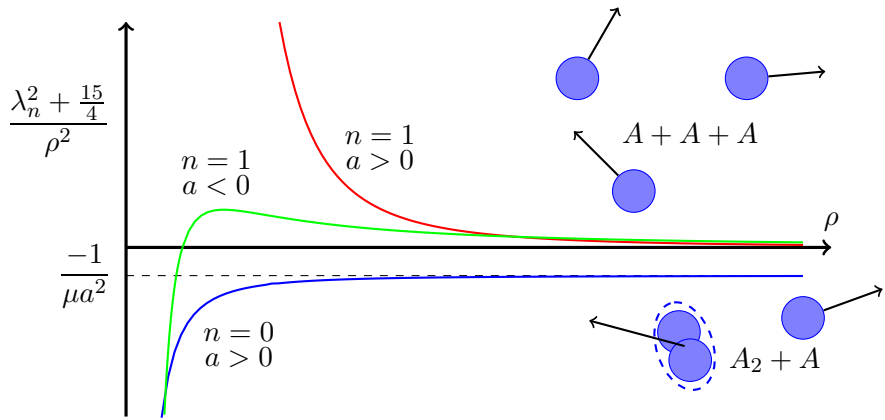
\begin{figure}[t!]
  \centering
  \tikzsetnextfilename{fig2_5}
  \begin{tikzpicture}[scale=0.75,inner sep=2mm]
\tikzstyle{particle} = [circle,fill,shading=ball,ball color=blue!50!red]
\tikzstyle{tarrows}  = [->,>=triangle 45,very thick]
\tikzstyle{particle} = [circle,draw=blue,fill=blue!50]

\draw[very thick,->] (0,0) -- (12.5,0) node[above] {$\rho$};
\draw[very thick,->] (0,-3) -- node[left,near end] {$\dfrac{\lambda_n^2+\frac{15}{4}}{\rho^2}$}(0,4);

\pgfmathsetmacro{\a}{2}
\pgfmathsetmacro{\c}{12*sqrt(0.5)*\a/pi}
\pgfmathsetmacro{\d}{(3*sqrt(3)+2*pi)*8*\a*\a/2/sqrt(3)/pi/pi}

\draw[dashed] (0,-2/\a/\a) node[left] {$\dfrac{-1}{\mu a^2}$} -- +(12,0);
\node at (2,-1.5) {$n=0$};
\node at (2,-1.95) {$a>0$};
\node at (4.5,2) {$n=1$};
\node at (4.5,1.55) {$a>0$};
\node at (1.6,1.55) {$n=1$};
\node at (1.6,1.1) {$a<0$};

\draw[thick,red]  plot[domain=2.72:12,samples=100] (\x,{(2+\c/\x+\d/\x/\x)^2/\x/\x});
\draw[thick,blue] plot[domain=0.63:12,samples=100] (\x,{(-2*\x*\x/\a/\a-1)/\x/\x});
\draw[thick,green]  plot[domain=1:12,samples=100] (\x,{5*(2*pow(\x,1/3)-2)/pow(\x,2)});
\draw[thick,green]  plot[domain=.68:1,samples=100] (\x,{-5*(2*pow(\x,1/3)-2)/pow(\x,2)});

\draw[particle]
                 (9,1)      node[particle] (d1) {}
                 (8,3)      node[particle] (d2) {}
                 (11,3)     node[particle] (d3) {}
                 (9.8,-1.5) node[particle] (c1) {}
                 (10,-2)    node[particle] (c2) {}
                 (12,-1.25) node[particle] (c3) {};

\draw[thick,->] (d1) -- +(135:1.5cm);
\draw[thick,->] (d2) -- +(60:1.45cm);
\draw[thick,->] (d3) -- +(5:1.4cm);

\draw[thick,->] ($0.5*(c1)+0.5*(c2)$) -- +(165:1.75cm);
\draw[thick,->] (c3) -- +(20:1.5cm);
\draw[dashed,thick,blue,rotate=108] ($0.5*(c1)+0.5*(c2)$) circle (0.8 and 0.5);

\node at (11.5,-2.0) {$A_2+A$};
\node at (10,2) {$A+A+A$};

\end{tikzpicture}
  \caption[Adiabatic potentials]{The adiabatic potentials for positive
  (the blue and red lines) and negative (the green line) scattering length
  $a$. The upper channels, $n=1$, correspond, at large $\rho$, to three
  free particles, $A+A+A$, far from one another for both positive and
  negative $a$. The lower channel, $n=0$, corresponds, at large $\rho$, to
  one free particles, $A$, and one dimer (shallow bound state of two
  particles), $A_2$. The asymptotic value for large $\rho$ of the $n=0$
  potential corresponds to the dimer binding energy. For negative $a$
  there is no similar potential.}
  \label{figure:2.5}
\end{figure}

Some interesting analytical solutions are found in the limit of $\rho\gg
|a|$. This corresponds to either all three particles far from each other
or two of the particles bound in a sub-system and the third particle far
away (the latter scenario only for positive $a$).

\fig{figure:2.5} shows the adiabatic potentials for $n=0$ and $n=1$ for
both positive and negative scattering length, $a$. The interpretation of a
dimer and a free particle is given in the channel marked with $n=0$ and
three free particles in the channels marked with $n=1$. The asymptotic
value of the $n=0$ potential is the binding energy of the dimer. The
channels will also be called adiabatic channels or adiabatic potentials.

\emph{Adiabatic channel $n=0$}

In the asymptotic limit of large $\rho$ compared to $a>0$ an analytical
solution is available. The lowest solution, denoted $n=0$ in the adiabatic
expansion \eqr{eq:2.38}, is found when the eigenvalue $\nu_0$ is
completely imaginary. Assume in the large $\rho/a$ limit that
$\nu_0=ic\rho$ with $c\in\mathbb R$
\begin{equation}
  \frac\rho{\sqrt\mu a}=\frac{ic\rho\cosh\left(c\rho\dfrac\pi2\right)-
    \dfrac8{\sqrt3}i\sinh\left(c\rho\dfrac\pi6\right)}
    {i\sinh\left(c\rho\dfrac\pi2\right)}
    \overset{\rho\rightarrow\infty}{\rightarrow}|c|\rho\;,
  \label{eq:2.56}
\end{equation}
where now $i$ is the complex unit. The right hand side is always positive,
so a solution is obtained only when the scattering length, $a$, is also
positive. Thus for large values of $\rho/a$ when $a>0$
\begin{equation}
  \nu_0(\rho)\overset{\rho\rightarrow\infty}=\frac{i\rho}{\sqrt{\mu}a}\;.
  \label{eq:2.57}
\end{equation}
In terms of the radial potential of \eqr{eq:2.42} this solution yields the
lowest potential with an asymptotic value of (in units of $\hbar=m=1$)
\begin{equation}
\frac{\lambda_0+\frac{15}{4}}{\rho^2}=
  \frac{\nu_0^2-\frac{1}{4}}{\rho^2}\overset{\rho\rightarrow\infty}{=}
    \frac{-1}{\mu a^2}\;.
  \label{eq:2.58}
\end{equation}
This value corresponds to the binding energy of a two-body subsystem in
the $n=0$ channel.

\emph{Adiabatic channel $n=1$}

The next solution is obtained by noting that $\nu=2$ yields a left hand
side that is infinite and thus also corresponds to the large $\rho$ limit.
Assuming the form
\begin{equation}
  \nu_1=2+\frac{c}{x}+\frac{d}{x^2}\;,
  \label{eq:2.59}
\end{equation}
where $x=\frac{\rho}{\sqrt{\mu}a}$ and $c$ and $d$ are constants. To
obtain $c$ and $d$ we put \eqr{eq:2.59} into \eqr{eq:2.55}.
Taylor-expansion of $\frac{1}{x}$ around $0$ and comparison of terms in
like powers of $x$ yields
\begin{equation}
  c=\frac{12}{\pi}\;,\quad d=\frac{8(3\sqrt{3}+2\pi)}{\pi^2\sqrt{3}}\;.
  \label{eq:2.60}
\end{equation}
This expression for $\nu_1$ is accurate to within $1\%$ for $x>5$. It is
furthermore valid for the scattering length, $a$, both positive and
negative.

A practical application of these asymptotic limits \eqr{eq:2.57} and
\eqr{eq:2.59} is described in appendix \ref{appendixA} where they are used
as initial guesses for the numerical solver routine.

\emph{The $a=\infty$ limit}

Precisely on resonance the scattering length diverges to either plus or
minus infinity. This situation is known as the universal limit and is the
origin of one of the most interesting effects in three-body physics,
namely the Efimov effect. The eigenvalue equation in this limit is
\begin{equation}
  \nu_n\cos\left(\frac{\nu_n\pi}2\right)-
    \frac8{\sqrt3}\sin\left(\frac{\nu_n\pi}6\right)=0\;.
  \label{eq:2.61}
\end{equation}
Again the solution for lowest the potential is imaginary $\nu_0=is_0$
where $s_0=1.0062378$ whereas the next solution is $\nu_1=4.46529$.

\subsection{Efimov and Thomas effects}
\label{efimov_and_thomas_effects}
With the basic properties of the angular eigenvalues established we now
turn to the radial equation \eqref{eq:2.42}. Whereas the zero-range
approximation yielded some useful analytical results the radial equation
is not quite so easy to work with. In the universal limit $a=\infty$, or
correspondingly $\rho=0$, however, an important analytical result can be
derived.

Using the value $\nu_0=is_0$ in the radial equation \eqref{eq:2.42} we get
\begin{equation}
  \left(-\frac{d^2}{d\rho^2}+\frac{-s_0^2-\frac{1}{4}}{\rho^2}-
    \frac{2mE}{\hbar^2}\right)f_0(\rho)=0\;.
  \label{eq:2.62}
\end{equation}
At very short distances, $\rho\ll k^{-1}$ where $k^2=2mE/\hbar^2$, the
energy term can be neglected. With an ansatz for the wave function of the
form $f_0=\rho^n$, a solution is found for $n=\frac{1}{2}\pm is_0$ or
\begin{equation}
  f_0(\rho)=\rho^{\frac12\pm is_0}=\sqrt\rho\exp(\pm is_0\ln\rho)\;.
  \label{eq:2.63}
\end{equation}
When $\rho\rightarrow0$ the exponential term will oscillate indefinitely,
corresponding to an infinite number of bound states. This is known as the
Thomas effect \cite{thomas} and is caused by a breakdown of the
short-range assumption of the contact interaction potential. Physical
potentials do obviously not have an infinite attraction at zero extension
and do therefore not suffer from this breakdown.

An important implication of this feature is that solving the radial
equation \eqref{eq:2.42} cannot be done from $\rho=0$ without
modification. One approach is to simply not include the origin but start
at a finite value, $\rc$, with the boundary condition $f(\rc)=0$. This is
known as a regularization cut-off \cite{PhysRevA.63.063608} and can be
attributed to the fact that three-body physics cannot be determined by
two-body physics alone in the zero-range approximation.

The WKB-method can be used to estimate the trimer bound state energies in
the universal limit using the quantization condition (with $m=\hbar=1$)
\cite{griffiths}
\begin{equation}
  \int_{\rc}^{\rho_\textnormal{t}}d\rho\sqrt{2E_p-\frac{\nu_0(\rho)^2}
  {\rho^2}}=\pi\left(p-\frac{1}{4}\right)\;,
  \label{eq:2.64}
\end{equation}
where the integer $p=1,2,\ldots$ indicates the ground state, first
excited, $\ldots$ etc. The regularization cut-off $\rc$ is used as the
inner turning point of the WKB integral and correspondingly
$\rho_\textnormal{t}$ is the outer classical turning point, for which
$2E_p=\nu_0(\rho_\textnormal{t})^2/\rho_\textnormal{t}^2$.

Note that in the integrand in \eqr{eq:2.64} the so-called Langer
correction term has been included. When applying the WKB approximation to
radial potentials this has been proven to yield much better results
\cite{langer}. In standard quantum mechanics this is done by replacing
$l(l+1)$ in the centrifugal barrier term by $(l+\frac{1}{2})^2$ or
equivalent adding $1/4$ in the numerator. This is why the term
$\nu_0^2-1/4$ is simply $\nu_0^2$ in the above.

In the universal limit $a=\infty$, where $\rho_t=s_0/\sqrt{-2E_p}$,
\eqr{eq:2.64} can be solved approximately to give the trimer bound state
energy
\begin{equation}
  E_p\approx-\frac{2s_0^2}{\rc^2} \exp\left(-\frac{2\pi p}{s_0}+
    \frac{\pi}{2s_0}-2\right)\;.
  \label{eq:2.65}
\end{equation}
Here we again see the Thomas effect, since the binding energy diverges as
$\rc^2$ when $\rc\rightarrow0$. Since $\rc$ is the only available length
scale in this limit this is the only possible relation between the energy
and $\rc$.

Another important result from this calculation is the scaling of the
energy with state number $p$, namely $E_{p+1}=e^{-2\pi/s_0}E_p\approx
E_p/515$. This is precisely the Efimov effect: in the universal limit
there exists an infinite number of bound states with a geometric scaling
law for the binding energies. Note that without the Langer correction term
in \eqr{eq:2.64} the Efimov scaling in \eqr{eq:2.65} would not have been
correctly obtained, which justifies its inclusion.

\subsection{The two-channel model, hyper-spherical edition}
\label{two_channel_model_hyperspherical_edition}

Here we combine the two-channel model from section \ref{two_channel_model}
with the hyper-spherical formalism of section
\ref{zero_range_model_hyperspherical_edition} to obtain the two-channel
hyper-spherical model. This section follows from
\citepublications{SFJ12a}.

The two-component angular wave function is
$\Phi(\rho,\Omega_i)=\left[\begin{matrix}
  \Phi_\cl(\rho,\Omega_i)\\
  \Phi_{\op\;\;}(\rho,\Omega_i)
\end{matrix}\right]$ with the same 'open' and 'closed' notation as in
section \ref{two_channel_model}. The boundary condition is a mix of
\eqr{eq:2.17} and \eqr{eq:2.49}.
\begin{equation}
  \frac{\p}{\p\alpha_i}
    \begin{bmatrix}
      \alpha_i\Phi_\cl\\
      \alpha_i\Phi_{\op\;\;}
    \end{bmatrix}
  \bigg|_{\alpha_i=0}=\frac{\rho}{\sqrt{\mu_i}}
    \begin{bmatrix}
      \dfrac{-1}{a_{i,\op}} & \beta_i\\
      \beta_i               & \dfrac{-1}{a_{i,\cl}}
    \end{bmatrix}
    \begin{bmatrix}
      \alpha_i\Phi_\cl\\
      \alpha_i\Phi_{\op\;\;}
    \end{bmatrix}
  \bigg|_{\alpha_i=0}\;,
  \label{eq:2.66}
\end{equation}
where $\beta_i$ is a coupling parameter and index $i$ denotes the
Jacobi-set. The Jacobi-index is suppressed from here on, since only
identical particles are considered. The hyper-angular equation
\eqref{eq:2.39} becomes
\begin{equation}
  \left(\Lambda+\frac{2m\rho^2}{\hbar^2}\left[V+
    \begin{bmatrix}
      \epsilon & 0\\
      0        & 0
    \end{bmatrix}\right]\right)
    \Phi(\rho,\Omega)=\lambda(\rho)\Phi(\rho,\Omega)\;,
  \label{eq:2.67}
\end{equation}
where the scalar quantities, $\Lambda$ and $V$, are multiplied by the
$2\times2$-identity matrix. The resulting differential equations for the
two components are identical except for the replacement
$\tilde\lambda(\rho)=\lambda(\rho)-\kappa^2\rho^2$ with $\kappa^2=
2m\epsilon/\hbar^2$ in the equation for $\Phi_\cl$. The Faddeev components
in the two-channel model with zero-range two-body interactions are
\begin{subequations}\label{eq:2.68}
  \begin{align}
    \varphi_{\op\;\;}(\rho,\alpha)&=N(\rho)\sin\left[
      \nu(\rho)\left(\alpha-\frac\pi2\right)\right]\;,
    \label{eq:2.68a}\\
    \qquad\varphi_{\cl}(\rho,\alpha)&=\tilde N(\rho)\sin\left[
      \tilde\nu(\rho)\left(\alpha-\frac\pi2\right)\right]\;,
    \label{eq:2.68b}
  \end{align}
\end{subequations}
with $\tilde\nu^2=4+\tilde\lambda=\nu^2-\kappa^2\rho^2$.

After rotations into the same Jacobi coordinate set using \eqr{eq:2.50}
the boundary condition \eqr{eq:2.66} becomes the $2\times2$ system of
equations
\begin{equation}
  \begin{bmatrix}
    \dfrac\rho{\sqrt\mu}\beta\sin\left(\dfrac{\nu\pi}{2}\right)&
      f_\cl(\tilde\nu)\\
    f_\op(\nu)&\dfrac\rho{\sqrt\mu}\beta
      \sin\left(\dfrac{\tilde\nu\pi}{2}\right)
  \end{bmatrix}
  \begin{bmatrix}
    N\\\tilde N
  \end{bmatrix}=0\;,
  \label{eq:2.69}
\end{equation}
where
\begin{equation}
  f_{\op/\cl}(x)=x\cos\left(\frac{x\pi}{2}\right)-
    \frac8{\sqrt3}\sin\left(\frac{x\pi}{6}\right)-
    \frac\rho{\sqrt\mu}\frac1{a_{\op/\cl}}
      \sin\left(\frac{x\pi}{2}\right)\;.
  \label{eq:2.70}
\end{equation}
The two-channel version of \eqr{eq:2.20} is
\begin{equation}
  \frac{\rho^2\beta^2}{\mu}\sin\left(\frac{\nu\pi}{2}\right)\sin\left(
    \frac{\tilde\nu\pi}{2}\right)-f_\op(\nu)f_\cl(\tilde\nu)=0\;.
  \label{eq:2.71}
\end{equation}

To find the effective scattering length, $a_\textnormal{eff}$, of this
two-level system, we note that the lowest eigenvalue in the single-channel
zero-range model, $\nu_0$, goes asymptotically as $\nu_0\rightarrow
i\rho/\sqrt{\mu}a$ for $\rho\rightarrow\infty$, see \eqr{eq:2.57}. Taking
the same limit in \eqr{eq:2.71} with the $\nu=\nu_0=i\rho/\sqrt\mu
a_\textnormal{eff}$ gives this equation for $a_\textnormal{eff}$
\begin{equation}
  \beta^2-\left(\frac{1}{a_\textnormal{eff}}-\frac{1}{a_\op}\right)
  \left(\sqrt{\frac{1}{a_\textnormal{eff}^2}+\mu\kappa^2}-
    \frac{1}{a_\cl}\right)=0\;.
  \label{eq:2.72}
\end{equation}
Given a set of model parameters, the effective scattering length can be
found from this equation. The general solution for $a_\textnormal{eff}$ as
a function of the parameters $\beta, \kappa, a_\cl, a_\op$ and $\mu$ is
not very handy. Near a Feshbach resonance, however, where $a$ diverges, we
have $a\gg\kappa^{-1}$, the square root in \eqr{eq:2.72} simplifies such
that an approximate solution can be obtained
\begin{equation}
  \frac1a_\textnormal{eff}\approx
    \frac1{a_\op}+\frac{\beta^2}{\sqrt\mu\kappa-\dfrac{1}{a_\cl}}\;.
  \label{eq:2.73}
\end{equation}
In spite of the apparent difference in derivations this looks quite
similar to \eqr{eq:2.22}.

Taking the expression for the effective range \eqr{eq:2.23} and making the
same replacement that leads to \eqr{eq:2.73} from \eqr{eq:2.23},
i.e. $\kappa\rightarrow \sqrt{\mu}\kappa$, gives
\begin{equation}
  R=\frac{-\beta^2}{\sqrt\mu\kappa\left(\sqrt\mu\kappa-
    \dfrac1{a_\cl}\right)^2}\;.
  \label{eq:2.74}
\end{equation}
Correspondingly \eqr{eq:2.25} becomes
\begin{equation}
    R_0=\frac{-1}{\sqrt\mu\kappa_0a_\op^2\beta^2}\;.
  \label{eq:2.75}
\end{equation}

For $a_\textnormal{eff}=\infty$, equivalently $\rho=0$, the solutions are
the same as for the single-channel zero-range model. This means that since
$\nu_0$ is imaginary, the Thomas effect persists even when the effective
interaction has a non-zero effective range. This can be understood by
noticing that the two-channel model consists of two single-channel
zero-range models that are coupled together by a coupling potential of
zero separation. This means that there is no scale coming from the
coupling that can regularize the three-body problem and in turn one still
needs to introduce a short-distance cut-off.

\subsection{Effective range expansion, hyper-spherical edition}
\label{boundary_condition_model_hyperspherical_edition}
Here the briefly mentioned effective range expansion of section
\ref{effective_range_expansion} is presented in the hyper-spherical
formalism. This model has the effective range, $R$, as an explicit
parameter and is thus simpler than the two-channel model. It is, however,
not based on any physical model but serves as a simpler way of including
the effective range. The simplicity allows for some analytical results to
be derived that are not readily available in the two-channel model but
coincide nicely with the corresponding numerical calculations.

The effective range expansion model uses \eqr{eq:2.8} as the boundary
condition. There, $R$ is just a parameter, independent off $a$. Later,
however, the dependency in \eqr{eq:2.24} will be adopted for actual
calculations. In reality both $a$ and $R$ depend on the external magnetic
field, $B$, but this dependency is implicitly fulfilled with the $R(a)$
expression.

The boundary condition \eqref{eq:2.49} becomes
\cite{FedorovJensen2001} (suppressing the index $i$)
\begin{equation}
  \frac{\partial(\alpha\Phi)}{\partial\alpha}\bigg|_{\alpha=0}=
    \frac{\rho}{\sqrt{\mu}}\left[-\frac1{a}+
    \frac12R\frac{\mu\nu^2}{\rho^2}\right]
    \alpha\Phi\bigg|_{\alpha=0}\;,
  \label{eq:2.76}
\end{equation}
where $R$ is the effective range between the particles, which are assumed
identical, and the momentum in \eqr{eq:2.8} is given by
$k=\nu/(\sqrt2\rho)$ \cite{FedorovJensen2001}. Again assuming that all
particles are identical, the eigenvalue \eqr{eq:2.55} becomes
\begin{equation}
  \frac{\nu\cos\left(\dfrac{\nu\pi}2\right)-
    \dfrac{8}{\sqrt3}\sin\left(\dfrac{\nu\pi}6\right)}
    {\sin\left(\dfrac{\nu\pi}2\right)}=
    \frac{\rho}{\sqrt\mu}\left(\frac{1}{a}-
    \frac{1}{2}R\frac{\mu\nu^2}{\rho^2}\right)\;.
  \label{eq:2.77}
\end{equation}
As before, the limit $\rho\rightarrow\infty$ grants a little insight. With
$\nu=i\rho/a_\textnormal{eff}$ one obtains for positive $a$
\begin{equation}
  a_\textnormal{eff}=\frac{a+\sqrt{a^2-2aR}}{2}\;.
  \label{eq:2.78}
\end{equation}
For $R=0$ this appropriately yields $a_\textnormal{eff}=a$. For $R\ll a$,
the dimer binding energy is
\begin{equation}
  E_D=\frac{1}{a_\textnormal{eff}^2}\approx
    \frac{-1}{a^2}\left(1+\frac{R}{a}\right)\;.
  \label{eq:2.79}
\end{equation}
Thus the dimer system can become more bound or less bound depending on the
sign of the effective range. In the case of atomic Feshbach resonances the
effective range is negative \cite{chin2010}, which lowers the binding
energy of the dimers, i.e. it becomes less negative.

\subsubsection{The short distance limit}
The small $\rho$ limit is distinct in this model. For $\rho\rightarrow0$
we write the lowest eigenvalue as $\nu_0=\sqrt{b\rho+c\rho^2}$ which
yields
\begin{equation}
  b=\frac{16\pi-12\sqrt3}{3\sqrt3\pi\sqrt\mu R},\qquad
  c=\frac{1458R+(256\pi^2+240\sqrt3\pi-972)a}{729a\mu R^2}\;,
  \label{eq:2.80}
\end{equation}
which is valid for $\rho\ll|R|$. Thus $\nu_0^2\propto\rho$ in the small
$\rho$ limit. This means that a regularization cut-off is not urgently
needed in this model. The calculation leading to \eqr{eq:2.63} would have
$s_0=0$ and $f(\rho)$ would not oscillate indefinitely. Also, the integral
in \eqr{eq:2.64} would be finite for $\rho_t=0$ since the divergent
integrand would be of the form $1/\sqrt{\rho}$ which can easily be
integrated. Thus the Thomas collapse is avoided. The Efimov effect
persists, however. Even though a cut-off is not needed in this model we
will still apply one since this allows us to use the cut-off as a fitting
parameter when comparing to experimental data.

The limit of $R\rightarrow0$ should correspond to the zero-range model.
This is indeed still the case despite the apparent divergence of the
expressions in \eqr{eq:2.80}. This is best seen in \fig{figure:3.1} where
the eigenvalues are plotted for several different effective ranges. For
small $|R|$ the effective range expansion model eigenvalues stay close to
the zero-range model eigenvalues until at $\rho\sim|R|$ where the
eigenvalues tend towards zero with a slope that increases for decreasing
$|R|$ precisely as in \eqr{eq:2.80}.

\subsubsection{Positive effective range}
There is no apparent reason that we cannot choose $R$ to be positive in
this model. However, if $R$ is chosen too large, for some specific value
of $a$, the eigenvalue equation \eqref{eq:2.77} cannot be solved, neither
for real nor complex eigenvalues. We will investigate the effects of the
sign of the effective range briefly in chapter
\ref{finite_range_effects_chapter}. To do so without the here mentioned
breakdown an additional term, proportional to $k^4$, must be included in
the effective range expansion model. The additional parameter, $P$, chosen
to be dimensionless, is known as the shape parameter
\cite{FedorovJensen2001}
\begin{equation}
  \lim_{k\rightarrow0}k\cot\delta(k)=-\frac{1}{a}+\frac12Rk^2+PR^3k^4\;.
\label{eq:2.81}
\end{equation}
The equations \eqref{eq:2.76} and \eqref{eq:2.77} get similar additional
terms. The shape parameter can be derived for the two-channel model if the
Taylor expansion of \eqr{eq:2.21} is carried out to fourth order in $k$.
For typical values of $a$ and $R$ this yields $P\sim0.1$. A similar
calculation for the finite box potential also yields similar values of
$P$. $0.1$ will thus be universally applied in this thesis where relevant.


\section{Three-body recombination rate}
\label{recombination}
A very important quantity that will be investigated in great detail in
this thesis is the recombination rate which is currently the only
experimental way to obtain information about three-body physics.

By recombination in this context is understood the process of three free
particles $A+A+A$ interacting with the result that two of them become
bound (they form a dimer) and the third carries away excess energy and
momentum. The resulting dimer $A_2$ and free particle $A$ have increased
kinetic energy due to the overall increased binding energy in the system.
The result is a loss of all three particles from the trap generally
leading to atom number loss of the form \cite{Zaccanti}
\begin{equation}
  \dot n = -\rec n^3\;,
  \label{eq:2.82}
\end{equation}
where $n$ is the atom number density, $\rec$ is known as the recombination
coefficient and the dot indicates derivative with respect to time.

The overall dependency for the recombination coefficient $\rec$ can be
obtained by simple dimensional analysis. Combining the scattering length
(which is the only available length scale in the zero-range model) with
the particle mass $m$ and Planck's constant $\hbar$ to obtain a unit of
\emph{length}$^6/$\emph{time}, which is the unit of the recombination
coefficient, can be done in one way only
\begin{equation}
  \rec=C\frac{\hbar a^4}{m}
  \label{eq:2.83}
\end{equation}
where $C$ is a dimensionless proportionality factor that cannot be
determined by this analysis. It turns out that $C$ is log-periodic with
the property that $C(a)=C(22.7a)$ (in the case of equal mass particles).
Furthermore, the overall $a^4$ dependency is modulated by a series of
characteristic minima in the spectrum for positive $a$ and a corresponding
series of maxima for negative $a$. See for instance Figures
\ref{figure:3.6} and \ref{figure:5.5}.

Two alternative ways of calculating the recombination coefficient will be
presented here. Both rely on calculating the transition matrix element
from the initial state of three free particles to the final state of a
dimer and a free particle. What is meant by a bound dimer is slightly
different depending on the sign of the scattering length. As noted earlier
in section \ref{zero_range_model} there exist bound dimers for positive
scattering lengths only. Or rather, only \emph{weakly} bound dimers, with
binding energy of the order $1/a^2$, exist for positive scattering length.
Zero-range potentials can describe only these weakly bound states. For
physical potentials the weakly bound dimer is the state just below
threshold when the scattering length is large. Imagine tuning the
potential strength such that the scattering length diverges, the
divergence is a manifestation of the existence of the weakly bound dimer
state. Of course in the real potential deeply bound states may also exist
with binding energies of the order $1/r_e^2$ where $r_e$ is the physical
range of the potential, but they do not concern us at this stage.

For negative scattering length, $a$, only the deeply bound states exist.
These bound states cannot be described by zero-range potentials, neither
in the two-channel nor in the effective range expansion model. Here, the
effective range modifications apply only to the weakly-bound states.

These two very different meanings of bound dimers might suggest that the
processes by which the recombination occurs are very different, depending
on the sign of the scattering length. This is also clearly seen in the
recombination coefficient, as noted above, with troughs/peaks for
positive/negative scattering lengths. Two approaches to the problem will
be taken. The methods share some of the same basic ideas but with some
clear differences.

\subsection{Hidden crossing theory}
\label{hidden_crossing_theory}
As noted in section \ref{zero_range_model_hyperspherical_edition} the
adiabatic potentials for positive scattering length correspond, at large
hyper-radius, to either three free particles, $n=1$, or one free particle
and the other two bound in a weak dimer, $n=0$. The transition from the
former state to the latter is known as recombination. The usual way of
attacking this problem consists of solving the coupled set of equations in
\eqr{eq:2.40} with a high number of adiabatic channels. This involves
calculating not only the $\lambda$'s but also $P$'s and $Q$'s before the
actual radial equations. This has been done \cite{wang11} and we do not
seek to reproduce the results of this method. Instead a much simpler
approach is taken, namely the method of hidden crossings.

\begin{figure}[ht]
  \centering
  \tikzsetnextfilename{fig2_6}
  \begin{tikzpicture}
\definecolor{currentcolor}{rgb}{1,0.82745,0}

\draw[fill=currentcolor,very thick] plot[domain=-4:4,samples=150] (\x,{exp(-0.55*(\x)^2)*2});

\begin{scope}[yshift=0.5cm]
	\clip (0,.5) ellipse (4 and 1.4);
	\draw[very thick,blue] plot[smooth] file {data/chap2/tunneling.dat};
\end{scope}

\draw[very thick,->] (-4,0) -- (4,0) node[right] {$x$};

\draw[dashed,thick] (-0.84,0) node[below] {$x_a$} -- ++(0,2);
\draw[dashed,thick] (0.84,0) node[below] {$x_b$} -- ++(0,2);

\draw[fill=currentcolor,very thick] (2.5,2) rectangle (3,2.25);
\node at (3.5,2.1) {$V(x)$};
\draw[->,very thick,red] (2.5,1.25) -- node[above,black] {${\color{red}T}e^{ikx}$} (4,1.25);
\draw[->,very thick] (-4,1.75) -- node[above,black] {${e^{ikx}+\color{orange}R}e^{-ikx}$} (-2,1.75);
\draw[<-,very thick, orange] (-3,1.6) -- (-2,1.6);

\draw[very thick, magenta,dashed] (-2,1.35) -- (2,1.35) node[above left]{$E$};
\end{tikzpicture}
  \caption[WKB tunnelling]{Tunnelling through a potential barrier. The
  points $x_a$ and $x_b$ indicate the classical turning points
  corresponding to the energy $E$ such that $V(x_a)=V(x_b)=E$. To the left
  of the potential the wave consists of an incoming wave with amplitude 1
  and a reflected wave with amplitude $R$. To the right of the potential
  there is only the transmitted wave with amplitude $T$.}
  \label{figure:2.6}
\end{figure}
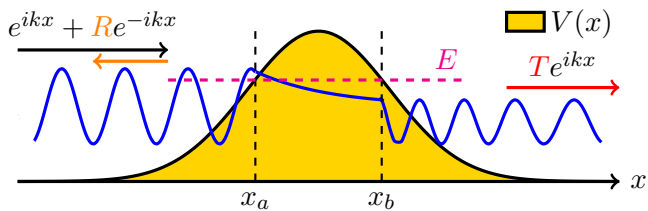

Introductory quantum mechanics \cite{griffiths} presents the simple WKB
approximation that has been already used in \eqr{eq:2.64} to estimate the
bound state energies. A similar estimation can be done for the tunnelling
probability through a potential barrier, as illustrated in
\fig{figure:2.6}. The transmission amplitude, $T$, for tunnelling through
the barrier given by $V(x)$ is $T\approx e^{-2\gamma}$ where
\begin{equation}
  \gamma = \frac{1}{\hbar}\int_{x_a}^{x_b}|p(x)|dx
         = \frac{1}{\hbar}\int_{x_a}^{x_b}\sqrt{2m(E-V(x))}dx\;,
  \label{eq:2.84}
\end{equation}
where $x_a$ and $x_b$ denote the classical turning points that correspond
to the energy $E$ such that $E=V(x_a)=V(x_b)$. \fig{figure:2.6}
illustrates the set-up, the wave incoming from the left is partially
transmitted with the amplitude $T$ and partially reflected with the
amplitude $R$.

Now imagine that the wave incoming from the left describes three free
particles in the hyper-spherical picture while the transmitted wave
describes a dimer and a free particle, i.e. the left wave lies on the
adiabatic channel $n=1$ in \fig{figure:2.5} and the right wave on channel
$n=0$. An integration path from $n=1$ to $n=0$ does not seem readily
available, however, this is only the case on the real line.

\begin{figure}
  \centering
  \input{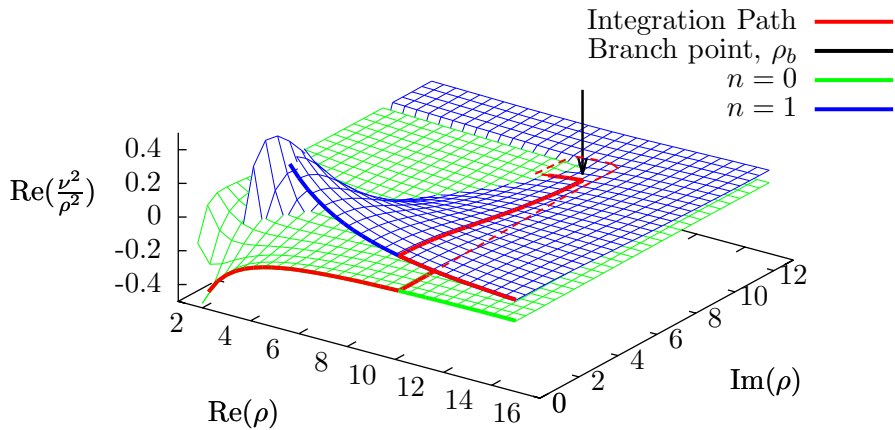}
  \caption[Hidden crossing]{A visualization of a hidden crossing in the
  complex $\rho$-plane. The adiabatic radial potentials are extended to
  complex $\rho$ which results in a sheet for each channel. The sheets
  intersect at the branch point, $\rho_b$, indicated by the arrow. An
  integration path (the red line) starting on one branch and encircling
  the branch point will return to the real axis on another branch.}
  \label{figure:2.7}
\end{figure}

\fig{figure:2.7} is an extension of \fig{figure:2.5} into the complex
$\rho$-plane. For $\textnormal{Im}(\rho)=0$ the figure is just
\fig{figure:2.5} at a tilted angle. Something interesting happens out in
the complex plane. The potential curves do no get extended into separate
sheets but are in fact part of the same multi-layered surface. The
integration path marked in red is obtained by taking a small continuous
step all the way, yet starting from one adiabatic potential, $n=1$, and
ending out in another, $n=0$. This is because the shown path encircles the
branch point marked by an arrow. This effect is akin to the square root
function of complex numbers which is not defined on the negative real
axis. Correspondingly, going around the branch point twice before heading
back to the real line would lead back to the initial curve.

Near the branch point the eigenvalue $\nu(\rho)$ behaves like a square
root type function of $\rho$. The branch point is found by solving
\cite{Nielsen2001}
\begin{equation}
  \frac{d\rho}{d\nu}\bigg\vert_{\nu_\text b}=0,
  \label{eq:2.85}
\end{equation}
for complex $\nu_b$ and evaluate $\rho_b=\rho(\nu_b)$ through
\eqr{eq:2.55}. In the single-channel zero-range model the branch-point is
$\rho_b\approx(2.592 + 2.974i)\sqrt\mu a$. For the two-channel model
$\rho(\nu)$ is given only implicitly and \eqr{eq:2.85} must be solved
numerically for each set of two-channel parameters. The above value for
$\rho_b$ is, however, still approximately correct. Likewise for the
effective range expansion model.

The WKB integration starts at the outermost classical turning point in the
adiabatic channel $n=1$, goes towards the origin but stops at $\rho_b$,
then goes out into the complex plane, around $\rho_b$ and back to the real
axis, now on channel $n=0$ and down to the innermost turning point. As
noted in section \ref{efimov_and_thomas_effects} the lower adiabatic
channel diverges as $\rho^{-2}$ for small $\rho$. Here enters the
regularization cut-off, which puts the lower turning point at an infinite
potential barrier at short distance. The integral takes the form
\begin{equation}
  \Delta+iS=
    \int_\textnormal{path}d\rho\sqrt{k^2-\frac{\nu(\rho)^2}{\rho^2}}\;,
  \label{eq:2.86}
\end{equation}
where 'path' is the integration path just described and
$\Delta+iS$ is a generalization of $\gamma$ in \eqr{eq:2.84} with $\Delta$
and $S$ purely real. Note that again the Langer correction is applied as
in \eqr{eq:2.64}. The transition probability is given by
\begin{equation}
  P(k)=4e^{-2S}\sin^2\Delta\;,
  \label{eq:2.87}
\end{equation}
from which the recombination coefficient is found as \cite{Macek1999}
\begin{equation}
  \rec=8(2\pi)^23\sqrt3\frac\hbar{m}\lim_{k\rightarrow0}
    \frac{P(k)}{k^4}\;,
\label{eq:2.88}
\end{equation}
with the wave number $k$ defined by $E=\hbar^2k^2/2m$.

The described integration path is actually the effective sum of two
separate paths. One starts on the branch $n=1$ at large $\rho$, goes
around the branch point and ends up on $n=0$, goes in towards the
innermost turning point and then out to large $\rho$ on the $n=0$ branch
with opposite sign in the integrand. The second path starts also on the
branch $n=1$, goes in to the classical turning point \emph{on the
n${}={}$1 branch}, then heads towards $\rho_b$ and goes around it, back to
the real axis on branch $n=0$ and then again out to large $\rho$. The
coherent sum of the results of these two integration paths is exactly the
same as for the 'path' described above \cite{Nielsen2001}.

The more familiar form $\rec=C(a)\hbar a^4/m$ (as for instance found in
\cite{braaten2006} or from the dimensional analysis argument above) can be
found from the above equations in the universal limit, $a = \infty$, using
the single-channel zero-range model where $\nu_0(\rho) = is_0$. Split the
integral into two parts: one from the cut-off, $\rc$, to the real part of
the branch point, $\rho_b$, and another for the rest. Denote the first
part by $\Delta_1+iS_1$, then the result is
$\Delta_1=s_0\log\left(\dfrac{\textnormal{Re}(\rho_b)}{\rc}\right)$ and
$S_1=0$ since the potential is negative in the lower branch and $k^2$ is
positive and thus the integrand is purely real. From \eqr{eq:2.85} we have
$\textnormal{Re}(\rho_b)\propto a$. When plugging this into \eqr{eq:2.87}
the log-periodic dependence is established. The rest of the integration
path only leads to a constant phase shift $\Delta_2$ independent of $a$
since $\nu(\rho)$ only depends on the ratio $\rho/a$.

That the limit for $k\rightarrow0$ exists can be seen from the integration
on the upper branch. Here $\nu_1=2$ asymptotically and the outer turning
point is given by $\rho_t=2/k$. On this branch the energy is lower than
the potential and a purely imaginary contribution is found with the value
$S_2=2\ln\dfrac{\rho_t}{\textnormal{Re}(\rho_b)}$ where the factor of $2$
is due to the value of $\nu$. Exponentiation leads to $P(k)\propto
\left(\frac{\textnormal{Re}(\rho_b)}{\rho_t}\right)^2\propto k^4a^4$ since
again $\textnormal{Re}(\rho_b)\propto a$ and $\rho_t=2/k$. The Langer
correction was again essential to obtain the correct behaviour.

The method for calculating the recombination coefficient when $a$ is
negative will be described in chapter \ref{optical_model}.


\chapter{Finite range effects}
\label{finite_range_effects_chapter}

\emph{Effects of the finite effective range in the two-channel and
effective range expansion models are investigated by comparing their
predictions for the trimer bound state spectrum and recombination rates
with the predictions from the single-channel zero-range model. A
comparison to experimental data is also performed.}

\noindent\rule{\textwidth}{0.4pt}

\noindent In this chapter the effects of the effective range are
investigated by comparing results for the single-channel zero-range model
of section \ref{zero_range_model_hyperspherical_edition} to the
two-channel model from section
\ref{two_channel_model_hyperspherical_edition} and the effective range
expansion model of section
\ref{boundary_condition_model_hyperspherical_edition}. The main effect is
that the characteristic Efimov scaling factor, which equals 22.7 for
identical particles when the effects of finite effective range are not
included, depends on both the value and sign of the effective range. As
noted previously in section \ref{two_channel_model} the two-channel model
has negative effective range, so only the effective range expansion model
is applied when investigating dependency on the sign of the effective
range. The results are based on work from \citepublications{SFJZ13}.

The comparison between the models is done first by plotting the adiabatic
potentials for the different models and noting the differences. These
simple plots provide good explanations for the effects observed in the
calculated observables.

The dependency of the Efimov scaling factor on the effective range is
found first by calculating the recombination coefficient for positive
scattering lengths using the method of hidden crossing described in
section \ref{hidden_crossing_theory}. The effect is observed as a
reduction of the distance between consecutive minima in the recombination
spectrum.

On the negative $a$ side this method cannot be used, instead the trimer
bound state spectrum is investigated. Specifically we look at a certain
threshold value of $a$ where the lowest Efimov trimer state appears from
the three-body continuum. This threshold value of $a$, denoted as
$a^{(-)}$ in most papers, is known as the three-body parameter, which in
recent studies \cite{berninger2011, naidon2011, chin2011, wang2012,
schmidt2012} has an apparent universal relation to the van der Waals
length of the inter-atomic two-particle potential. This effect is
investigated further in the next chapter.

\section{Model Comparison}
\label{Model_comparison}

\begin{figure}[ht]
  \centering
  \input{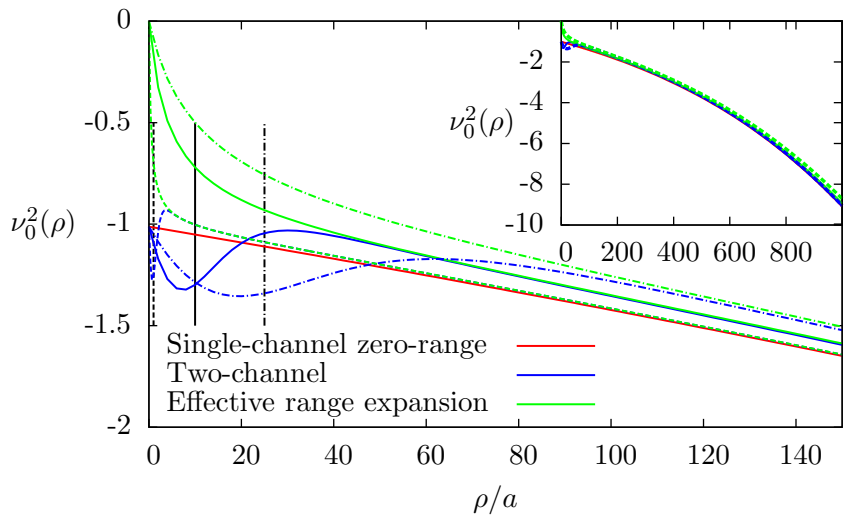}
  \caption[Lowest eigenvalue for small and large $\rho$]{Small and large
  $\rho$ behaviour of the hyper-angular eigenvalues for $n=0$, for the
  three models with $a=500a_0$, $R_0=-a_0$ (short dash), $R_0=-10a_0$
  (solid lines) and $R_0=-25a_0$ (dash dot) where $a_0$ is the Bohr
  radius. The vertical black lines indicate $|R_0|$ at these respective
  values. Notice how the eigenvalues for the effective range expansion
  model tend to 0 for $\rho\rightarrow0$. As the discussion around
  \eqr{eq:2.80} suggests, this means that the model does not require a
  cut-off.}
  \label{figure:3.1}
\end{figure}

Before considering the recombination rates and the binding energies in the
different models, we first make a comparison of the models in terms of the
adiabatic eigenvalues $\nu_0(\rho)$ that provide the effective potential
for the three-body system in the hyper-radial equation \eqref{eq:2.42}.
The models are compared in \fig{figure:3.1} by explicitly plotting their
associated eigenvalues. At large $\rho$ all models have the same
asymptotic value $\nu^2=-2\rho^2/a^2$ as noted in \eqr{eq:2.57} (note
that $\mu=\frac{1}{2}$ in this chapter). This is more clearly
illustrated in the inset of \fig{figure:3.1}, where the horizontal axis
extends up to $\rho/a=1000$. For intermediate distances
$\rho\gtrsim2|R_0|$, where $R_0$ is the effective range at $a=\infty$
from \eqr{eq:2.14}, the finite-range models show surprisingly similar
forms given their quite different formalism. The two-channel model has
both a barrier with respect to the single-channel zero-range model as
well as an inner pocket region. This feature is key to understanding
some results of the full calculations in the later sections. The
effective range expansion model only has a barrier with respect to (i.e.
it is strictly larger than) the single-channel zero-range model.
\begin{figure}[ht]
  \centering
  \input{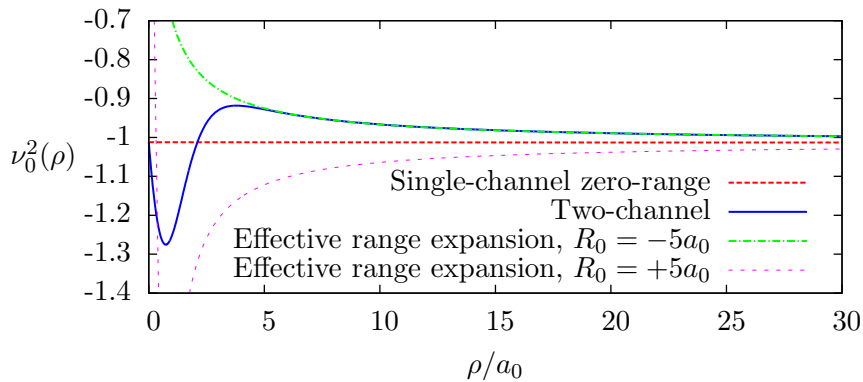}
  \caption[Lowest eigenvalue for small and large $\rho$]{The same
    eigenvalues as in \fig{figure:3.1} for $a=\infty$. Positive effective
    range is included for the effective range expansion model. The single
    channel solution has the constant value $-s_0^2\approx-1.012$.}
  \label{figure:3.2}
\end{figure}

\fig{figure:3.2} is similar, but plotted for $a=\infty$, where the same
features are seen. Additionally the eigenvalue solution for the effective
range expansion model is also plotted for positive effective range with
the shape parameter from \eqr{eq:2.81} $P=0.1$. It appears to have almost
mirror symmetry around the single-channel zero-range solution, except for
very short distances where both the positive and negative effective range
solutions go to zero. For positive effective range only a pocket region is
observed. The eigenvalues for $n=1$ would show very similar tendencies if
they were plotted as in \fig{figure:3.1} and \fig{figure:3.2}.

\section{Bound trimers}
\label{boundTrimers}
Effects of the effective range are now considered by studying the trimer
bound state spectrum for the different models. When the scattering length,
$a$, is large, the WKB expression \eqref{eq:2.64} yields the same trimer
binding energy, $E_T(a)$, as the radial equation \eqref{eq:2.42}. However,
for small $a$ or energies close to $0$, the radial equation provides the
best results. The boundary conditions for the radial solutions are
$f(\rc)=0$, following the regularization procedure, and
$f(\rho_\textnormal{max})=0$ for some large $\rho_\textnormal{max}$,
chosen such that the bound state energy has converged to the desired
degree of accuracy. Notice that we also regularize the effective range
expansion model so that we can move the trimer bound state energies while
keeping the effective range fixed.

\begin{figure}[ht]
  \centering
  \input{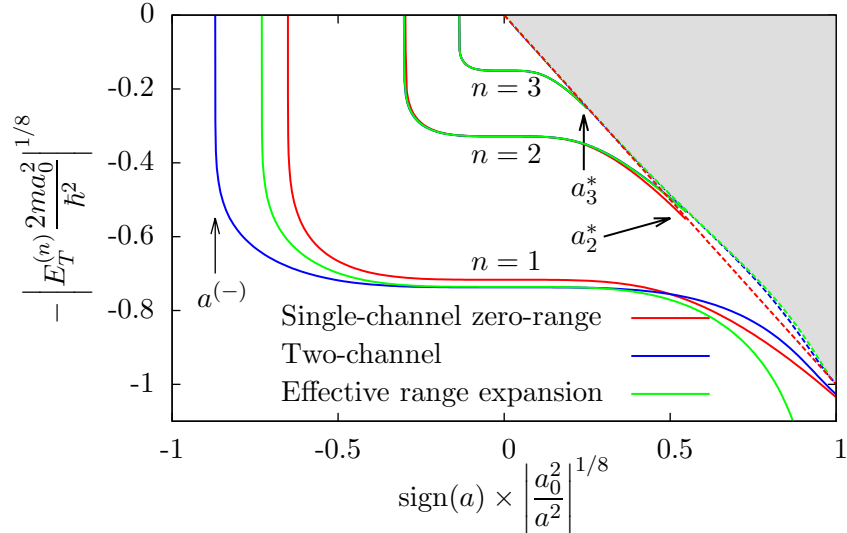}
  \caption[Trimer bound state energies]{The trimer bound state energy,
  $E_T^{(n)}$, versus inverse scattering length, $a$, squared for
  $R_0=-5a_0$ where $a_0$ is the Bohr radius. The superscript $n=1,2,3$
  indicates the lowest, first excited and second excited trimer states,
  respectively. Both axes are scaled to the power $1/8$ to reasonably fit
  the entire spectrum in the plot. Dashed lines indicate the atom-dimer
  threshold for positive scattering lengths. The annotated points $a_i^*$
  indicate where the trimer bound states disappear into the atom-dimer
  continuum, shown as the light grey area above the dashed lines. Likewise
  $\aminus$ indicates the threshold for the lowest trimer disappearing
  into the three-body continuum for negative $a$.}
  \label{figure:3.3}
\end{figure}

\fig{figure:3.3} shows the three lowest trimer bound state energies,
$E_T^{(n)}$, where $n$ indicates the level of excitation, as function of
the scattering length for each of the three models. For positive $a$ the
dashed lines indicate the atom-dimer threshold which is given by the dimer
binding energy $-1/a^2$ for the single-channel zero-range model and by
\eqr{eq:2.79} for the effective range expansion model. For the two-channel
model this can be calculated only numerically, yet it agrees surprisingly
well with the analytical formula for the effective range expansion model.
It can be hard to discern in the figure, but the green and blue dashed
curved are practically identical. The regularization cut-offs were chosen
such that the excited trimer energies, $n=2$, coincide with each other for
all the models at $|a|=\infty$. The three spectra for $n=3$ are virtually
identical. This is reasonable since the binding energy is very small for
the $n=3$ states and hence the wave function lives at very large
hyper-radius and is almost completely insensitive to short-range effects.
However, for the lowest state, $n=1$, a clear distinction between the
models appears. At $|a|=\infty$ the finite-range models give practically
the same trimer energy, a factor of $\sim25.3^2$ times larger than the
$n=2$ state (for this particular choice of effective range). In comparison
the single-channel zero-range model trimer energy is only a factor of
$22.7^2$ times larger, which is the usual Efimov scaling factor.

\subsection{Threshold for trimer creation}
\label{trimer_threshold}
For negative scattering length, $a$, the value of $\aminus$ indicates the
threshold scattering length for creation of the lowest Efimov trimer. When
given in units of the van der Waals length, $\vdW$, this quantity is the
subject of much recent discussion since it seems to have the universal
value of $\aminus\sim -9.8\vdW$ for different cold atomic systems
\cite{berninger2011}. The relation between $\aminus$ and $\vdW$ is
discussed in the next chapter. In this section we focus on finite range
effects in $\aminus$ for the two models. Some other works that address
finite range effects on the threshold value $\aminus$ can be found in
\cite{tfj08c} and \cite{naidon2011}.

\begin{figure}[ht]
  \centering
  \input{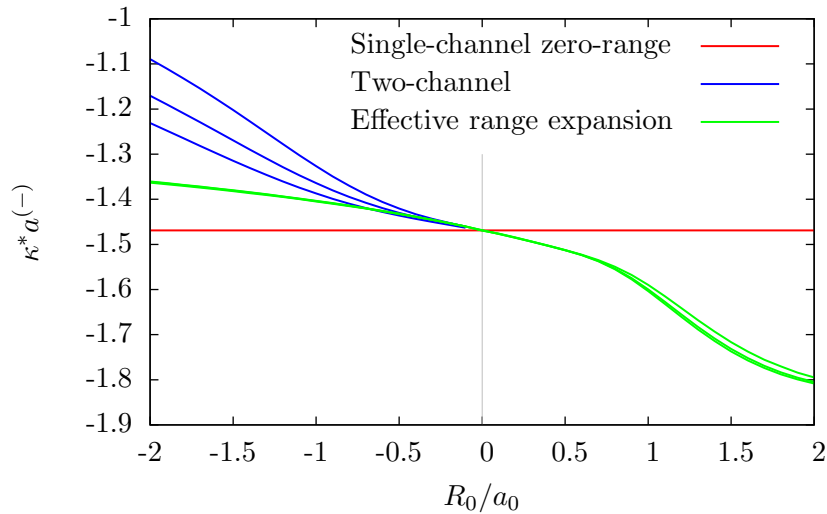}
  \caption["Universal" product $\kappa^*\aminus$]{The product
    $\kappa^*\aminus$ as a function of effective range, $R_0$, where
    $(\kappa^*)^2=-2mE_T^{(1*)}/\hbar^2$ with $E_T^{(1*)}$ being the
    binding energy of the lowest trimer at $|a|=\infty$ and $\aminus$ is
    the threshold scattering length for trimer creation as indicated in
    \fig{figure:3.3}. The universal value $\kappa^*\aminus=-1.5076$
    \cite{PhysRevLett.100.140404}for the single-channel zero-range model
    is not correct for the lowest trimer, we find the value $-1.469$
    instead, independent of cut-off. The two-channel model curves are for
    different value coordinate-space cut-offs on the hyper-radial
    potential. The cut-off is $0.5,\,0.6$ and $0.7$ in units of
    $a_\textnormal{bg}$ for the top, middle and bottom blue curves. For
    the effective range expansion model the dependency on the cut-off is
    insignificant. Note that the two-channel model only allows $R_0<0$.
  Here $a_0$ is the Bohr radius and $R_0$ is the effective range on
resonance from \eqr{eq:2.14}.}
    \label{figure:3.4}
\end{figure}

The results within the different models for $\aminus$ as a function of
$R_0$ are shown in \fig{figure:3.4}. Most noticeable is the decrease in
$\aminus$ for the finite-range models compared to the single-channel
zero-range model for negative $R_0$. This is partly due to the lower
binding energy $E_T^{(1*)}\equiv E_T^{(1)}|_{|a|=\infty}$. The product
$\kappa^*\aminus$ (where $(\kappa^{*})^2=-2mE_T^{(1*)}/\hbar^2$) is
universal in the single-channel zero-range model
\cite{PhysRevLett.100.140404}. Thus increasing $|E_T^{(1*)}|$ will reduce
$|\aminus|$. However, this effect is not enough to account for the
deviation from the single-channel zero-range result. The product
$|\kappa^*\aminus|$ is further reduced for decreasing $R_0$, indicating a
lower value of $\aminus$.

Both finite-range models show the same trend for negative $R_0$. However,
the value of the change is different for the two models when $|R_0|$ gets
sufficiently large. The plot also shows that for the two-channel model the
effect depends on the cut-off. The cut-off, chosen for the purpose of
illustration, but with reasonable values, is $0.5a_\textnormal{bg},\,
0.6a_\textnormal{bg}$ and $0.7a_\textnormal{bg}$ for the top, middle and
bottom blue curves in \fig{figure:3.4}, where $a_\textnormal{bg}$ is the
background scattering length far from resonance. The effective range
expansion model shows only a very small dependency on the cut-off (the
three green curves are almost identical).

A similar calculation of $\kappa^*\aminus$ was done in reference~\cite{
schmidt2012}. However, they find that the trimer binding energies for
$|a|=\infty$ get smaller compared to the single-channel zero-range value
and the value of $|\aminus|$ gets larger for larger effective range. As
this is exactly the opposite of the present calculation they must be using
a finite range potential with a positive effective range. To obtain
results for positive effective range the boundary condition with the shape
parameter \eqr{eq:2.81} in the effective range expansion model is used.
Indeed the opposite behaviour is obtained as seen in \fig{figure:3.4}.

\begin{figure}[ht!]
  \centering
  \input{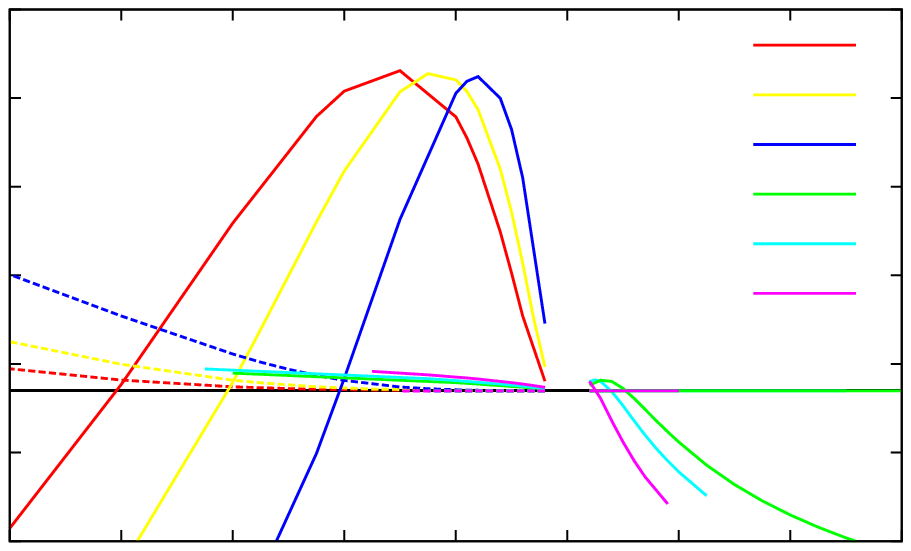}
  \caption[Ratio of bound state energies on resonance]{The ratio of the
  trimer bound state energies on resonance, $E_T^{(n*)}$, as a
  function of the effective range for several different cut-offs. The
  solid lines show the ratio of energies for the first and second state,
  the dashed lines for the second and third state. Red, yellow and blue
  curves (top three in legend) are for the two-channel model while green,
  cyan and magenta curves (bottom three in legend) are for the effective
  range expansion model. The reference value is the horizontal black line
  which lies at $22.7$, the value for the single-channel zero-range model.
  }
  \label{figure:3.5}
\end{figure}

In order to better understand the behaviour of the trimer energies $E_T$
and $\aminus$, \fig{figure:3.5} shows the ratio of the trimer bound state
energies on resonance for the two-channel and the effective range
expansion models for three different cut-offs, chosen for the purpose of
illustration. The most noticeable feature is the change of the Efimov
scaling ratio $E_T^{(1*)}/E_T^{(2*)}$ (which equals $22.7^2\approx515$ in
the single-channel zero-range model) between the two lowest trimer states
in the two-channel model. This non-monotonous behaviour can be understood
if one assumes a three-body wave function that lives at large hyper-radii,
$\rho$. When the effective range is decreased from zero ($|R_0|$
increases) the barrier in \fig{figure:3.1} initially decreases the binding
energy with respect to the pure single-channel zero-range model. As the
effective range increases, the wave function will leak into the attractive
pocket at small $\rho$, which will again increase the binding energy
compared to the single-channel zero-range result. This effect is strong
for the ratio of the two lowest trimers but becomes weaker for the ratio
of the two highest trimers. This is understandable since the least bound
trimers reside at very large hyper-radii and are largely insensitive to
the short-range changes in the hyper-radial potential. The effect is seen
for all cut-offs, however with different absolute values.


\section{Finite-range effects in the recombination rate}
\label{recRate}
\begin{figure}[ht]
  \centering
  \input{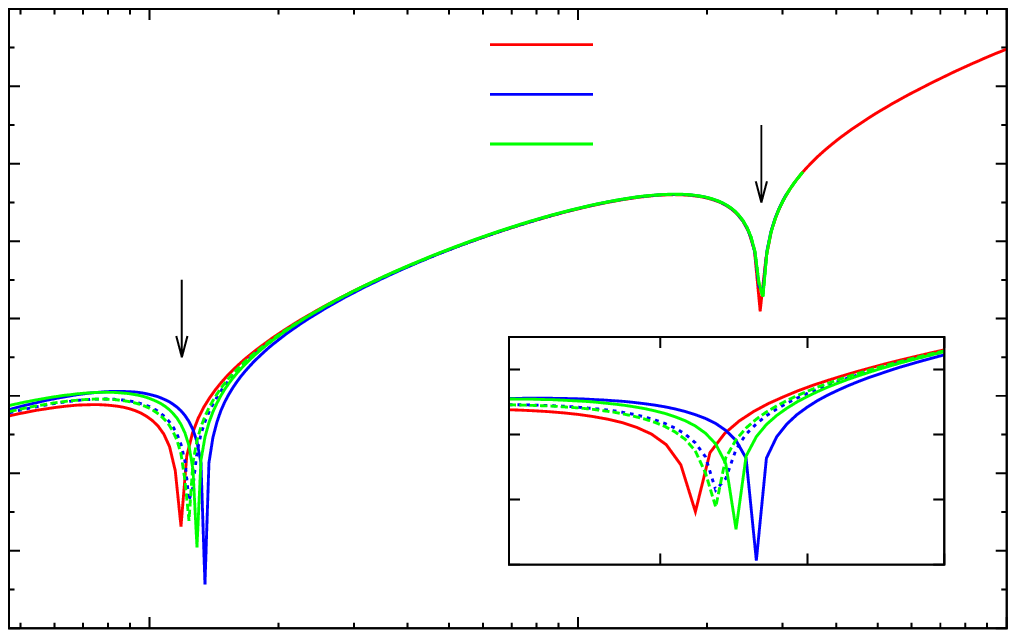}
  \caption[Recombination coefficient $\rec$ for three models]{The
  recombination coefficient $\rec$ from \eqr{eq:2.88} for the
  single-channel zero-range model, the effective range expansion model and
  the two-channel model with $R_0=-3a_0$ (dotted) and $R_0=-10a_0$
  (solid). The inset shows a closer look at the minimum near $a_1^*$. Note
  that the cut-off is such that all models reproduce the minimum at
  $a_{2}^{*}$. This allows us to study the effects of the effective range
  at the other minimum.}
  \label{figure:3.6}
\end{figure}

We now proceed to consider three-body recombination on the positive $a$
side of the Feshbach resonance. On this side of the resonance the
recombination takes place by transition of the three particles into the
channel with a bound two-body dimer with the universal binding energy
proportional to $-1/a^2$. On the $a<0$ side there is no bound dimer and
the decay goes directly into some strongly bound two-body state of the
atom-atom potential that depends on the short-range details. This latter
case is investigated in chapter \ref{optical_model} where recombination
into deep dimers is modelled using optical potentials.

\begin{figure}[ht]
  \centering
  \input{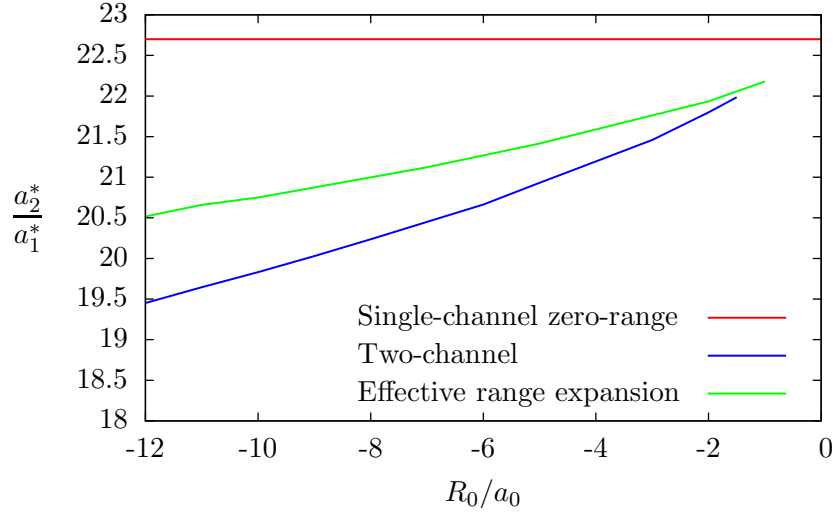}
  \caption[Ratio of recombination minima]{The ratio $a_2^*/a_1^*$ for the
  single-channel zero-range model, the effective range expansion model and
  the two-channel model as a function of effective range $R_0$. For the
  single-channel zero-range model the ratio is 22.7. The scale factor is
  reduced when the effective range is increased, in absolute value, to
  become more negative.}
  \label{figure:3.7}
\end{figure}

The recombination coefficients into shallow dimers for different values of
the effective ranges and different models are shown in \fig{figure:3.6}.
The scattering length values $a_1^*$ and $a_2^*$ indicate locations of
minima in the recombination rate. The minima are caused by the vanishing
of bound trimers into the atom-dimer continuum as shown in
\fig{figure:3.3}. The difference in the Efimov scale factor as compared to
the single-channel zero-range model is directly related to the difference
in the location where the trimer bound states vanish into this continuum.
The cut-offs were chosen such that the minimum at $a_2^*$ is the same for
all models and the comparison can then be made by looking at the minimum
at $a_1^*$. For the single-channel zero-range model, the ratio of $a_2^*$
to $a_1^*$ is $22.7$, showing that this calculation scheme agrees with the
universal result. For the other models this ratio is reduced, the minimum
at $a_1^*$ moves towards higher $a$. In order to make this more clear the
ratio of the minima as a function of the effective range on resonance,
$R_0$, is plotted in \fig{figure:3.7}. The two-channel and effective range
expansion models give similar qualitative predictions but there are small
quantitative differences. The curves cannot be extended all the way to
$R_0=0$ due to numerical issues, but the trends should be clear. The scale
factor reduces quite drastically at large negative $R_0$ for both models.
This corresponds to narrow Feshbach resonances, where there are currently
not enough experimental data to make a proper comparison.

\section{Comparison to experiment}
\label{experimental_comparison}

In this section a brief comparison to some of the available experimental
data is presented. Generally the models fit quite well to the available
data. However, the available data consists only of systems with large
resonance widths and thus small effective ranges. This makes  it difficult
to observe range effects. Furthermore, some datasets have only a single
recombination minimum wherefore no comparison with model predictions can
be made.

The two-channel models are compared to the experimental data for the cold
atomic gases listed in \tab{table:5.1}. The effective ranges are
calculated using the formula \eqref{eq:2.14}.
\begin{table}[ht]
\setlength{\tabcolsep}{4.5pt}
  \centering
  \begin{tabular}{rlccccc}
    \hline\noalign{\smallskip}
                 &                    & $B_0$ [G] & $\Delta B$ [G] & $\delta\mu$\,[$\mu_\text B$] & $a_{bg}\,[a_0]$ & $R_\text{eff}\,[|a_{bg}|]$\\[3pt]
    \hline\noalign{\smallskip}
    ${}^{23}$Na  & \cite{stenger}     & 907   & 0.70   & 3.8  & 63   & -21\\
    ${}^{133}$Cs & \cite{kraemer2006} & -11.7 & 28.7   & 2.3  & 1720 & -1.99$\times10^{-4}$\\
    ${}^{39}$K   & \cite{Zaccanti}    & 402.4 & -52    & 1.5  & -29  & -2.02\\
    ${}^{7}$Li   & \cite{pollack2009} & 736.8 & -192.3 & 1.93 & -25  & -3.17\\
    \noalign{\smallskip}\hline
  \end{tabular}
  \caption[Experimental data for Feshbach resonances]{Experimental data
    for Feshbach resonances for four atomic gases. $\mu_B$ is the Bohr
    magneton and $a_0$ the Bohr radius.}
  \label{table:5.1}
\end{table}

\begin{figure}
  \centering
  \input{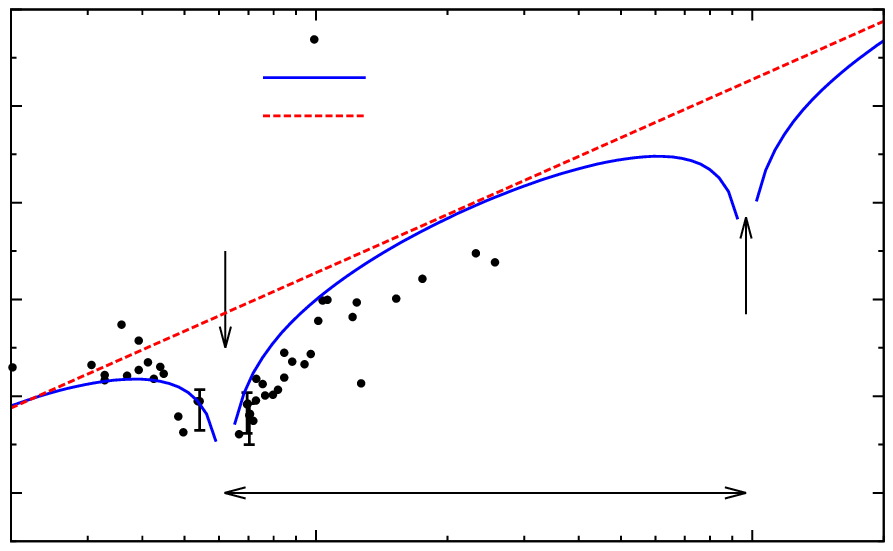}
  \caption[Recombination for ${}^{23}$Na]{The recombination coefficient
    $\rec$, \eqr{eq:2.88}, from the two-channel model for ${}^{23}$Na as a
    function of scattering length $a$  compared with the experimental data
    from \cite{stenger}. The theory predicts the next minimum to be around
    $a_2^*\approx1000a_0$.}
  \label{figure:3.8}
\end{figure}

In \fig{figure:3.8} the result from the two-channel model is shown
together with the experimental data for ${}^{23}$Na. The cut-off is fixed
by the experimental minimum at $a_1^*=62a_0$. The rather large effective
range could make the finite range effect, i.e. the reduction of the
scaling factor down to 15.7, clearly noticeable. However, at least one
additional minimum is needed to make a proper comparison. Experimental
data is not yet available in this range.

\begin{figure}
  \centering
  \input{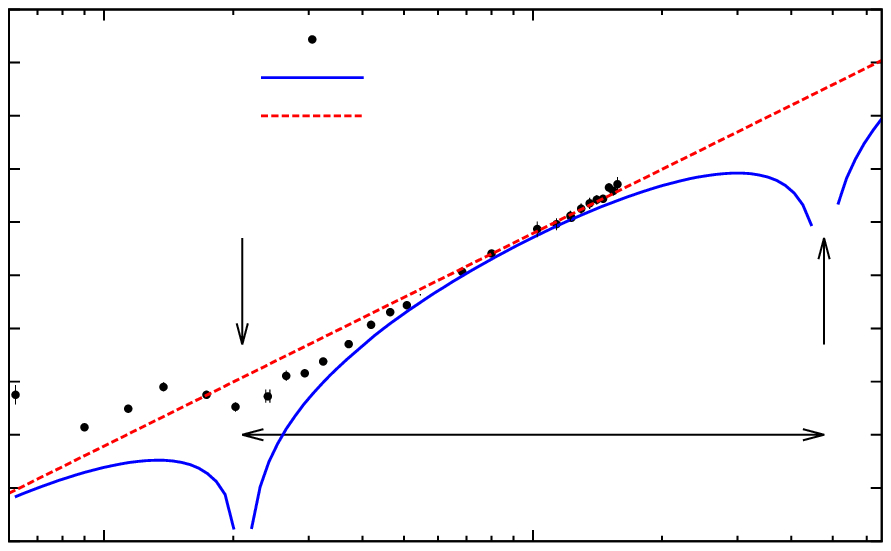}
  \caption[Recombination for ${}^{133}$Cs]{The recombination coefficient
    $\alpha$, \eqr{eq:2.88}, from the two-channel model for ${}^{133}$Cs
    as a function of the scattering length $a$ compared with the
    experimental data from \cite{kraemer2006}. The theory predicts the
    next minimum to be around $a_2^*\approx4770a_0$.}
  \label{figure:3.9}
\end{figure}

In \fig{figure:3.9} the result from the two-channel model is shown
together with the experimental data for ${}^{133}$Cs. The cut-off is fixed
by the experimental minimum at $a_1^*\approx210a_0$. The effective range
is very small indeed and the results from the two-channel model are
virtually indistinguishable from the single-channel zero-range model with
the scaling factor of 22.7. The next minimum should be found at
$a_2^*\approx4770a_0$.

\begin{figure}
  \centering
  \input{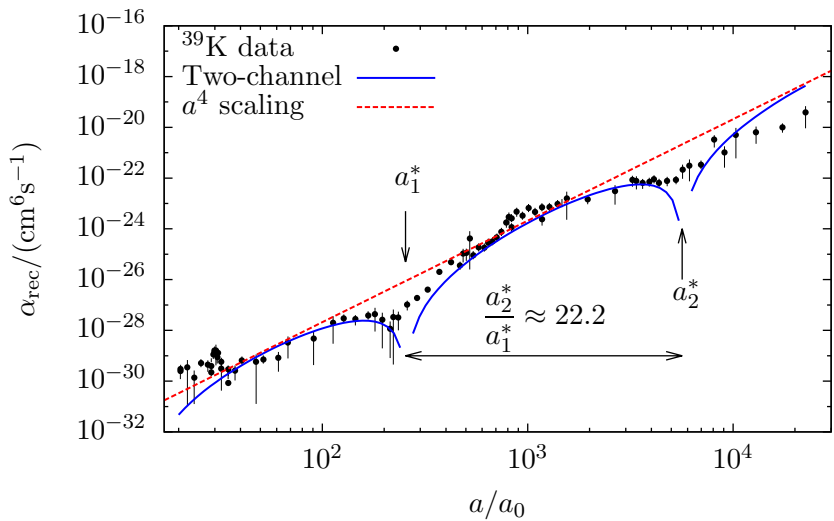}
  \caption[Recombination for ${}^{39}$K]{The recombination coefficient
  $\alpha$, \eqr{eq:2.88}, from the two-channel model for ${}^{39}$K as a
  function of the scattering length $a$ compared with the experimental
  data from \cite{Zaccanti}. The location of the minima are reasonably
  well described by the two-channel model.}
  \label{figure:3.10}
\end{figure}

\fig{figure:3.10} shows the recombination coefficient for ${}^{39}$K. The
cut-off parameter is chosen to fit the recombination minimum
$a_2^*=5650\pm900$. The two-channel model gives $a_1^*=254a_0$, with the
experimental value of $a_1^*=(224\pm7)a_0$. Overall, the two-channel model
fits the data quite well. Notably the scaling is correct compared to
experiment. The ratio of minima from the two-channel model is 22.2,
whereas the experimental value is $25.2\pm4.1$. The result lies within the
experimental uncertainty.

\begin{figure}
  \centering
  \input{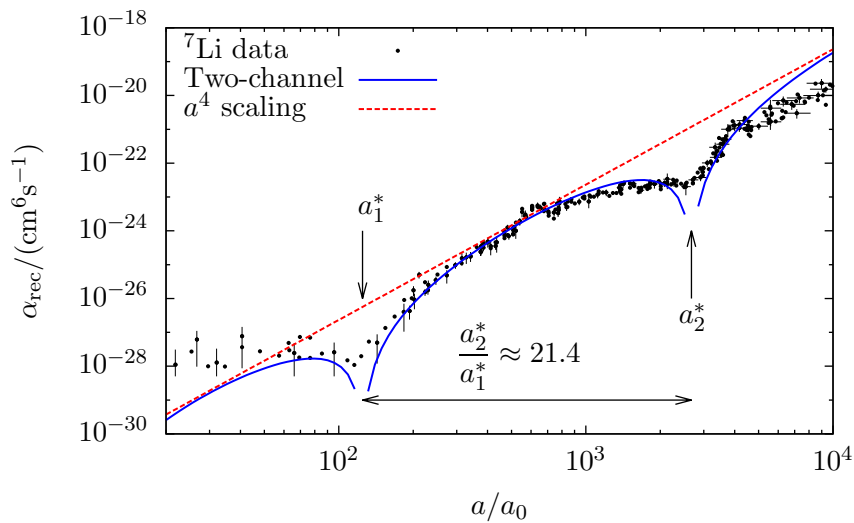}
  \caption[Recombination for ${}^{7}$Li]{The recombination coefficient,
    $\alpha$, \eqr{eq:2.86}, from the two-channel model for ${}^{7}$Li as
    a function of the scattering length, $a$, compared with the
    experimental data from \cite{pollack2009}. Theory and experiment agree
    very well.}
  \label{figure:3.11}
\end{figure}

The recombination coefficient for ${}^7$Li is shown in \fig{figure:3.11}.
The two minima are at $a_1^*=(119\pm11)a_0$ and $a_2^*=(2676\pm195)a_0$.
The cut-off is fixed by $a_2^*$, giving the two-channel prediction
$a_1^*=125a_0$. Again the theory describes the experimental data very
well. The two-channel model ratio of minima is 21.4 while the experimental
value is $22.5\pm2.6$ and again the result lies within the experimental
uncertainty.

\section{Conclusions}
\label{conclusion3}

In this chapter we investigate finite range effects in three-body
recombination rates in cold atomic gases near Feshbach resonances as well
as finite range effects in the trimer bound state energy spectrum. We use
two models which include the finite range effects and compare their
results with the single-channel zero-range model. The first model is the
effective range expansion model which is a straightforward extension of
the single-channel zero-range model. Here the effective range is included
directly in the boundary condition on the three-body wave function
following the effective range expansion of standard scattering theory.
Variation of the scattering length through the Feshbach resonance is done
phenomenologically as in the single-channel zero-range model. This model
can also be used for positive effective range calculations. The second
model is a two-channel contact interaction model which naturally includes
both the finite effective range and the variation of the scattering length
through the Feshbach resonance.

We show that with these well-tested two-body interaction models the
three-body physics can display complicated non-monotonic behaviour as the
effective range is varied. In particular, we find that the geometric
scaling factor of 22.7 for equal mass particles changes when including
effective range corrections, and that it can become both larger and
smaller than this value depending on the magnitude and sign of the
effective range.

In the current set-up this can be understood based on the functional form
of the effective hyper-radial potential. On resonance where the scattering
length diverges, the lowest trimer bound state has the strongest
dependency on the effective range since it lives at small hyper-radius,
whereas the excited states live at much larger hyper-radii and the
effective range contribution is much less profound. The adiabatic
potential of the effective range expansion model is raised and lowered
relative to the single-channel zero-range model potentials when the
effective range is negative and positive respectively. This leads to bound
states being less bound or more bound respectively. For the two-channel
model the effective range is always negative which can only be achieved by
using two-body potentials with an outer barrier. The hyper-radial
potential reflects this fact and develops a pocket at small hyper-radii
that the lowest states will eventually leak into. This feature is similar
to the effective range expansion model for the case of positive effective
range.

Our results demonstrate that effective range corrections within the
framework of single-channel zero-range model potentials can lead to
non-trivial behaviour of the trimer energies, thresholds and interference
features in recombination rates. Effective range corrections are expected
to be important for the case of narrow Feshbach resonances
\cite{chin2011}. The experimental data on Efimov states for narrow
resonance systems is sparse and more measurements are needed in order to
fully discriminate between different models that include finite range
corrections. However, what we can conclude is that care must be taken when
a particular two-body scattering model is used for the trimer states that
have the largest binding energies in a universal set-up, i.e. for the
lowest states that have binding energies related to the background
short-range length scales. For higher lying trimers it is less important
since the states are largely insensitive to the short-distance behaviour
of the effective three-body potential.


\chapter{Universal Three-Body Parameter}
\label{3BP}
\emph{This chapter investigates the universal relation between the
three-body para\-meter, $\aminus$, and the van der Waals length, $\vdW$. A
simple two-body inter\-action model is used to relate the number of bound
states in the two-body potential to the three-body parameter.}

\noindent\rule{\textwidth}{0.4pt}

\noindent When using zero-range potentials as described in the previous
chapters, two-body variables like the scattering length and effective
range are not enough to predict three-body observables without additional
parameters. The additional parameter needed is known as the three-body
parameter, or 3BP. In the previous chapter it manifested itself in terms
of a short-range regularization cut-off, $\rc$. The 3BP is more commonly
given by the threshold for creation of the lowest three-body bound state
on the negative $a$ side, denoted as $\aminus$ in \fig{figure:3.3}. In
the single-channel zero-range model $\rc$ and $\aminus$ are directly
related by a simple expression. A surprising result that has turned up in
later years is the fact the $\aminus$ is apparently related to the
two-body van der Waals length in alkali atoms with
$|\aminus|/\vdW\sim9.8$, thus relating the 3BP to two-body physics
\cite{berninger2011}.

In this chapter the relation between the two-body van der Waals length and
the three-body parameter $\aminus$ is investigated using a simple model of
the two-body potential, namely a pure van der Waals $1/r^6$ attraction
with a hard-core cut-off. The number of bound states in such a potential
is easily derived which is used to relate the 3BP to the number of bound
states in the two-body potential. Reasonable agreement with experimental
data is found. Other two-body potentials like the Lennard-Jones and Morse
potentials give similar predictions and thus the exact form of the
two-body potential is not of any qualitative importance. Furthermore we
investigate how the effective range affects the value 9.8. Results of this
chapter are also found in \citepublications{SFJZ12}.

This chapter will utilize the so-called resonance strength
$s_\textnormal{res}$ given by \cite{chin2010}
\begin{equation}
  s_\textnormal{res}\equiv\frac{\vdW}{|R_0|}\;,
  \label{eq:4.2}
\end{equation}
where $\vdW$ is the van der Walls length which is introduced in the next
section. Unfortunately all experimental data lie in the regime of broad
resonances, $s_\textnormal{res}\gg1$, and corresponding short effective
ranges, and range effects are thus hard to see in these systems.


\section{Trimer threshold value revisited}
\label{trimer_threshold_value_revisited}

In \fig{figure:3.3} of the previous chapter the threshold value $\aminus$
for appearance of the lowest Efimov trimer bound state for negative
scattering lengths is shown for the zero-range model and the two effective
range models. In this section the dependency of $\aminus$ on the cut-off
and effective range is investigated. The deviation between the
single-channel zero-range and the two-channel curve\footnote{The effective
range expansion model is not considered in this discussion, but similar
conclusions would follow if it had been.} can be split into two
parts. First, the trimer bound state energy $E_T$ on resonance is lower
for the two effective range models compared to the single-channel
zero-range value given that the states with $n=2$ have been fixed to the
same energy for all models. Reducing the cut-off in the single-channel
zero-range model increases the binding energy on resonance as per
\eqr{eq:2.65} and Ref.~\cite{efi70}, resulting in a decreased value of
$|\aminus|$ according to \fig{figure:3.3}. This relation between $\rc$ and
$\aminus$ is linear and we find it numerically to be
\begin{equation}
  \aminus=-\delta\rc\;,\qquad\delta\approx31.756\;,
  \label{eq:4.1}
\end{equation}
for the single-channel zero-range model. A linear relation is the only
possibility since $\rc$ is the sole input length scale.

For the two-channel model $|\aminus|$ is further reduced when the
effective range is increases, i.e. $|R|$ increases. A simple linear
relation as the one above is, however, not obtainable as the dependency is
more intricate.

\begin{figure}[ht!]
  \centering
  \input{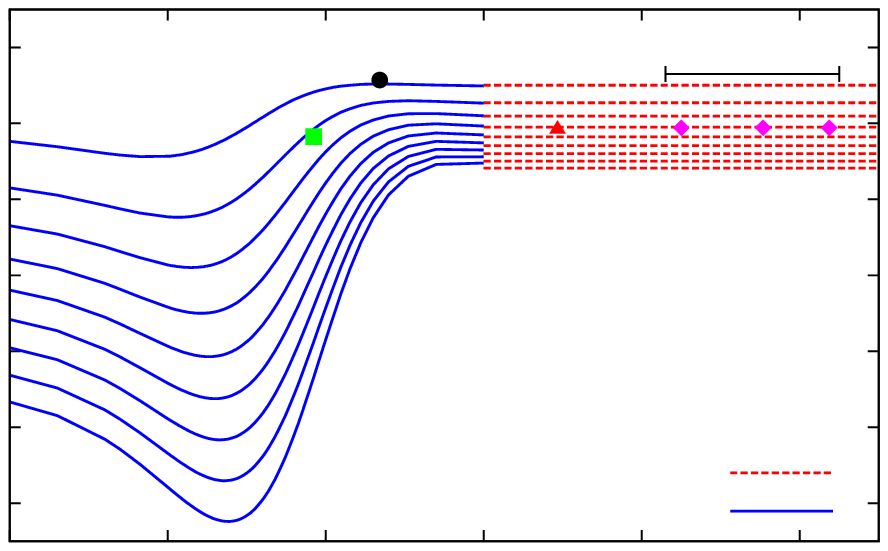}
  \caption[$\aminus$ threshold behaviour]{The threshold scattering length,
  $\aminus$, at which the lowest universal Efimov trimer merges with the
  three-atom continuum, for negative $a$, plotted against the strength,
  $s_\text{res}$, of the Feshbach resonance. The right-hand side
  corresponds to broad resonances. The different curves show results with
  different three-body parameters, $\rc$, in units of the atomic van der
  Waals length, $r_\textnormal{vdW}$. The values of $\rc$ decrease from
  top to bottom from $1.20$ down to $0.38$ with the best fit value for
  broad resonances being $0.58$. Experimental values are for $^{133}$Cs
  \cite{kraemer2006, berninger2011}, $^7$Li \cite{pollack2009}, $^{39}$K
  \cite{Zaccanti} and $^{85}$Rb \cite{wild2012}.}
  \label{figure:4.1}
\end{figure}

A systematic study of the influence of both $\rc$ and $R_0$ (in terms of
$s_\textnormal{res}$) is shown in \fig{figure:4.1}. The values of
$\rc/r_\textnormal{vdW}$ from top to bottom are 1.20, 0.82, 0.66, 0.58,
0.51, 0.47, 0.42, 0.40, and 0.38. They correspond to $n=0$ to $8$ in
\eqr{eq:4.6} below. Both models agree for $s_\textnormal{res}\gg 1$. The
two-channel model results are shown only in the region where they deviate
from the single-channel zero-range results. To reproduce the experimental
data for $s_\textnormal{res}\gg 1$, the cut-off value of
$\rc/r_\textnormal{vdW}= 0.58$ provides the best fit. However, for small
$s_\textnormal{res}$, the same cut-off does not reproduce the used data
point coming from $^{7}$Li \cite{pollack2009}(other measurements have
slightly smaller $|\aminus|$ \cite{gross2009, gross2010}, which increases
the ratio $|\vdW/\aminus|$ by about $5\%$). The increase toward the
$^{39}$K data point at small $s_\textnormal{res}$ cannot be accommodated
for the same $\rc$.

The non-monotonic behaviour observed is exactly the same as for the trimer
bound state energies as discussed near \fig{figure:3.4} and
\fig{figure:3.5}. When the effective range becomes more negative the
trimer binding energy decreases (becomes less negative) with a
corresponding increase in $|\aminus|$. Further increasing the effective
range eventually increases the binding energy and $|\aminus|$ decreases
yet again.

The opposite behaviour was found in \cite{schmidt2012}. However, their
approach seems to rely on finite range potentials with a positive
effective range, thus the opposite direction of change is no surprise.
Feshbach resonances usually have negative effective ranges
\cite{chin2010}, so the here presented method is preferable.

\section{Two-body potentials}
\label{connection}
The zero-range models do not carry any inherent information about the van
der Waals length. However, the three-body parameter or cut-off, $\rc$, has
a physical meaning as it provides a hard-core repulsion in the
hyper-spherical three-body coordinates. To connect the formalism to the
experimental data, it is therefore necessary to find a relation between
the two-body atomic physics and, $\rc$,. In this section a relation
between the two-body cut-off, $r_c$, and the van der Waals length $\vdW$
is obtained using a simple two-body van der Waals potential. In the next
sections the two and three-body cut-offs are related and finally the van
der Waals length is related to $\aminus$.

\setlength{\tabcolsep}{15pt}
\begin{table}[ht]
  \centering
  \begin{tabular}{llrrrrr}
    \hline\noalign{\smallskip}
    & Unit & \Li7 & \K & \Rb & \Cs \\
    \noalign{\smallskip}\hline\noalign{\smallskip}
    $C_6$  & $a_0^6E_h$ & 1393 & 3897 & 4691 & 6851 \\
    $\vdW$ & $a_0$      & 65   & 129  & 164  & 202  \\
  \end{tabular}
  \caption[Van der Waals data]{The Van der Waals coefficient, $C_6$, and
    length, $\vdW$, for some alkali atoms. $a_0$ is the Bohr radius and
    $E_h\approx27.2$ eV is the Hartree unit of energy.}
  \label{table:4.1}
\end{table}

Interactions between neutral atoms are often described using a potential
with a long-range tail of the form $C_6/r^6$ where the coefficient $C_6$
is known as the van der Waals coefficient \cite{pethicksmith}. This long
range behaviour originates from small fluctuations in the electron clouds
causing mutual polarization of the atoms. The resulting effect is known as
van der Waals forces\footnote{Which quite ingeniously the gecko has
utilized in its ability to climb glass surfaces.}. From the $C_6$
coefficient the van der Waals length can be constructed
\begin{equation}
  \vdW=\left(\frac{mC_6}{\hbar^2}\right)^{1/4}\;,
  \label{eq:4.3}
\end{equation}
where $m$ is the mass of the atoms. The coefficient $C_6$ can be found by
chemical calculations \cite{pethicksmith}. A list of alkali atoms and
their corresponding van der Waals coefficients are given in
\tab{table:4.1}.

\begin{figure}[ht!]
  \centering
  \input{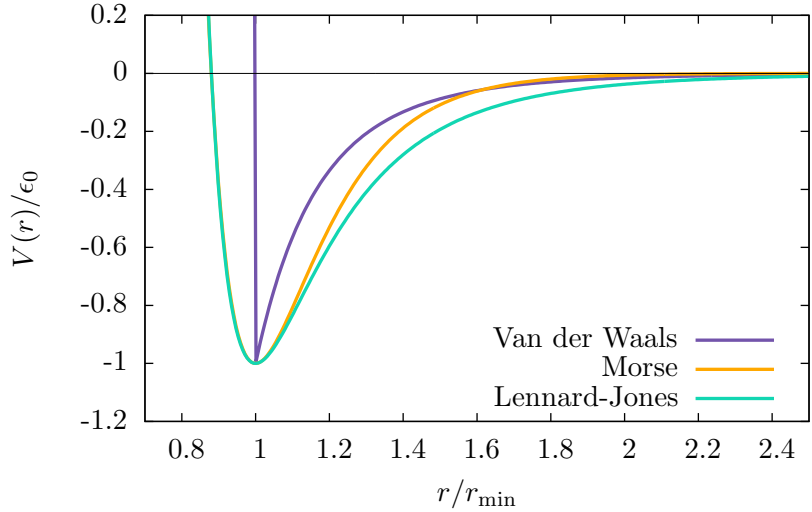}
  \caption[Typical two-body potentials]{A small assortment of two-body
    neutral atom potentials from \eqr{eq:4.4}, \eqr{eq:4.10} and
    \eqr{eq:4.11}. $\epsilon_0$ indicates the minimal value of the
    potential and $r_\textnormal{min}$ is the location of the minimum. For
    the van der Waals potential in \eqr{eq:4.4} $r_\textnormal{min}=r_c$.}
  \label{figure:4.2}
\end{figure}

Since the $1/r^6$ behaviour cannot continue to $r=0$, a cut-off is applied
at some small distance $r_c$. This acts like a strong short-range
repulsion which is also seen in physical potentials. The potential is
\begin{equation}
  V(r) =
  \begin{cases}
    \phantom{-}\infty&\textnormal{for } r<r_c\;,\\
    -\dfrac{C_6}{r^6}=-\epsilon_0\left(\dfrac{r_c}{r}\right)^6&
      \text{for }r\geq r_c\;,
  \end{cases}
  \label{eq:4.4}
\end{equation}
where $\epsilon_0=C_6/r_c^6$ is the minimal value of the potential. For
graphical illustration this potential, along with the Morse and
Lennard-Jones potentials discussed later, is plotted in \fig{figure:4.2}
such that the minimal value $\epsilon_0$ and minimum location
$r_\textnormal{min}$ coincide for all models. For \eqr{eq:4.4} we have of
course $r_\textnormal{min}=r_c$.

The scattering length of the potential in \eqr{eq:4.4} can be found
analytically by rewriting the corresponding Schr\"o{}dinger equation to
the Bessel equation of order $1/4$. Details can be found in
\cite{pethicksmith} and will not be reproduced here. The result for the
scattering length is (see also \cite{PhysRevA.48.1993,
PhysRevA.59.1998})
\begin{equation}
  a=\vdW\frac{2\pi}{\Gamma(\frac{1}{4})^2}[1-\tan\left(\Phi-3\pi/8\right)]
    \approx0.478\vdW[1-\tan\left(\Phi-3\pi/8\right)]\;,
  \label{eq:4.5}
\end{equation}
where  $\Phi=\vdW^2/2r_c^2$. Whenever the scattering length diverges the
potential is able to support yet another bound state. Thus counting the
number of divergences in $a$ as a function of $\Phi$ yields the number of
bound states. The scattering length $a$ diverges when $\Phi-\frac{3\pi}{8}
=(n+\frac12)\pi$ for integer $n$, thus the number of $s$-wave bound states
in the potential is
\begin{equation}
  n=\left\lceil\frac1{2\pi}\left(\frac{r_\textnormal{vdW}}{r_c}\right)^2
    -\frac78\right\rceil\;,
  \label{eq:4.6}
\end{equation}
where the "bracket" indicates round off to highest nearby integer.
$s$-wave states are all we are interested in as discussed in chapter
\ref{theory}.

\subsection{Relating two- and three-body cut-offs}
To relate the van der Waals length, which is a two-body parameter, to the
3BP, $\aminus$, a simple link between the two- and three-body cut-offs has
to be established. We do this by presenting this a simple geometric
interpretation.

The atom-atom two-body potential has a steep repulsive inner core which is
here modelled by a hard inner wall, as in \eqr{eq:4.4}. In this case the
boundary condition is simply that the two-body wave function must be zero
at $r_c$ and below. This must then be translated into the three-body
problem where it implies that the total wave function must be zero
whenever any of the relative distances between two of the three atoms is
less than or equal to $r_c$. Any penetration of the wave function into the
wall would cost an infinite amount of energy and is thus forbidden.

A neat and elegant way to obtain a cut-off condition on $\rho$ is the
following. For three equal mass particles the hyper-radius from
\eqr{eq:2.36} can be written as
\begin{align}
  \rho^2=\frac{1}{3}\sum_{i<k}\left(\bm r_i-\bm r_k\right)^2
        =\frac{1}{2}\bm r^{2}_{12}+\frac{2}{3}\bm r^{2}_{12,3}
        =\bm x^2+\bm y^2\;,
\label{eq:4.7}
\end{align}
where $\bm r_{12}=\bm r_1-\bm r_2=\sqrt{2}\bm x$ and $\bm r_{12,3}=\bm
r_3-(\bm r_1+\bm r_2)/2=\sqrt{\frac{3}{2}}\bm y$ are respectively the
relative vector from particles 2 to 1 and the relative vector from the
center of mass of particle 1 and particle 2 to particle 3.

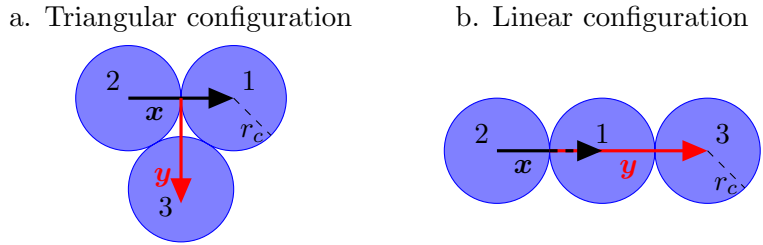
\begin{figure}[ht]
  \centering
  \tikzsetnextfilename{fig4_3}
  \begin{tikzpicture}[inner sep=1mm,scale=0.7]
  \tikzstyle{particle} = [circle,fill,shading=ball,ball color=blue!50!red]
  \tikzstyle{tarrows} = [->,>=triangle 45,very thick]
  \tikzstyle{particle} = [circle,draw=blue,fill=blue!50]

  \node at (1,1.5) {a. Triangular configuration};
  \draw[fill=blue!50,draw=blue]
    (2,0) circle(1cm) node[above right] {1} 
    (0,0) circle(1cm) node[above left] {2}
    (1,-1.732) circle(1cm) node[below left] {3};
  \draw[tarrows] (0,0)     -- node[below,near start] {$\bm x$} (2,0);
  \draw[tarrows,red] (1,0) -- node[left, near end] {$\bm y$} (1,-2);
  \draw[dashed] (2,0)      -- node[below]{$r_c$} ++(-45:1);

\begin{scope}[xshift=2cm]
  \node at (7,1.5) {b. Linear configuration};
  \draw[fill=blue!50,draw=blue]
    (7,-1) circle(1cm) node[above]       {1}
    (5,-1) circle(1cm) node[above left]  {2}
    (9,-1) circle(1cm) node[above right] {3};
  \draw[tarrows,-] (6,-1) -- node[below] {$\bm x$}(5,-1);
  \draw[tarrows,red] (6,-1) -- node[below] {$\bm y$} (9,-1);
  \draw[tarrows,dashed] (6,-1) -- (7,-1);
  \draw[dashed] (9,-1)      -- node[below]{$r_c$} ++(-45:1);
\end{scope}
\end{tikzpicture}
  \caption[Linear and triangular Jacobi coordinates]{Schematic drawing of
  the triangular, a, and linear, b, configurations for an equal mass
  three-body system. $\bm x$ and $\bm y$ indicate the Jacobi coordinates.}
  \label{figure:4.3}
\end{figure}

For the close packed triangular configuration in \fig{figure:4.3}a,
\eqr{eq:4.7} yields $\rho=r_c$. A linear configuration (as in
\fig{figure:4.3}b but not exactly the one shown), can however, have $\bm
y=0$ and $|\bm x|=r_c$ wherefore $|\bm r_{12}|=\sqrt{2}r_c$, which is
allowed. However, particle 3 lies in between particles 1 and 2,
overlapping with the hardcore cut-off, which is not allowed.

Consider instead the linear configuration with $r_{12}=r_c$ and impose the
requirement $r_{23}\geq r_c$, where $\bm r_{23}=\bm r_2-\bm r_3$. Since
$\bm r_{12,3}=\bm r_{23}+\bm r_{12}/2$ we get
\begin{align}
  \rho^2\geq r_{c}^{2}\left( \frac{1}{2} +
  \frac{2}{3}[1+\frac{1}{2}]^2\right)=2r_{c}^{2}.
  \label{eq:4.8}
\end{align}

The condition $\rho>\sqrt{2}r_c$ ensures that both the triangular and the
linear configurations are outside of the hard-core regions. Since these
configurations are extremal, the condition implies that no regions with
infinite potential are reached by the hyper-radial three-body wave
function.

The rigorous formal argument for the validity of the relation
$\rc=\sqrt{2}r_c$ using the hyper-spherical approach can be found in
Ref.~\cite{jen97}, where the relation is derived using a square well
potential. The asymptotic region is precisely $\rho>\sqrt{2}r_c$ as found
above. Here a hard-core potential was assumed for simplicity which gives
the factor of $\sqrt{2}$. For a real atom-atom potential, the hard-core is
slightly softer (typically of the $1/r^{10}$ as in \eqr{eq:4.10} below)
which may lead to a minor change in the factor $\sqrt{2}$.

\subsection{Relating \texorpdfstring{$\aminus$}{aminus} to
\texorpdfstring{$\vdW$}{rvdW}}
Combining the results of the previous sections, namely \eqr{eq:4.1},
\eqr{eq:4.6} and \eqr{eq:4.8} is a trivial matter but nevertheless yields
a very important result
\begin{equation}
  \frac{\aminus}{\vdW}=
    -\frac{2\delta}{\sqrt{\left(n+\frac78\right)\pi}}\;,
  \label{eq:4.9}
\end{equation}
where $n$ is the number of bound states. This semi-analytical expression
for the threshold in terms of the number of bound states very elegantly
relates the two-body van der Waals length, $\vdW$, with the three-body
parameter, $\aminus$.

\begin{figure}[ht!]
  \centering
  \input{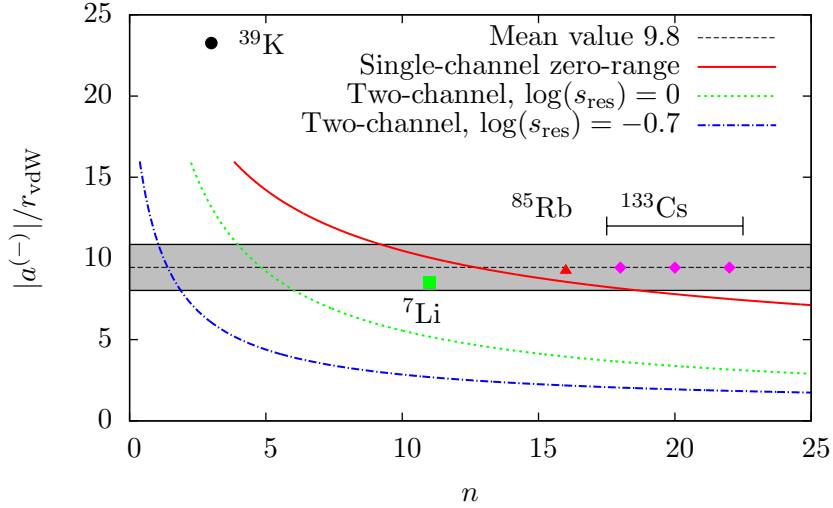}
  \caption[Semi analytic results for the three-body parameter
  $\aminus$]{Semi-analytic results for the three-body parameter $\aminus$
    plotted against the number of bound states in the two-body van der
    Waals plus hard-core potential, \eqr{eq:4.4}. The horizontal position
    of the experimental data is arbitrary. The grey band indicates a 15\%
    margin around the value $\sim9.8$.}
  \label{figure:4.4}
\end{figure}

The relation in \eqr{eq:4.9} is plotted in \fig{figure:4.4} along with the
experimental data and the numerical results obtained from the two-channel
model for different values of $s_\textnormal{res}$. For the two-channel
model $\delta$ varies with effective range. The single-channel zero-range
model is consistent with the data for $n\sim 10-20$ and reproduces the
universal ratio of Ref.~\cite{chin2011} for $n=13$. This is also
consistent with the findings of Ref.~\cite{wang2012}, although their
results only goes to $n=10$. The $n^{-1/2}$ behaviour seems to also appear
in Ref.~\cite{wang2012}, where an extension to higher $n$ could confirm
this prediction.

The results from the two-channel model with small $s_\textnormal{res}$,
i.e. narrow reso\-nances and large effective ranges, indicate that
$|\aminus|$ drops faster with $n$ (blue curve in \fig{figure:4.4}) than
for $s_\textnormal{res}\gg 1$ (red curve in \fig{figure:4.4}). This is
seen in the experimental data for \Li7 which is slightly below the \Rb and
\Cs points, but the model overestimates this trend. More results on narrow
resonance systems are required to address the question of effective range
corrections.

\section{Number of dimer bound states}
\label{number_of_dimer_bound_states}
The other potentials in \fig{figure:4.2} are the more realistic
Lennard-Jones (LJ) potential
\begin{equation}
  V_\textnormal{LJ}(r)=\frac{C_{10}}{r^{10}}-\frac{C_6}{r^6}=\epsilon_0
    \left[\frac{3}{2} \left(\frac{r_\textnormal{min}}{r}\right)^{10}-
      \frac{5}{2}\left(\frac{r_\textnormal{min}}{r}\right)^6\right]\;,
  \label{eq:4.10}
\end{equation}
(sometimes also written with the power $12$ in the first term, in which
case the rightmost expression would of course also have to be changed) and
the Morse (M) potential, given by
\begin{equation}
    V_\textnormal{M}(r)=\epsilon_0\left[e^{-2\alpha(r-r_\textnormal{min})}
      -2e^{-\alpha(r-r_\textnormal{min})}\right]\;,\\
  \label{eq:4.11}
\end{equation}
where $r_\textnormal{min}$ is the location of the potential minimum. They
have a smoother behaviour at the inner barriers. This implies only minor
quantitative corrections. More importantly the number of bound states in
the two-body alkali dimer potential need to be addressed.

The number of $s$-wave bound states in the Lennard-Jones and Morse
potentials can be estimated analytically and yields \cite{MahanLapp}
\begin{equation}
  n_{LJ}=\left\lceil0.361\sqrt\beta-\tfrac{5}{8}\right\rceil\;,
  \label{eq:4.11a}
\end{equation}
and
\begin{equation}
  n_{M}=\left\lceil0.245\sqrt\beta-\tfrac{1}{2}\right\rceil\;,
  \label{eq:4.11b}
\end{equation}
where $\beta=\dfrac{m r_\textnormal{min}^{2}\epsilon_0}{2\hbar^2}$ with
$r_\textnormal{min}$ the radius at which the potential takes its minimal
value, $\epsilon_0$. For comparison, the expression in \eqr{eq:4.6} can be
written $n=0.225\sqrt{\beta}-\tfrac{7}{8}$, with
$r_\textnormal{min}\leftrightarrow r_c$ such that
$\beta=\frac{\vdW^4}{2r_c^4}$. The similarity of
these expressions makes it clear that the behaviour seen in
\fig{figure:4.4} is generic and does not depend on the choice of two-body
potential. The difference in constant in front of $\sqrt\beta$ provides
only a minor quantitative change in the numbers.

An important question, however, remains about the number of bound states,
$n$, in a real alkali dimer system. This is estimated using the bond
lengths $r_\textnormal{min}$ and dissociation energies $\epsilon_0$ of
Ref.~\cite{Igel-Mann} listed in \tab{table:4.2} where also the estimates
for $n$ are listed. The estimated number of bound states is outside the
axis in \fig{figure:4.4} and also much beyond the results shown in
Ref.~\cite{wang2012}. The agreement with theory at a rather limited number
of bound states ($n\sim 10-20$) is then quite surprising.

\setlength{\tabcolsep}{15pt}
\begin{table}[ht]
  \centering
  \begin{tabular}{llrrrrr}
    \hline\noalign{\smallskip}
    & Unit & Li & K & Rb & Cs \\
    \noalign{\smallskip}\hline\noalign{\smallskip}
    $r_\textnormal{min}$ & \AA        & 2.67 & 3.92 & 4.18 & 4.65 \\
    $\epsilon_0$         & eV         & 1.06 & 0.52 & 0.49 & 0.45 \\
    \noalign{\smallskip}\hline\noalign{\smallskip}
    $n_\textnormal{vdW}$ &            & 18   & 43   & 67   & 88   \\
    $n_\textnormal{LJ}$  &            & 29   & 70   & 108  & 142  \\
    $n_\textnormal{M}$   &            & 20   & 47   & 73   & 96   \\
  \end{tabular}
  \caption[Two-body potential parameters]{The two-body inter-atomic
    potential parameters for some alkali atoms. The first two rows
    tabulate bond lengths and strengths \cite{Igel-Mann}. The last three
    rows tabulate the number of bound $s$-wave states in the van der Waals
    plus cut-off, Lennard-Jones and Morse potentials using the respective
    bond lengths and strengths.}
  \label{table:4.2}
\end{table}

A number of important observations can be made. First, the decrease of
$|\aminus|$ with $n$ is weak, and a shift of the length scale in
\fig{figure:4.4} would therefore place the single-channel zero-range model
within the experimental range for larger $n$ and it would stay within the
15\% deviation from the mean for a larger interval (since the slope at
larger $n$ decreases even faster). Second, the experimental data might
indicate that only a certain number of bound states play an active role.
Equivalently, even if the two-body potential is very deep, only the upper
part of the two-body potential and the bound states closest to threshold
set the scale of the three-body problem. This appears to be very
reasonable since we are considering universal Efimov trimers here and not
strongly bound three-body states. Third, the case of small
$s_\textnormal{res}$ has $|\aminus|\propto n^{-r}$ with $r>1/2$ as seen in
\fig{figure:4.4}. This implies that narrow resonance systems should be
even less sensitive to $n$ beyond a certain lower limit.

A quantitative argument for the lack of sensitivity to the many deep bound
states in the van der Waals potential is as follows: The number of bound
states in the potential $V(r)$ with energies larger than $E$, can be
estimated using the WKB approximation
\begin{equation}
  \int_{r_i}^{r_o}\sqrt{E-V(r)}\mathrm dr =
    \pi\hbar\left(n(E)-\frac{1}{4}\right)\;,
  \label{eq:4.1234}
\end{equation}
where $r_i$ and $r_o$ are the inner and outer classical turning points
such that $E=V(r_i)=V(r_o)$. For $E=0$ this yields the total number of
bound states $n_\textnormal{total}$ (given by $n$ in \tab{table:4.2}),
essentially by counting the total number of oscillations of the
zero-energy wave function. For finite energy, $E<0$, the integral is not
analytically solvable when $V(r)$ is the van der Waals potential from
\eqr{eq:4.4} (in which case the inner turning point is the cut-off $r_c$).
However, we find numerically that the number of bound states with the
energy in the interval $[E,0]$, to a good approximation, is given by
$n(E)=n_\textnormal{total}(|E|/E_\textnormal{vdW})^{1/3}$ for
$-E_\textnormal{vdW}<E<0$, where $E_\textnormal{vdW}=\hbar^2/m\vdW^2$ and
the energy, $E$, is now measured from $0$ downwards.
For the $s_\textnormal{res}\gg 1$ cases ($^{85}$Rb and $^{133}$Cs),
$n(E)/n_\textnormal{total}\sim 0.10-0.20$ which implies
$|E|/E_\textnormal{vdW}\sim 0.001-0.01$. Numerically we find a three-body
energy on resonance $E_T=0.006E_\textnormal{vdW}$ (using
$\rc=0.58r_\textnormal{vdW}$). However, universality relates
$E_T=\hbar^2\kappa^2/m$ and $\aminus\kappa\sim -1.51$ \cite{efi70,
braaten2006, PhysRevLett.100.140404} as noted in the previous chapter. The
energy scale at the continuum threshold is given by $\aminus$ through
$|E|\sim \hbar^2/m(\aminus)^{2}=0.003E_\textnormal{vdW}$, in agreement
with the interval above. In the case of $^{7}$Li, $E_T$ is similar but
this is compensated by a smaller $n_\textnormal{total}$ so this case can
also be explained. For the heavier alkali atoms at a narrow resonance, our
two-channel results predict a smaller $|\aminus|/r_\textnormal{vdW}$ than
9.8, which is a good experimental test of our theory.

\section{Conclusions}
\label{conclusion4}
In this chapter the three-body cut-off was expressed as the three-body
para\-meter $\aminus$ in the single-channel zero-range model via a simple
linear relation. Using this relation the number of bound states in a
semi-realistic van der Waals potential was related to the ratio of the
three-body parameter and the van der Waals length. This ratio has the
universal value of $\sim9.8$ across several different atomic species. A
comparison between the experimental data and the here presented model
yielded reasonable agreement. The case of narrow Feshbach resonances,
corresponding to small strengths $s_\textnormal{res}$ and large effective
ranges, was investigated using the two-channel model. Our model predicts
that the universal ratio $9.8$ should decrease when the effective range
increases (becomes more negative).




\chapter{Recombination for negative scattering lengths}
\label{optical_model}

\emph{This chapter investigates recombination for negative scattering
lengths using optical potentials to emulate the presence of deep dimers in
the two-body potentials. Additionally, the effects of finite temperature
are included.}

\noindent\rule{\textwidth}{0.4pt}

\noindent In this chapter the recombination rate is investigated for
negative scattering lengths, $a$, as presented in
\citepublications{SFJZ13b}. The physical method by which particles
recombine is different than for positive scattering lengths as there are
no weakly bound states for negative $a$. The particles must recombine into
deeply bound states in the two-body potentials. This regime is not
immediately available using zero-range models. We will therefore emulate
the existence of deeply bound states by letting the three-body potential
have a complex value in a restricted region of hyper-space when all three
particles are close to one another. The imaginary value of the potential
acts as a probability sink that particles can disappear into. The method
reproduces experimental data quite well. Additionally, we investigate the
effects of finite temperature in the systems which follows naturally from
the method used to describe the recombination. Universal scaling of the
form $\rec=C(a)a^4$ (which is still valid for negative $a$, but with a
different form for the $C(a)$ coefficient) is obtained only for
sufficiently low temperatures. This is known as the unitarity limit
\cite{dincao2004}.


\section{The optical model}
\label{optical_potentials}

The hidden crossing method of calculating the recombination coefficient
for positive scattering lengths relies on going from one adiabatic channel
to another via a path in the complex $\rho$-plane. Therefore, it cannot be
used for negative $a$ since recombination must go into deeply bound
dimers, which are not in scope of the zero-range models.

\begin{figure}[ht!]
  \centering
  \input{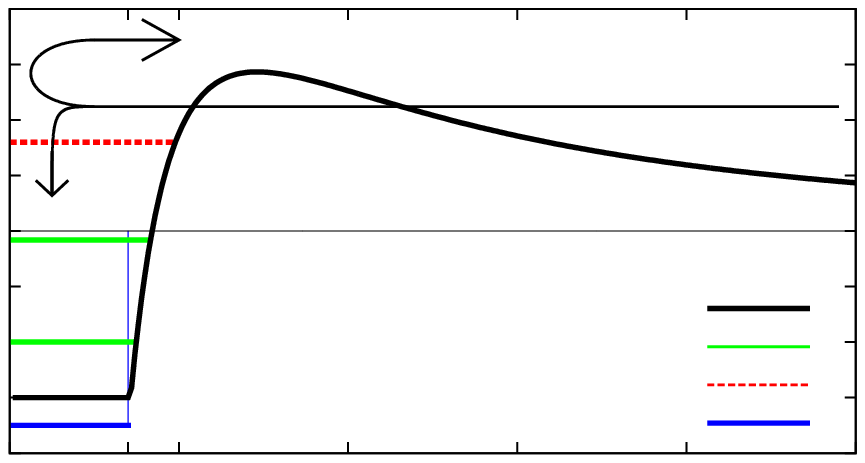}
  \caption[Three-body potential barrier]{The three-body radial potential
  for negative scattering length, $a$, as a function of the hyper-radius,
    $\rho$. The $\rim$ used in calculations is much smaller than
    illustrated. The potential drops as $1/\rho^2$ at large distances,
    diverges as $-1/\rho^2$ for small distances (for $\rho>\rim$) and has
    a constant (complex) value $\Vim$ for $\rho<\rim$. The split arrow
    indicates that the wave function, $f(\rho)$, is both reflected and
    absorbed through the barrier and via the complex potential. The green
    full lines illustrate bound states in the potential while the red
    dotted line indicates a resonance.}
  \label{figure:5.1}
\end{figure}

Instead we solve the differential equation \eqr{eq:2.42}. The transition
ampli-tude is estimated using only the lowest adiabatic channel, $n=1$,
schematically shown in \fig{figure:5.1}. The potential has a barrier that
the wave function must tunnel through. The barrier maximum is at
$\rho\approx1.46|a|$ with a maximum value of $0.143\hbar^2/ma^2$. The
potential furthermore crosses 0 at $\rho\approx0.84|a|$ which is found by
setting $\nu=1/2$ in \eqr{eq:2.55}.

At large hyper-radii we decompose the wave function into an incoming and
an outgoing free wave $f(\rho) = He^{-ik\rho}+Ge^{ik\rho}$ where
$k^2=2mE/\hbar^2$ and $H$ and $G$ are parameters that depend on energy.
The transition amplitude of inelastic scattering is the ratio of
amplitudes for the outgoing, $G$, to incoming, $H$, components, thus the
probability of recombination is $P(k) = 1-|G/H|^2$. For a one-dimensional
purely real potential this would yield $P=0$ identically due to
conservation of probability. Therefore, a constant imaginary value, of
magnitude $|\Vim|$, is added to the potential for distances smaller than
the value $\rim$. In this region the real part of the potential is also
held constant at $V(\rho<\rim)=V(\rim)$. The boundary condition is
correspondingly $f(0)=0$. This is in stark contrast to the
previous regularization cut-off method where the potential is set to
infinity for small $\rho$-values.

The parameters $\rim$ and $\Vim$ are chosen to fit the experimental data
by using the resonance location to determine $\rim$ and the resonance
shape (or rather width) to determine $\Vim$. Both parameters are short
range parameters reflecting that recombination requires all three
particles to be close to one another for the recombination to occur. These
parameters describe the short-range physics and are a way of including the
deeply bound states, that are otherwise unreachable using zero-range
models.

\section{The recombination coefficient}
\label{negative_a_rec_rate}
The loss of probability due to the complex potential is quantified using a
complex phase shift, $\theta+i\gamma$, between the incoming and outgoing
waves
\begin{equation}
  G = e^{2i(\theta+i\gamma)}H\;.
  \label{eq:5.1}
\end{equation}
The recombination probability is then $P(k) = 1-e^{-4\gamma}$ where
$\gamma$, which depends on energy, parametrizes the recombination. The
recombination coefficient, $\rec$, is obtained using \eqr{eq:2.88}
\begin{equation}
  \rec(a,E) = 4(2\pi)^23\sqrt3\frac{\hbar^5}{m^3}
    \frac{1-e^{-4\gamma}}{E^2}\;,
  \label{eq:5.2}
\end{equation}
where the wave number, $k$, has been replaced by the energy, $E$, for
convenience in the following discussion. At energies well below the
barrier height the complex phase shift, $\gamma$, is proportional to $E^2$
and the limit $E\rightarrow0$ can be safely taken (see appendix
\ref{appendixB} for details).

When the energy, $E$, of the incoming wave corresponds to the energy,
$E_0$, of a resonance state behind the barrier, the tunnelling rate is
greatly enhanced due to constructive interference \cite{sakurai}. This
means that there is an increased probability to reach the imaginary
potential where absorption occurs, correspondingly the recombination rate
has a resonant peak. At such a resonance the real part of the phase shift,
$\theta$, has an abrupt change in $\pi$, indicating the presence of a
resonance \cite{sakurai}. When the energy of the incoming wave is fixed
the only way to obtain resonant absorption is by changing the resonant
state energy $E_0$, which is done by changing the scattering length $a$.
For certain values of $a$, dubbed $\aminus_i$, the resonance energy $E_0$
lies at $0$. This corresponds exactly to the threshold value for creating
a bound trimer as discussed in the previous chapter. When the energy, $E$,
is very low it is at $\aminus$ that the recombination rate spectrum has
resonant peaks.

This picture of resonances leads us to an expression for the recombination
coefficient for finite energy parametrized by the Breit-Wigner
distribution
\begin{equation}
  \rec(a,E)=4(2\pi)^23\sqrt3\frac{\hbar^5}{m^3}
    \frac{K}{\left[E-E_0(a)\right]^2+\frac{1}{4}\Gamma^2\left(a\right)}\;,
  \label{eq:5.3}
\end{equation}
where the numerical factor is chosen for easy comparison to \eqr{eq:5.2}.
This expression exhibits the physical interpretation of tunnelling through
the barrier and subsequently subject to absorption and reflection at short
distance. The dimensionless constant $K$ depends only on the imaginary
potential. It is worth noting that the width $\Gamma$ cannot solely be
related to the lifetime of the resonance due to the barrier but also has a
contribution from decay due to the complex potential. When the barrier
height is negligible compared to the energy, $E\gg E_0,E\gg\Gamma$, all
resonance features are lost. In the opposite limit of small energy an
upper limit for the recombination coefficient is obtained. This is known
as the unitarity limit \cite{dincao2004}.

\begin{figure}[ht!]
  \centering
  \input{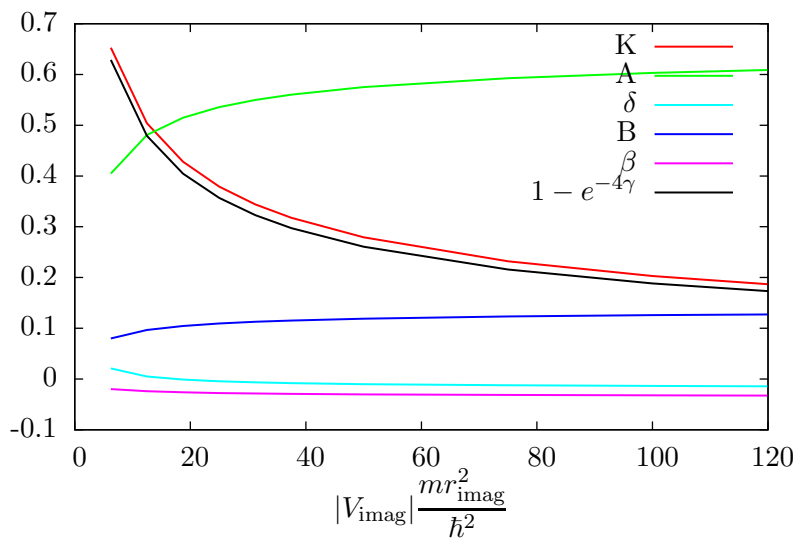}
  \caption[Optical potential model parameters]{The parameters of
  \eqr{eq:5.4} and \eqr{eq:5.5} as functions of the strength of the
  imaginary square-well potential. All the plotted quantities are
  dimensionless.}
  \label{figure:5.2}
\end{figure}

The next step in the parametrization is to find $K$, $E_0$, and $\Gamma$.
The choice of the numerical factor in \eqr{eq:5.3} immediately gives the
high-energy limit, $K\to 1-\exp(-4\gamma)$, where $E$ has to be large
compared to the other terms in the denominator of \eqr{eq:5.3}.
Numerically we find that the peaks, $\aminus_i$, in the recombination
coefficient follow nicely the Efimov scaling relations and that the
overall $a^4$ tendency is obeyed. This leads to the parametrization of
$\Gamma$ and $E_0$ as
\begin{align}
  \Gamma^2(a)\frac{m^2a^4}{\hbar^4}&=A\sin^2\!
    \left[s_0\ln\left(\frac{a}{a^{(-)}}\right)\right]+\delta\;,
  \label{eq:5.4}\\
  E_0(a)\frac{ma^2}{\hbar^2}&=B\sin^{\phantom{2}}\!
    \left[s_0\ln\left(\frac{a}{a^{(-)}}\right)\right]+\beta\;,
  \label{eq:5.5}
\end{align}
where $A,B,\beta$ and $\delta$ are constants that depend weakly on the
imaginary potential. This form ensures that both the $a^4$-rule and the
Efimov scaling are obeyed with periodic $22.7$ peak-recurrence in $\rec$.
The parameters, that are obtained by fitting \eqr{eq:5.3}, \eqr{eq:5.4}
and \eqr{eq:5.5} to the results from \eqr{eq:5.2} at some finite energy,
are plotted in~\fig{figure:5.2} as functions of the imaginary strength,
$|\Vim|m\rim^2/\hbar^2$. The coefficient $B$ is much smaller than $A$
meaning that $E_0$ is of little significance compared to $\Gamma$. The
variables $\beta$ and $\delta$ are also insignificant, since they are at
least a factor of $10$ smaller than $A$ and $B$. The low-energy dependence
of $\rec$ on energy is thus primarily determined by $K/\Gamma^2$. The
variations of the imaginary strength between $10$ and $120$ amount to only
about $10-20$~\%, except for $K$ which decreases by about a factor of 2.
As will be seen below, the experiments constrain the imaginary strength
variation interval to $\sim 10-70$.

\begin{figure}[ht!]
  \centering
  \input{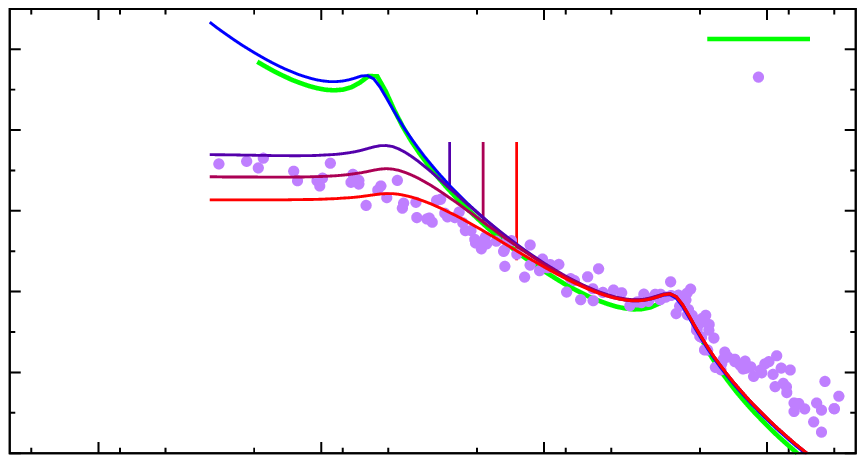}
  \caption[Finite temperature recombination in \Li7]{The recombination
  coefficient, $\rec$, at zero and finite temperature for \Li7 with the
  experimental data at a temperature of $1.5\mu$K \cite{hulet2013}. The
  scattering length, $a$, is in units of the Bohr radius, $a_0$. The
  parameter $a_C$ indicates the critical scattering length where the
  height of the barrier equals the mean energy of the atoms. At this value
  the spectrum starts to deviate from the $a^4$ behaviour, which
  corresponds quite nicely to the behaviour of the experimental data.}
  \label{figure:5.3}
\end{figure}

\subsection{Temperature effects}
\label{temperature_effects}
Since experiments are performed with fixed temperature, as opposed to
fixed energy, we average the finite energy calculations using the
normalized Boltzmann distribution for three particles, that is
\begin{equation}
  \langle\rec(a)\rangle_T=\frac{1}{2(k_BT)^3}\int
  E^2e^{-E/k_BT}\rec(a,E)\;dE\;,
  \label{eq:5.6}
\end{equation}
where the factor $E^2$ arises due to the phase-space for three particles.
The effect of temperature has been considered in other works such as
\cite{dincao2004} and \cite{braaten2008}. The integration can readily be
achieved with the parametrized expression in \eqr{eq:5.3}.

\begin{figure}[ht!]
  \centering
  \input{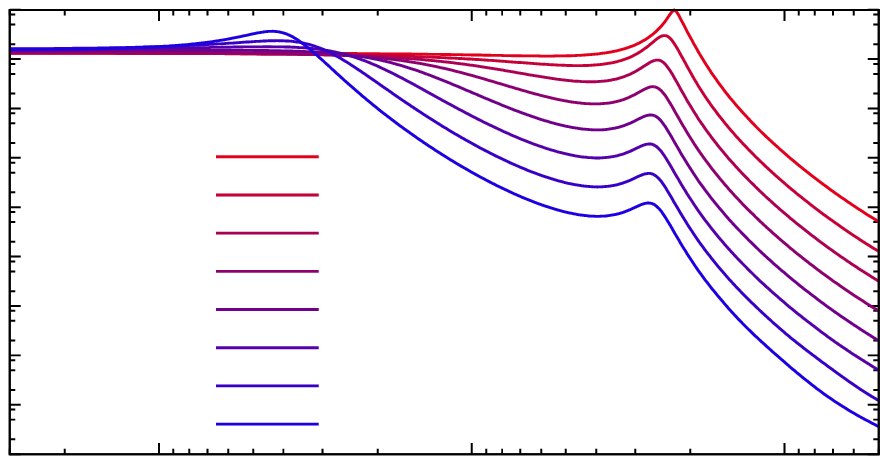}
  \caption[Decay parameter $\gamma$ for \Li7]{The decay parameter $\gamma$
  for the \Li7 system as a function of scattering length, $a$, for finite
    energies in temperature units. At large $|a|$ all curves have about
    the same value $\sim0.14$, independent of energy.}
  \label{figure:5.4}
\end{figure}

When the value of $|a|$ is increased the barrier location moves to large
$\rho$ while the barrier height is reduced. This means that the
high-energy limit is approached and an $a$-independent recombination rate
is obtained. The energy dependence in this limit is $1/E^2$ and the value
of $K$ determines the limiting value of $\rec$. We show in
\fig{figure:5.4} the calculated values of $\gamma$ as a function of $a$
for several finite energies (displayed in temperature units, however, a
temperature averaging of the form \eqr{eq:5.6} has not been performed
here). For small $|a|$ all $\gamma$-values are lowered when the energy
in increased, for large $|a|$ the energy and scattering length
independent constant of about $0.14$ is reached. This value depends on
the strength of the imaginary potential $m|\Vim|\rim^2/\hbar^2$, which
controls the height and shape of the absorption peaks as functions of
both $E$ and $a$. This numerical value is deceivingly similar to the
$\eta^-$ of \cite{hulet2013} used to fit the peak in \fig{figure:5.3}.
Formally there is also a connection although $\eta^-$ is more
complicated and derived through multiple scattering theory for zero
energy \cite{braaten2006}. The physical meaning is different from our
$\gamma$ and the expressions are not one-to-one related.

\begin{figure}[ht!]
  \centering
  \input{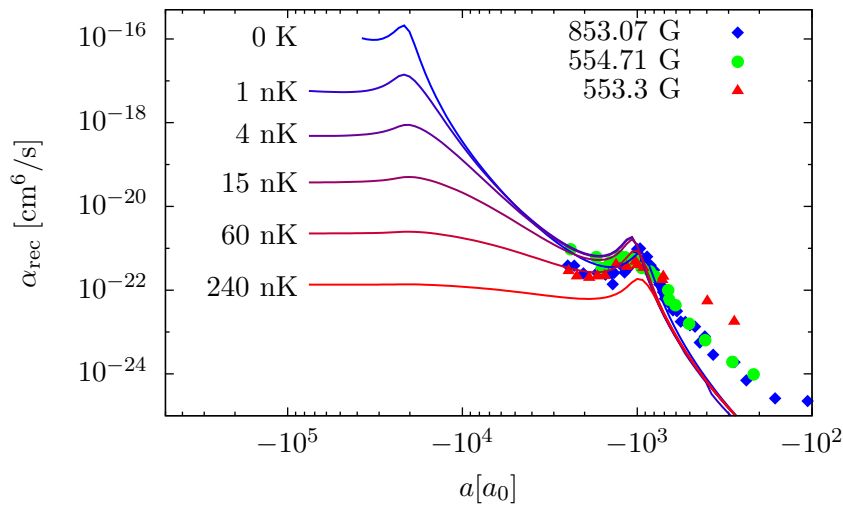}
  \caption[Recombination for finite temperature in \Cs]{The recombination
    coefficient $\rec$ at zero and finite temperatures for \Cs~with
    the experimental data taken at a temperature of about $15$~nK
    \cite{berninger2011}.}
  \label{figure:5.5}
\end{figure}

\section{Comparison to experimental data}
We now compare the available experimental data with numerical calculations
from the optical potential model. The numerical results and the
parametrization from \eqr{eq:5.3} are virtually indistinguishable. The
experimental recombination data for \Li7 \cite{hulet2013} along with the
calculations from our model at zero and finite temperatures are shown in
\fig{figure:5.3}. The only pronounced measured peak at $a \approx -280
a_0$ (where $a_0$ is the Bohr radius) is well described by our model. The
peak position is fitted with $\rim=0.41 a_0$ and the overall shape of the
peak is fitted with $\Vim=-68\hbar^2/ma_{0}^{2}$. For zero temperature we
find, for all $a$, almost precisely the same as the zero-energy formula of
Ref.~\cite{braaten2006} where $\eta^-=0.12$ and $a^{(-)}=-241a_0$
\cite{hulet2013}. At finite temperatures, we find the observed lowering of
recombination rates for large negative $a$. This flattening of $\rec$
appears for temperatures exceeding the barrier height, in other words for
$a^2>a_C^2\equiv0.143\hbar^2a_0^2/(mTk_B)$ as shown in \fig{figure:5.3}
for the indicated temperatures.

Recombination rates are also measured for \Cs~at $T\sim 15$~nK for three
different Feshbach resonances \cite{berninger2011} which show very similar
behaviour. These are shown in \fig{figure:5.5} along with our calculations
for different temperatures using $\rim=1.58 a_0$ and
$\Vim=-10\hbar^2/ma_{0}^{2}$. Our model reproduces the data for all three
resonances with the same model parameters. No data exists at
$a\sim-2\times10^4a_0$ where we predict another peak. From
\fig{figure:5.5} we conclude that a temperature below $\sim2$~nK seems to
be required to observe this peak clearly.

\section{Conclusions}
\label{conclusion6}
We present a simple and physically transparent model of three-body
recombination for negative scattering lengths that does not require a
short-range three-body cut-off. Instead it includes an imaginary potential
at short distance that takes decay into deeply bound dimers into account.
Full numerical solutions of the three-body equations were used to obtain
the recombination rate and subsequently a parametrization in terms of the
Breit-Wigner resonance formula was presented and shown to display the
expected scaling behaviour. Finally, it was shown how this new model
reproduces the experimental data on \Li7 and \Cs. If we express the radius
of the imaginary potential in units of the van der Waals length we find
$\rim/r_\textrm{vdW}=0.0063$ and $\rim/r_\textrm{vdW}=0.0078$
respectively, while the strength is $|\Vim|/V_\textrm{vdW}=2.87\cdot 10^5$
and $|\Vim|/V_\textrm{vdW}=4.08\cdot 10^5$ where
$V_\textrm{vdW}=\hbar^2/mr_\textrm{vdW}^2$. The similarity of $\rim$ and
$\Vim$ in van der Waals units indicates that there could be universality
hidden in these parameter. The differences that we find is most likely
related to the difference in deeply bound states of the two-body
potentials of \Li7 and \Cs.


\chapter{Mass-imbalanced systems}
\label{mass_imbalanced_systems}

\emph{This chapter considers recombination in systems of mixed species of
atoms for negative scattering lengths. The eigenvalue equation is
generalized to non-equal mass systems and the method of optical potentials
is applied.}

\noindent\rule{\textwidth}{0.4pt}

Systems of mixed species atoms like K-Rb \cite{PhysRevA.76.020701} and
Cs-Li \cite{PhysRevA.87.010701} have gotten quite a bit on interest lately
as they provide yet another window into the realm of few-body physics.
Furthermore, they posses some quite interesting features worthy of
investigation. Most notably, the Efimov scaling effect persists in these
system, however, the scaling parameter is reduced from the value 22.7 for
the three equal mass case. We show in the case of the Cs-Li system that
the scale factor is only 4.85, practically doubling the frequency of peaks
in the recombination coefficient as a function of the scattering length,
$a$. Being able to observe more than a single resonance peak is essential
to efficiently study the effects of the effective range as discussed
hitherto. Results in this chapter are, as of this writing, still in the
initial investigation phase.

\section{The mass-imbalanced eigenvalue equation}
\label{mass_imba}
\begin{figure}[h!]
  \centering
  \tikzsetnextfilename{fig6_1}
  \begin{tikzpicture}[inner sep=2mm,scale=1.0]
  \usetikzlibrary{calc}
  \usetikzlibrary{arrows}

  \coordinate (p1) at (.5,0.0);
  \coordinate (p2) at (0,-1.7);
  \coordinate (p3) at (2,-1.0);

  \draw[shading=ball,ball color=blue!50!red  ] (p1) node [label=180:$1$] {} circle (.25);
  \draw[shading=ball,ball color=blue!50!green] (p2) node [label=135:$2$] {} circle (.50);
  \draw[shading=ball,ball color=blue!50!green] (p3) node [label= 45:$3$] {} circle (.50);

  \draw[-triangle 60,very thick] ($0.5*(p2)+0.5*(p3)$) -- node  [left,black] {$\bm y_1$}  (p1);
  \draw[-triangle 60,very thick] (p2) -- node  [below,black] {$\bm x_1$} (p3);
\end{tikzpicture}
  \caption[A system of two heavy and one light atom]{The system of one
  light (small sphere) and two heavy (large spheres) atoms along with the
  Jacobi coordinates $\bm x_1$ and $\bm y_1$.}
  \label{figure:6.1}
\end{figure}
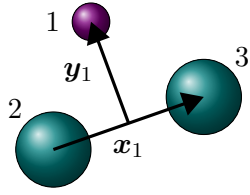

The derivation of \eqr{eq:2.55} relied on the assumption that all three
particles were identical, with the implication that the rotation angle
from \eqr{eq:2.51} was equal to $\frac{\pi}{3}$. Here we relax that
assumption and instead assume that the system consists of two kinds of
particles, one light particle and two heavy particles as shown in
\fig{figure:6.1}. For this system of particles the Jacobi-coordinates are
still given by \eqr{eq:2.34} but with, say, $m_2=m_3$. The arbitrary
scaling mass, $m$, is chosen to be the mass of the lighter of the atoms,
that is $m=m_1$.

The generalization of \eqr{eq:2.55} to mass-imbalanced systems is given by
setting the determinant of the matrix $M$ equal to zero where the elements
of $M$ are given by \cite{FedorovJensen2001}
\begin{subequations}\label{eq:6.1}
  \begin{align}
    M_{ii}&=\nu\cos\left(\frac{\nu\pi}{2}\right)+
      \frac{\rho}{\sqrt{\mu_i}a_i}\sin\left(\frac{\nu\pi}{2}\right)\;,\\
    M_{ij} &= \frac{2\sin\left[\nu\left(\phi_{ij}-\frac{\pi}{2}\right)
      \right]}{\sin\left(2\phi_{ij}\right)}\;,\qquad i\neq j\;,
  \end{align}
\end{subequations}
where $\phi_{ij}$ is given by \eqr{eq:2.51} and $a_i$ is the scattering
length between particles $j$ and $k$. The radial equation remains
unchanged. We will treat specifically the mixed system of one \Li6 atom
and two \Cs atoms. This implies $a_2=a_3$ following \fig{figure:6.1}. In
this case the matrix $M$ reduces to a $2\times2$-matrix. Furthermore, near
the Feshbach resonance of the Cs-Li system at $\sim850$~G
\cite{PhysRevA.87.010701} the interaction between the Cs atoms is
comparatively negligible and we set $a_1=0$ for simplicity and the
resulting eigenvalue equation becomes
\begin{equation}
  \frac{\nu\cos\left(\dfrac{\nu\pi}{2}\right)}
    {\sin\left(\dfrac{\nu\pi}{2}\right)} +
  \frac{2\sin\left[\nu\left(\phi-\dfrac{\pi}{2}\right)\right]}
    {\sin\left(2\phi\right)\sin\left(\dfrac{\nu\pi}{2}\right)}=
  \frac{\rho}{\sqrt\mu a}\;,
  \label{eq:6.2}
\end{equation}
where now $a=a_2$ is the scattering length between the sub-system of one
light and one heavy atom, $\mu=\frac{1}{m_1}\frac{m_1m_2}{m_1+m_2}$ is the
reduced mass of this subsystem and $\phi=\phi_{12}$ from \eqr{eq:2.51}.

\subsection{Properties of the mass-imbalanced potentials}
\label{mass_imbalanced_properties}
To investigate the effect of changing the masses we look at the location
of the zero-crossing of the radial potential $\frac{\nu_1^2-1/4}{\rho^2}$;
the location and height of the potential; the tail-behaviour of the
potential and of course the geometric scaling factor. We only consider
negative scattering lengths in this chapter. For positive scattering
lengths there is, of course, no barrier.

\begin{figure}[h!]
  \centering
  \input{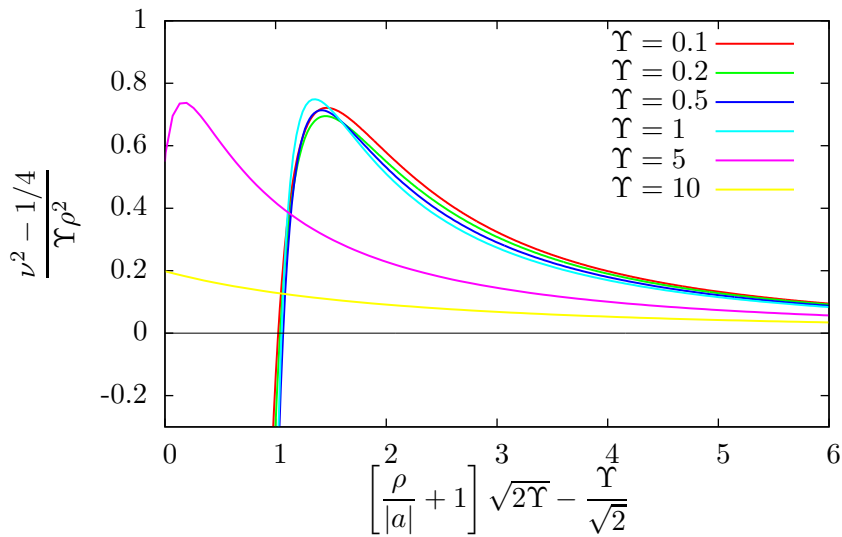}
  \caption[Adiabatic potentials for non-equal masses]{The hyper-radial
    adiabatic potentials for several mass ratios, $\R=m_1/m_2$. The
    somewhat unusual scaling of the abscissa comes from \eqr{eq:6.4} and
    ensures that the zero-crossing of the potentials are independent of
    mass ratios for $\R\leq1$. However, the denominator $\sqrt{2}$ in the
    final term does not come from \eqr{eq:6.4} but is included as it
    provides a better overall independence of $\R$ all the way up to $1$.
    For $\R>1$ the scaling is clearly not good, nor should it be expected
    to be so. Notice also how the height of the potentials scale quite
    well with $\R$.}
  \label{figure:6.2}
\end{figure}

The zero-crossing, $\rho_0$ for which $V(\rho_0)=0$, of the potential can
be found by inserting $\nu=\frac{1}{2}$ into \eqr{eq:6.2} with the result
\begin{equation}
  \frac{\rho_0}{\sqrt\mu a}=\frac{1}{2}+\frac{2\sqrt2}{\sin(2\phi)}
    \sin\left(\dfrac{\phi}{2}-\dfrac{\pi}{4}\right)\;.
  \label{eq:6.3}
\end{equation}
We define the ratio of masses to be $\R=m_1/m_2$. In the limit $\R\ll1$,
i.e. the mass of the light atom much smaller than the masses of the two
heavy atoms, the location of the zero-crossing is given approximately as
\begin{equation}
  \frac{\rho_0}{|a|}\approx\frac{1}{\sqrt{2\R}}-1+\sqrt{\frac{\R}{2}}\;,
  \label{eq:6.4}
\end{equation}
where $\mu=1/(1+\R)$ has been used. In \fig{figure:6.2} the adiabatic
potentials are plotted for several different mass ratios. The abscissa has
been transformed according to \eqr{eq:6.4} in such a way that all
potentials with small $\R$ cross zero at an $x$-value of about 1. The
expression \ref{eq:6.4} is accurate for $\R$ values up to unity. It must
be noted, however, that the mass ratio of 1 in this context does not
correspond to a system of three identical particles as described in the
previous chapters, since the interaction between two of the particles has
been specifically put to zero in this chapter.

The corresponding limit for large $\R$ is
\begin{equation}
  \frac{\rho_0}{|a|}=\frac{\sqrt{2}-1}{2 \sqrt{\R}}
  \left(1-\frac{1}{\sqrt{2\R}}\right)\;,
  \label{eq:6.5}
\end{equation}
but we do not consider these systems presently. One reason why these
systems are not as attractive is the fact that the Efimov scaling factor
actually increases, making it even more difficult to observe more than a
single recombination peak.

As seen in \fig{figure:6.2}, transforming the hyper-radial coordinate,
$\rho$, according to \eqr{eq:6.4} and simultaneously scaling the height of
the potential with the mass ratio $\R$ yields an almost mass-independent
curve for $\R\leq1$. This enables us to make some general observations
with regards to how the recombination coefficient must behave when the
mass ratio is changed.

\begin{figure}[h!]
  \centering
  \input{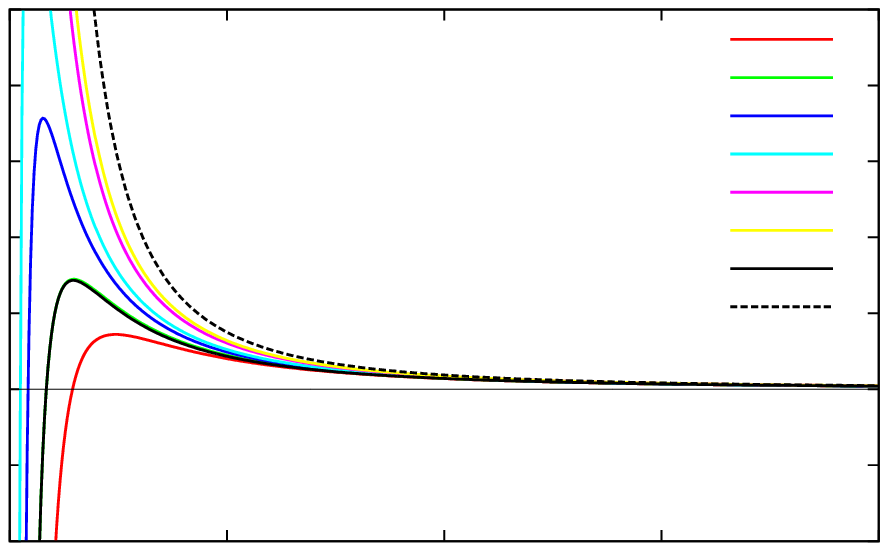}
  \caption[Adiabatic potential tails for non-equal masses]{The tail of the
    adiabatic potentials. Notice how all tails have the same asymptotic
    behaviour $V(\rho)=\dfrac{15}{4\rho^2}$. The curve for identical
    particles including the interaction between particles $2$ and $3$ in
    \fig{figure:6.1}, apparently corresponds quite well to the mass ratio
    $\R=0.2084$ in a system where the heavy particles do not interact.}
  \label{figure:6.3}
\end{figure}

\fig{figure:6.3} shows the tail behaviour of the adiabatic potentials. We
find that the long-range tail still has the form $V(\rho)=\dfrac{15}
{4\rho^2}$. Corrections due to the mass ratio show up only in terms of
order $1/\rho^3$ and higher. This means that for instance the probability
for tunnelling through the barrier in \fig{figure:6.2} still goes as $E^2$
for small energies.

The last step in our initial analysis of the mass-scaling properties of
the system in \fig{figure:6.1}, is to find the Efimov scaling factor,
which is found by solving \eqr{eq:6.2} for $\rho/a=0$ with $\nu=is_0$
where $s_0$ is real
\begin{equation}
  \cosh \left(\frac{\pi s_0}{2}\right)s_0\sin(2\phi)=
  2\sinh \left[s_0\left(\frac{\pi}{2}-\phi\right)\right]\;.
  \label{eq:6.6}
\end{equation}
The scale factor between consecutive minima then is $F=e^{\pi/s_0}$ as
stated in section \ref{efimov_and_thomas_effects}. In the case of the
Cs-Li system this yields $F=4.85$. For systems of three identical
particles the factor is $F=22.7$, however, this value is not obtained from
\eqr{eq:6.6} as this equation assumes no interaction between the two heavy
particles. If all three particles are assumed to interact, the first $2$
in the right hand side should be a $4$ instead. Since
$4.85^2=23.5\sim22.7$ there should be about twice as many resonance peaks
in the recombination coefficient for a given scattering length range for
the mixed Cs-Li system compared to a system of identical particles.

\section{Results}
\label{results6}
The overall $a^4$ scaling can easily be attributed, using a WKB
calculation, to the inner turning point given by \eqr{eq:6.3}. This is
seen by the following: assume for simplicity that the potential has the
form $V(\rho)=\nu_1^2/\rho^2$ for $\rho_0\leq\rho<\infty$ (the Langer
correction has been included) with $\nu_1=2$. This simple form is quite
reasonable as the potentials, as seen in \fig{figure:6.2}, drop quite
rapidly once the potential has reached its maximum value. The probability
of tunnelling through this barrier can be estimated using \eqr{eq:2.84}.
In the limit of $E\rightarrow0$ the integral takes the form (with
$m=\hbar=1$)
\begin{equation}
\gamma=\int_{\rho_0}^{\rho_t}\sqrt{\frac{\nu_1^2}{\rho^2}}\mathrm{d}\rho=
    \nu_1\ln\left(\frac{\rho_t}{\rho_0}\right)\;,
  \label{eq:6.123}
\end{equation}
where the outer classical turning point is $\rho_t=\nu_1/\sqrt{2E}$. The
tunnelling probability then reads
\begin{equation}
  P=\exp(-2\gamma)=\left(\frac{\rho_0}{\rho_t}\right)^2\propto
    E^2\rho_0^4\;.
  \label{eq:6.125}
\end{equation}
Since $\rho_0$ is proportional to the scattering length, $a$, this
establishes the $a^4$ relation. This also shows that the $a^4$ dependency
does not change for the mass-imbalanced systems. Moreover, the overall
rate of recombination is expected to increase for smaller mass ratios,
$\R$. For instance, for $\R=0.1$, \eqr{eq:6.4} yields
$\rho_0\approx1.46|a|$ and $1.46^4=4.5$, i.e a 4.5-fold increase in the
rate of recombination for any given scattering length.

This is, of course, a very simple analysis and more factors have to be
taken into account to obtain accurate results. For instance, the numerical
factor in \eqr{eq:5.2} might not be valid for the mass-imbalanced systems.
This is yet to be determined.

\begin{figure}[h!]
  \centering
  \input{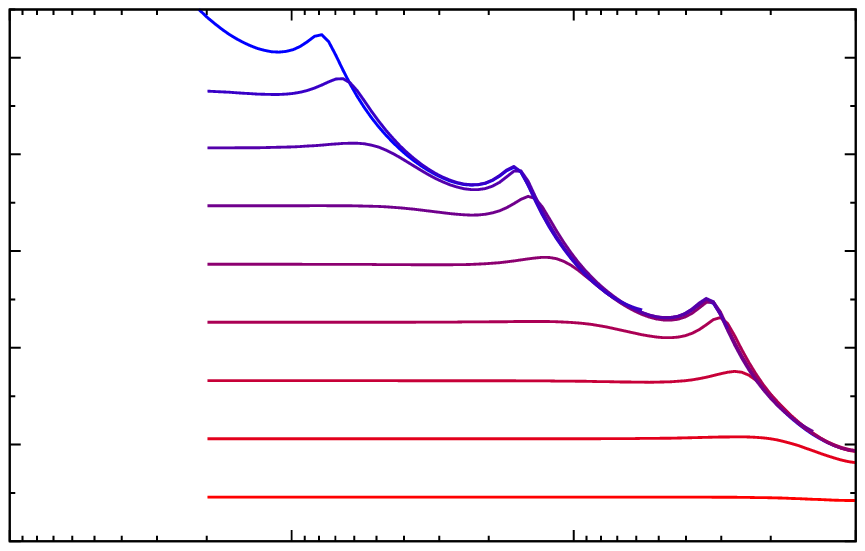}
  \caption[Recombination in a \Cs$-$\Li6 mixture]{The recombination
  coefficient, $\rec$, in a system of \Cs and \Li6 for negative scattering
  lengths, $a$, using \eqr{eq:5.2} and \eqr{eq:5.6} at zero and finite
  temperatures.}
  \label{figure:6.4}
\end{figure}

\fig{figure:6.4} shows an exploratory calculation of the recombination
coefficient for the Cs-Li system using the optical potential model from
the previous chapter as well as the energy averaging from
\eqr{eq:5.6} for several finite temperatures. The values of the depth,
$\Vim$, and width, $\rim$, of the imaginary potential have been reused
from \fig{figure:5.3}.

\section{Conclusions}
\label{conclusions6}
In this chapter we investigated the recombination coefficient for
mass-imbalanced systems. The hyper-radial adiabatic potential was
generalized to systems with two heavy, non-interacting atoms and one light
atom. Simple properties of these potentials led us to the prediction that
the overall magnitude of the rate of recombination should increase for
decreasing mass ratios $\R=m_1/m_2$. Of greater interest is the fact that
the Efimov scaling factor is reduced down from 22.7 to a mere 4.85 in the
case of the Cs-Li system. This makes the effective range dependency on the
scale factor a lot easier to verify experimentally as more resonance peaks
can be seen in a given range of scattering lengths as compared to systems
of three identical particles.

Additional study is required in this field, especially since experimental
work is progressing strongly in this direction in these years.


\chapter{Summary and Outlook}
\label{summary_and_outlook}
\noindent\rule{\textwidth}{0.4pt}
\\

\noindent This thesis has investigated effective range effects in
three-body recombination, bound state spectra and the three-body
parameter. The effective range is a correction to the well-known
scattering length approximation that is often used in cold gas physics. In
this thesis the effective range was implemented using a two-channel model
with zero-range interactions. This model was built on the physics of
Feshbach resonances and among other thing described the relation between
the scattering length, $a$, and applied magnetic field strength, $B$. The
effective range of a system of interacting particles is inversely
proportional to the width of the Feshbach resonance that is used to tune
the interaction strength in the system. For broad resonances the effective
range is rather small with corresponding small modifications to the
universal single-channel zero-range predictions. Feshbach resonances with
narrow widths are therefore required to be studied in order to accurately
confirm (or refute) the predictions of the effective range models in this
thesis.

The effective range takes into account the physical range of the
interaction instead of merely approximating it with a zero-range
potential. Investigating such effects will become more important in the
coming years as more and more experimental data with ever increasing
accuracy will become available. Thus deviations from the universal
zero-range theories are expected to become observable.
\\

Specifically, as we found in chapter \ref{finite_range_effects_chapter},
the two models that include the effective range, namely the two-channel
model and the effective range expansion model, predict that for the
recombination coefficient $\rec$ for positive scattering lengths the ratio
of scattering length values of consecutive minima is less than the
universal value which equals 22.7 for identical particles. We found the
origin of this effect to be a change in the atom-dimer threshold when the
effective range was included. The threshold obtains effective range
corrections that modify the values of the scattering length, $a^*$, where
the trimer and dimer energies coincide.

The method applied to calculate the recombination coefficient, namely the
hidden crossing method, has, to our knowledge, not been compared to the
experimental data before. Our comparison with available experimental data
proved that the method is quite good at providing reasonably accurate
predictions, in spite of the simplicity. Unfortunately the available data
is for systems of wide resonances and correspondingly short effective
ranges, wherefore the effects of the effective range are to small to see.
More data for narrow Feshbach resonance systems are required.
\\

Chapter \ref{3BP} was dedicated to the three-body parameter $\aminus$. For
zero-range interactions an additional length scale is required to prevent
the system from becoming infinitely bound. This length scale is known as
the three-body parameter and has been thought to depend on short-range
details of the physical potential. Quite surprisingly a universal relation
between the three-body parameter, $\aminus$, and the two-body van der
Waals length, $\vdW$, appears to exists across several different atomic
species. This relation was investigated using the single-channel
zero-range model and our two-channel model to include the effective range.
We found that the available experimental data is modelled well by our
calculations. We also found that for large effective ranges the ratio
$\aminus/\vdW$ is expected to be lower than the universal value of
$\sim9.8$. Again more study in narrow resonance systems is required to
confirm this prediction.
\\

In chapter \ref{optical_model} we investigated the recombination
coefficient for negative scattering lengths, $a$. The hidden crossing
method does not work for $a<0$ since weakly bound dimers do not exist for
negative $a$. Instead we employed an optical potential model where the
adiabatic potential, which is used in the radial differential equation,
gets a complex value $\Vim$ for small $\rho<\rim$. This trick effectively
emulates recombination into deep dimers and provides a quite accurate
reproduction of the experimental data. The model furthermore included
finite temperature effects which we showed did also corresponds nicely to
trends in the experimental data. To be able to see a second peak in the
data for \Li7 we predict that a temperature in the sub $\mu$K regime is
required.

The basic parameters of our optical model were the imaginary potential
strength, $\Vim$, and range, $\rim$. The parameters obtained from fitting
to the data revealed the quite surprising results $\rim/\vdW=0.0063$ for
\Li7 and $\rim/\vdW=0.0078$ for \Cs, the similarity in these numbers could
indicate some universal feature in the recombination process. Likewise for
the strengths we obtained $|\Vim|/V_\textnormal{vdW}=2.87\cdot10^5$ for
\Li7 and $|\Vim|/V_\textnormal{vdW}=4.08\cdot10^5$ for \Cs, where
$V_\textnormal{vdW}=\hbar^2/m\vdW^2$.

This chapter did not include effective
range effects but it is reasonable to assume that the conclusions from
chapter \ref{finite_range_effects_chapter} still hold true.
\\

Finally, chapter \ref{mass_imbalanced_systems} treated our most recent
results obtained for a system of non-identical particles. Specifically,
systems of one light atom and two heavy atoms, where the interaction
between the heavy atoms is negligible compared to the light-heavy
interaction, as is the case for \Li6 and \Cs, was treated. There are no
experimental data for recombination in such systems yet, but many groups
are reporting observations of Feshbach resonances in several mixed atom
systems \cite{PhysRevA.76.020701, PhysRevA.87.010701}. It is therefore
only a matter of time before the first experimental evidence of
recombination is obtained. We show that the most interesting feature in
these systems is that the geometric scaling factor, which equals 22.7 for
identical particle systems, is greatly reduced. For the Cs-Li system the
factor is only 4.85, which would allow twice as many trimer states in a
given range of scattering lengths as compared to systems of identical
particles. Thus more peaks/troughs in the recombination coefficient, are
expected to be present in a given range of scattering lengths. This is a
great boon as the scaling factor 22.7 is far too large to experimentally
observe enough resonances within currently attainable scattering length
ranges.
\\

The future clearly lies with the mass-imbalanced systems. The basics
outlined in chapter \ref{mass_imbalanced_systems} provide a stepping stone
for future works in this field. Combining the outlined method of optical
potentials with the two-channel and effective range expansion models
should be relatively straight forward such that the effective range
effects can be probed in the mixed species systems. Another thing to take
properly into account is the interactions between the heavy atoms, which
has been explicitly neglected for simplicity in the initial calculations.

\appendix

\chapter{Efficiently solving the eigenvalue equations}
\label{appendixA}
The eigenvalue equations \eqr{eq:2.55}, \eqr{eq:2.71} and \eqr{eq:2.77}
yield two equations in two unknowns when treating the eigenvalue as a
complex quantity, $\nu=\nu_\textrm{re}+i\nu_\textrm{im}$. The equations
need to be solved for $\rho$-values ranging from very large down to zero
and even for complex values when using the hidden crossing method of
section \ref{hidden_crossing_theory}.

\subsubsection{Quadratic extrapolation}
Efficiently solving these equations require good initial guesses for the
solver routine. Solving for $\nu(\rho)$ should always start at large
$\rho$ and work towards smaller $\rho$ since the asymptotic expressions
\eqr{eq:2.57} and \eqr{eq:2.59} are very accurate at large $\rho$. Once a
few points have been found, say $\nu_1,\,\nu_2$ and $\nu_3$ at
$\rho_1,\,\rho_2$ and $\rho_3$, respectively, a quadratic extrapolation
scheme can be used to guess the value of $\nu_4$.

Assume the following quadratic form
\begin{equation}
  \nu^2_\textrm{quad}(\rho)=a\rho^2+b\rho+c\;.
  \label{eq:b1}
\end{equation}
Since $\nu^2$ is a real value (except when also $\rho$ is complex) it is
better to extra\-polate the square of $\nu$ than $\nu$ itself. This is
straightforward to solve for $a,\,b$ and $c$ in terms of $\nu_1,\,\nu_2$
and $\nu_3$ and hence to obtain $\nu_4$. This method yields a much better
initial guess for the next step than simply using the previous value
($\nu_3$ in this case) and lowers the number of iterations required by a
factor of $\sim1.5$ with an overall speed increase of $\sim25\%$. Another
benefit is that the change in $\nu$ from purely real to purely imaginary,
i.e. $\lambda(\rho)$ changes form positive to negative or vice versa (only
possible when $a<0$) is handled gracefully by the numerical solver.

In the case of fixed step size, the quadratic extrapolation simply yields
\begin{equation}
  \nu_4^2 = \nu_1^2+3(\nu_3^2-\nu_2^2)\;,
  \label{eq:b2}
\end{equation}
which is easily calculated at a low extra cost in computation.

\subsubsection{Adaptive step size}
To further speed up the calculation an adaptive step size can be
implemented. In case of the effective range expansion and two-channel
models, the eigenvalue solutions show a sharp curvature feature when
$\rho\lesssim|R|$ as seen in \fig{figure:3.1}. To resolve this region
without manually having to specify an appropriate density of points an
adaptive step size controller is used.

The local error estimate is taken to be the absolute of the difference
between the actual solution $\nu(\rho)$ and the guess provided by
\eqr{eq:b1}. Note that errors will not accumulate, the solution is as good
at any one point than it is at any other since the guess is only used as
an initial value.

Given the local error, $\epsilon=|\nu_i-\nu_{i,\textrm{guess}}|$, and
tolerance, $\tau$, (typically $10^{-4}$ give good results) the size of the
next step, $h_\textnormal{new}$, given the current step size,
$h_\textnormal{old}$, is
\begin{equation}
  h_\textnormal{new} =
  \begin{cases}
    \min\left[2h_\textnormal{old},h_\textnormal{max}\right]&
      \epsilon<\tau\\
    \max\left[0.95h_\textnormal{old}\left(\dfrac{\tau}{\epsilon}\right)
      ^{1/4}, h_\textnormal{min}\right]&\epsilon>\tau
  \end{cases}
  \label{eq:b3}
\end{equation}
where $h_\textnormal{min}$ and $h_\textnormal{max}$ are chosen
to avoid inappropriately large or small step sizes.

The greatest benefit of these method is not so much an increase in speed
but instead an increase in the number of points where the solution varies
rapidly, resulting in smoother curves to be used for further numerical
work.


\chapter{Low-energy limit of recombination probability}
\label{appendixB}
In section \ref{negative_a_rec_rate} we took the limit $k\rightarrow0$ to
obtain the recombination rate at zero energy using numerical calculations
for finite energies. This limit was of the form
\begin{equation}
  \rec\propto\lim_{k\rightarrow0}\frac{1-e^{-4\gamma}}{k^4}\;.
  \label{eq:B.1}
\end{equation}
For this limit to be finite we must have $1-e^{-4\gamma}\propto k^4$, or
equivalently $\gamma\propto k^4$, for small $k$. In this appendix it is
proven that this limit exists by considering a simpler version of the
potential with the same qualitative properties.

The adiabatic potential in the $n=1$ channel is $V(\rho) =
\dfrac{15}{4\rho^2}$ for large $\rho$ since $\nu\rightarrow2$ in this
limit, according to \eqr{eq:2.59}. Consider the potential
\begin{equation}
  V(\rho) =
  \begin{cases}
    V_0                        & \textnormal{for } \rho<r_0\\[.5em]
    \dfrac{\nu^2-1/4}{2\rho^2} & \textnormal{for } \rho>r_0
  \end{cases}\;,
  \label{eq:B.2}
\end{equation}
where $\nu = 2$ in the present problem and $V_0$ is a complex constant.
The units are $\hbar = m = 1$. This is a simplified version of
\fig{figure:5.1} where the $1/\rho^2$ behaviour has been extended down to
$r_0$ (which plays the role of $\rim$ in the figure). For $\rho<r_0$ the
solution with the boundary condition $f(0)=0$ is
\begin{equation}
  f_<(\rho) = \sin(\kappa\rho)\;,\qquad\kappa = \sqrt{2(E-V_0)}\;.
  \label{eq:B.3}
\end{equation}
 For $\rho>r_0$ the general solution is
(\cite{abramowitz} 9.1.49)
\begin{equation}
  f_>(\rho) = \sqrt{\rho}\left(
      AH^{(2)}_\nu(k\rho)+BH^{(1)}_\nu(k\rho)\right)\;,
  \label{eq:B.4}
\end{equation}
where $H^{(1)}_\nu$ and $H^{(2)}_\nu$ are Hankel functions of the first
and second kind and $k = \sqrt{2E}$. In the limit of large $\rho$,
$\sqrt{\rho}H_\nu^{(1)}$ behaves like an outgoing wave $\propto
e^{ik\rho}$ while $\sqrt{\rho}H_\nu^{(2)}$ behaves like an incoming wave
$\propto e^{-ik\rho}$.

In the limit of small $z$ the Hankel functions behave like
\begin{align}
  H_\nu^{(1,2)}(z)&\approx
    \frac{1}{\Gamma(\nu+1)}\left(\frac{z}{2}\right)^\nu\pm
    \frac{\Gamma(\nu)}{\pi i}\left(\frac{2}{z}\right)^\nu\;.
  \label{eq:B.5}
\end{align}
To stitch together the solutions from \eqr{eq:B.3} and \eqr{eq:B.4} at the
boundary $\rho = r_0$ we need the derivative of the Hankel functions (from
\cite{abramowitz} 9.1.30)
\begin{equation}
  \frac{d\,\mathcal H_\nu(z)}{dz} = \mathcal
  H_{\nu-1}(z)-\frac{\nu}{z}\mathcal H_\nu(z)\;,
  \label{eq:B.6}
\end{equation}
where $\mathcal H$ is any of the Hankel functions (incidentally the
expression is valid for Bessel and Neumann functions as well).

By the requirements that the wave function and its derivative must be
continuous at the boundary $\rho=r_0$ we get the coefficients $A$ and $B$
as
\begin{align}
  A & = -\sqrt{r_0}\left[H_{\nu}^{(1)}\left(f_0-2f'r_0\right)+
f_0\left(H_{\nu-1}^{(1)}-H_{\nu+1}^{(1)}\right)r_0 k\right]D^{-1}\;,
  \label{eq:B.7}\\
  B & = \phantom{-}\sqrt{r_0}\left[H_{\nu}^{(2)}\left(f_0-2f'r_0\right)+
f_0\left(H_{\nu-1}^{(2)}-H_{\nu+1}^{(2)}\right)r_0 k\right]D^{-1}\;,
  \label{eq:B.8}
\end{align}
with
\begin{equation}
  D = \left[H_{\nu}^{(1)}\left(H_{\nu-1}^{(2)}-
    H_{\nu+1}^{(2)}\right)+\left(H_{\nu+1}^{(1)}-
    H_{\nu-1}^{(1)}\right)H_{\nu}^{(2)}\right]k\;,
  \label{eq:B.9}
\end{equation}
where $f_0 = f_<(r_0)$ and $f_0' =
\left.\dfrac{df_<}{d\rho}\right\vert_{r_0}$. The denominator $D$ is not
really needed to obtain the ratio of $B$ to $A$ but included nevertheless
for the sake of completeness.

All this was for a general $\nu$. Now we use the specific value $\nu = 2$
appropriate for our problem. With \eqr{eq:B.5} to get the low-energy limit
the probability of recombination is
\begin{equation}
  1-e^{-4\gamma} = 1-\left|\frac{B}{A}\right|^2 = 2\pi(r_0
  k)^4\textnormal{Im}\left(\frac{5f_0-2f_0'r_0}{48f_0+32f_0'r_0}\right)\;.
  \label{eq:B.10}
\end{equation}
At low energy the probability for recombination is proportional to $k^4$
and the limit \eqref{eq:B.1} can be safely taken. It is the asymptotic
form of the potential $V(\rho)=15/4\rho^2$ that gives this result, and the
conclusion is therefore also valid for the actual adiabatic potential, not
just for the toy model in \eqr{eq:B.2}.

Incidentally the power of 4 is due to $\nu=2$, the general form is
$\rec\propto k^{2\nu}$. We note also from \eqr{eq:B.10} that only if the
ratio $f_0'/f_0$ is complex will the probability be non-zero. This can
only happen if the parameter $V_0$ is itself complex.

Slightly less obvious is that for the probability for recombination to be
positive we need the imaginary part of $V_0$ to be negative. This can most
easily be seen by plotting the expression \eqr{eq:B.10} as function of the
imaginary depth. A positive value of the imaginary part of $V_0$ would
yield a negative probability, which in this case should be interpreted as
particles being created at short distance, i.e. the reverse process of
what we want to study. Therefore, we let $V_0$ have a negative imaginary
part.

\backmatter


\renewcommand\bibname{Bibliography}
\bibliographystyle{thesisstyle}
\bibliography{data/bibliography}

\end{document}